\documentclass[aps,prx,amsmath,amssymb,floatfix,twocolumn,superscriptaddress,nofootinbib,10pt]{revtex4-2}

\usepackage{graphicx}
\usepackage{dcolumn}
\usepackage{bm}
\usepackage{hyperref}

\DeclareFontFamily{U}{mathb}{\hyphenchar\font45}
\DeclareFontShape{U}{mathb}{m}{n}{
      <5> <6> <7> <8> <9> <10> gen * mathb
      <10.95> mathb10 <12> <14.4> <17.28> <20.74> <24.88> mathb12
      }{}
\DeclareSymbolFont{mathb}{U}{mathb}{m}{n}
\DeclareFontSubstitution{U}{mathb}{m}{n}
\DeclareMathSymbol{\circlearrowleft}       {3}{mathb}{"F6}
\DeclareMathSymbol{\circlearrowright}      {3}{mathb}{"F7}
\DeclareMathSymbol{\curvearrowleftright}   {3}{mathb}{"F2}

\newcommand{\inter}{\curvearrowleftright}

\newcommand{\intra}{{\mathchoice
    {
        \hspace{-0.7ex}
        {\mathrel{{\ooalign{\hss\raisebox{1.2ex}{
            \rotatebox{180}{$\circlearrowright$} 
            }\hss\cr\raisebox{1.2ex}{
            \rotatebox{180}{$\circlearrowleft$} 
            }}}}
        \hspace{-0.7ex}
        }
    }
    {
        \hspace{-0.7ex}
        {\mathrel{{\ooalign{\hss\raisebox{1.2ex}{
            \rotatebox{180}{$\circlearrowright$} 
            }\hss\cr\raisebox{1.2ex}{
            \rotatebox{180}{$\circlearrowleft$} 
            }}}}
        \hspace{-0.7ex}
        }
    }
    {
        \hspace{-1.89ex}
        {\mathrel{{\ooalign{\hss\raisebox{0.9ex}{
            \rotatebox{180}{
                \scalebox{0.64}{
                    $\circlearrowright$
                }
            } 
            }\hss\cr\raisebox{0.9ex}{
            \rotatebox{180}{
                \scalebox{0.64}{
                    $\circlearrowleft$
                }
            } 
            }}}}
        \hspace{-1.89ex}
        }
    }
    {
        \hspace{-1.3ex}
        {\mathrel{{\ooalign{\hss\raisebox{0.7ex}{
            \rotatebox{180}{
                \scalebox{0.6}{
                    $\circlearrowright$
                }
            } 
            }\hss\cr\raisebox{0.7ex}{
            \rotatebox{180}{
                \scalebox{0.6}{
                    $\circlearrowleft$
                }
            } 
            }}}}
        \hspace{-1.3ex}
        }
    }
}
} 

\usepackage{stackengine}  
\usepackage{amsmath} 
\usepackage{amssymb} 
\usepackage{mathrsfs} 
\usepackage{enumitem} 
\usepackage{hyperref}
\usepackage{color} 
\usepackage{braket}

\newcommand{\pH} { \mathfrak{p} }

\begin{document}

\title{Non-Hermitian  topological superconductivity with symmetry-enriched spectral and eigenstate features
}

\author{Chuo-Kai Chang}
\affiliation{Institute of Physics, Academia Sinica, Taipei 115201, Taiwan}
\affiliation{Department of Physics, National Taiwan Normal University,  Taipei 10677, Taiwan}

\author{Kazuma Saito}
\affiliation{Institute of Physics, Academia Sinica, Taipei 115201, Taiwan}
\affiliation{Department of Applied Physics, Tokyo University of Science, Katsushika, Tokyo 125-8585, Japan}

\author{Nobuyuki Okuma}
\affiliation{Department of Basic Sciences,
Kyushu Institute of Technology,
Sensui-cho 1-1, Tobata, Kitakyushu, Fukuoka, 804-8550, Japan
}

\author{Hsien-Chung Kao}
\affiliation{Department of Physics, National Taiwan Normal University,  Taipei 10677, Taiwan}

\author{Chen-Hsuan Hsu}
\affiliation{Institute of Physics, Academia Sinica, Taipei 115201, Taiwan}
\affiliation{Physics Division, National Center for Theoretical Sciences, Taipei 106319, Taiwan}
  
\date{\today}

\begin{abstract}

We investigate a one-dimensional superconducting lattice that realizes all internal symmetries permitted in non-Hermitian systems, characterized by nonreciprocal hopping, onsite dissipation, and $s$-wave singlet pairing in a Su-Schrieffer-Heeger-type structure. 
The combined presence of pseudo-Hermiticity and sublattice symmetry imposes constraints on the energy spectra. We identify parameter regimes featuring real spectra, purely imaginary spectra, complex flat bands, and Majorana zero modes, the latter emerging when a uniform transverse magnetic field suppresses the non-Hermitian skin effect. We show that a uniform component of the onsite dissipation is essential for stabilizing the zero modes, whereas a purely staggered dissipation destroys the topological superconductivity.
Through Hermitianization, we construct a spectral winding number as a topological invariant and demonstrate its correspondence with the gap closing conditions and appearance of the Majorana zero modes, allowing us to establish topological phase diagrams.
Moreover, we reveal nontrivial correlations between the particle-hole and spin components of left and right eigenstates, enforced by chiral symmetry, pseudo-Hermiticity, and their combination. Our results highlight how non-Hermiticity, sublattice structure, and superconductivity together enrich symmetry properties and give rise to novel topological phenomena.

\end{abstract}

\maketitle

\section{\label{sec:level1} Introduction }
 
The exploration of topological phases has become a cornerstone of modern condensed matter physics, driven by both fundamental interest and the promise of novel quantum technologies~\cite{Hasan:2010,Qi:2011}. Among these phases, topological superconductivity stands out for its potential to host Majorana zero modes~\cite{Kitaev:2001,Fu:2008,Sato:2009,Hasan:2010,Lutchyn:2010,Oreg:2010,Qi:2011,Alicea:2012,Klinovaja:2012,Sato:2017,Hsu:2018,HAIM:2019,Hsu:2021}, non-Abelian quasiparticles with promising applications in fault-tolerant quantum computation~\cite{Ivanov:2001,Nayak:2008,Alicea:2010,Jay:2010,Beenakker:2013,Sarma:2015,Beenakker:2020}. A paradigmatic example is the Kitaev chain~\cite{Kitaev:2001}, where $p$-wave pairing gives rise to boundary-localized Majorana modes protected by a topological invariant. Recent developments have shown that similar physics can be effectively engineered in quantum-dot-based architectures, sometimes referred to as ``poor man’s Majorana setups''~\cite{Leijnse:2012,Dvir:2023}, illustrating how lattice models can successfully emulate realistic nanoscale structures.

A complementary route to topological matter originates from nonsuperconducting systems. A prominent example is the Su-Schrieffer-Heeger (SSH) model~\cite{SSH:1979}, where nontrivial topology arises from chiral symmetry associated with sublattice degrees of freedom. This symmetry protects zero-energy edge modes and defines a winding number as a topological invariant~\cite{Sticlet:2014,Rachel:2018,Yu:2020}. The SSH model has since become a canonical framework for exploring symmetry-protected topological phases across various platforms. These developments reflect a deeper organizing principle: topological phases are classified according to their internal symmetries, as formalized by the Altland-Zirnbauer (AZ) and tenfold classification schemes for Hermitian Hamiltonians~\cite{Altland:1997,Ryu:2010}, highlighting the rich interplay between symmetry and topology. 
 
More recently, the landscape of topological phases has broadened to encompass non-Hermitian systems~\cite{Bender:2003,Bender:2007,Gong:2018,Hamazaki:2019,Kawabata:2019,Yokomizo:2019,Borgnia:2020,Lee:2020,Torres:2020,Delplace:2021,Garcia-Garcia:2022,Sakaguchi2022,Yoshida:2022,Okuma:2023,Yoshida:2023,Yoshida:2024,Monkman:2025,YPW:2025}, which provide a natural framework for describing quantum systems with asymmetric (nonreciprocal) hopping~\cite{Hatano:1996,Hatano:1997}, open-system dynamics involving energy or particle exchange with the environment~\cite{Zeuner:2015,Zhen:2015,Hayden:2022,Hamanaka:2024}, or evolution conditioned on measurement postselection~\cite{Gopalakrishnan:2021}. This extension has uncovered a range of phenomena with no Hermitian counterparts. A hallmark example is the non-Hermitian skin effect~\cite{Yao:2018,Borgnia:2020,Okuma:2020,Zhang:2021}, 
in which a macroscopic number of eigenstates accumulate near system boundaries under open boundary conditions (OBC), violating the conventional bulk-boundary correspondence~\cite{Xiong:2018}.

To account for these unconventional features, new classification frameworks have been developed, incorporating topological invariants defined over complex energy spectra~\cite{Kawabata:2019,YPW:2025} and formulated within the non-Bloch band theory~\cite{Yokomizo:2019}. These tools enable a robust classification of non-Hermitian topological phases, now actively studied across diverse physical platforms~\cite{Rotter:2015,Yoshida:2019,Budich:2020,Helbig:2020,Midya:2021}. In parallel, extensions of the SSH model to non-Hermitian regimes have revealed rich connections between symmetry and topology, particularly in relation to parity-time   symmetry and the winding of complex energy spectra around exceptional points~\cite{Rotter:2009,Halder:2022,Ye:2024}.

A remarkable development in recent years is the integration of topological superconductivity and non-Hermiticity, offering a fertile ground for quantum states that are inaccessible in Hermitian counterparts~\cite{Okuma:2019,Lieu:2019,Avila:2019,Zhou:2020,Sakaguchi2022,Gandhi:2024,Cayao:2025}.  
In particular, Ref.~\cite{Okuma:2019} extended the Hatano-Nelson model~\cite{Hatano:1996,Hatano:1997} to include dissipation and superconducting pairing. This minimal setup revealed the existence of Majorana zero modes, demonstrating that non-Hermiticity and particle-hole symmetry can jointly lead to unusual boundary-localized states. Such findings raise important theoretical questions regarding the classification and stability of topological excitations in non-Hermitian superconductors.

Building on this foundation, it becomes natural to ask the question of how the interplay of non-Hermiticity, superconductivity, and additional internal symmetries shapes the spectral and topological structure of quantum systems. This question becomes particularly compelling when one considers sublattice symmetry (SLS) and/or chiral symmetry (CS), which are known to take on qualitatively new roles when non-Hermiticity is introduced. 

In this work, we investigate a lattice model {\it that incorporates all internal symmetries} permitted in non-Hermitian systems. Specifically, the system features sublattice degrees of freedom and onsite $s$-wave spin-singlet pairing, with the normal (nonsuperconducting) part described by a generalized non-Hermitian SSH model that includes both nonreciprocal hopping and onsite dissipation.
Including an onsite transverse magnetic field, we identify the system as belonging to the BDI symmetry class in the real AZ classification. The interplay between pseudo-Hermiticity (pH) and SLS imposes constraints on the energy spectrum, enabling a systematic analysis of the non-Hermitian topology.

\begin{figure}[t]
\centering
\includegraphics[width=\linewidth]{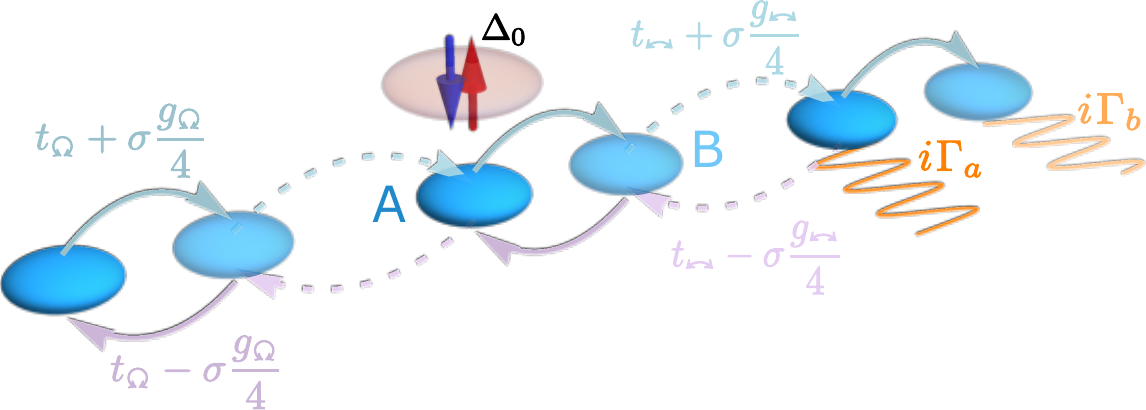}
\caption{Illustration of the setup described by Eq.~\eqref{Eq:H_nHSC}: a one-dimensional lattice  with sublattice sites (labeled by A and B) and onsite $s$-wave singlet pairing (ellipse) with pairing strength $\Delta_0$. 
The curly arrows indicate the spin-dependent nonreciprocal hopping strengths for particles moving to the right ($t+\sigma g /4$) and to the left ($t-\sigma g /4$) for spin $\sigma \in \{\uparrow \, {\rm (blue \, arrow)}, \, \downarrow \, {\rm (red \, arrow)} \}$.
The subscripts $\protect\intra$
and $\inter$ denote intra- and inter-unit cell processes, respectively.
The wavy curves represent onsite dissipation terms, with $\Gamma_{a}/2$ and $\Gamma_{b}/2$ corresponding to sublattices A and B, respectively.
}
\label{Fig:illustration_nHSC}
\end{figure}

By exploring a broad parameter space, we uncover a variety of spectral features, including regions with real or purely imaginary spectra, complex flat bands, and the emergence of Majorana zero modes when the skin effect is suppressed by the transverse magnetic field. Our analysis reveals that a uniform onsite dissipation component is essential for stabilizing Majorana zero modes, while a purely staggered dissipation component suppresses the corresponding topological region. Furthermore, we demonstrate a correlated structure between the particle and hole components of the left and right eigenstates of the zero modes, reflecting the constraints imposed by CS, pH, and their combination. 
Finally, we establish the winding number through the Hermitianizaed Hamiltonian as a topological invariant, mapping out the phase diagrams in broad parameter space.  
Our findings highlight how the combination of non-Hermiticity, sublattice structure, and superconductivity leads to enriched topological properties in spectral and eigenstate features.  

The rest of this article is organized as follows. In Sec.~\ref{sec:Hamiltonian}, we introduce the Hamiltonian, examine its symmetries, and classify it within the framework of non-Hermitian topology. In Sec.~\ref{Sec:PBC}, we discuss spectral properties under the periodic boundary condition (PBC), including the conditions under which the gap closes and those where the system exhibits gapless superconductivity. We also describe spectral features such as complex flat bands and real or purely imaginary bands, and identify the conditions for their appearance. In Sec.~\ref{sec:OBC spectra}, we analyze the OBC spectra and demonstrate the emergence of Majorana zero modes when the system is subject to an onsite magnetic field. 
We further construct a winding number, connected to the appearance of the zero modes. 
In Sec.~\ref{Sec:density}, we investigate the density profile of the Majorana zero modes, and uncover the symmetry-enabled relation between the components of the right and left eigenstate wavefunctions. 
In Sec.~\ref{sec:non-uniform}, we consider nonuniform onsite dissipation terms and establish the gap closing conditions.
In Sec.~\ref{Sec:winding_general}, we consider general settings and construct topological phase diagrams in broad parameter space. 
Finally, in Sec.~\ref{Sec:discussion}, we summarize our findings and provide additional discussions, including the connection of our model and experimental realizations.
In Appendix~\ref{Appendix:no_hx}, we present details about the symmetries and energy spectra in the absence of the onsite transverse magnetic field. 
In Appendix~\ref{Appendix:IGT}, we perform the imaginary gauge transformation and derive the energy spectra under both PBC and OBC in the absence of onsite transverse magnetic fields.  
In Appendix~\ref{Appendix:Winding}, we show the spectral trajectories of a block
Hamiltonian and demonstrate that the results are consistent with the introduced winding number.
In Appendix~\ref{Appendix:parameter}, we summarize the adopted parameter values throughout this work in Table~\ref{Table:Parameters}.

\section{Hamiltonian}
\label{sec:Hamiltonian}

We introduce our system illustrated in Fig.~\ref{Fig:illustration_nHSC}, which describes a non-Hermitian superconductor with spin-dependent nonreciprocal hopping, onsite dissipation and pairing,  
\begin{widetext}
\begin{equation}
\begin{split}
\mathcal{H}_{\rm nHSC} &=\sum_{j,\sigma=\pm}
\Bigg[ 
-\left(t_{\intra}-\sigma\frac{g_{\intra}}{4} \right) a_{j,\sigma}^{\dagger}b_{j,\sigma}
-\left(t_{\intra}+\sigma\frac{g_{\intra}}{4} \right) b_{j,\sigma}^{\dagger}a_{j,\sigma}
-\left(t_{\inter}+\sigma\frac{g_{\inter}}{4} \right) a_{j+1,\sigma}^{\dagger}b_{j,\sigma}
-\left(t_{\inter}-\sigma\frac{g_{\inter}}{4} \right) b_{j,\sigma}^{\dagger}a_{j+1,\sigma}
\\
& \hspace{36pt}
-\frac{i}{2} \left( \Gamma_{a} a_{j,\sigma}^{\dagger}a_{j,\sigma}+ \Gamma_{b} b_{j,\sigma}^{\dagger}b_{j,\sigma} \right) 
\Bigg]
+\sum_{j}\left(\Delta_{0} a_{j,\uparrow}^{\dagger}a_{j,\downarrow}^{\dagger}+\Delta_{0} b_{j,\uparrow}^{\dagger}b_{j,\downarrow}^{\dagger}+\mathrm{H. c.}\right) . 
\label{Eq:H_nHSC}
\end{split}
\end{equation}
\end{widetext}
In the above, $a$ and $b$ ($a^{\dagger}$ and $ b^{\dagger}$) are the annihilation (creation) operators for electrons on sublattices $A$ and $B$, respectively. The real parameters
$t_{\intra}$ and $g_{\intra}$ denote symmetric and antisymmetric hopping strengths within a unit cell, respectively, whereas  $t_{\inter}$ and $g_{\inter}$ denote those between sites in different unit cells. Finally, we introduce onsite dissipation terms with $\Gamma_{a,b}$ (for sublattices A and B, respectively) and $s$-wave singlet pairing with pairing strength $\Delta_0$.
Without loss of generality, we will assume $\Delta_0, \Gamma_a \geqslant 0$ , while retaining the sign freedom of the other parameters in the following analysis.

In terms of experimental realization, ultracold Fermi gas systems provide a promising platform when the required ingredients are assembled properly in one-dimensional lattices with sublattice structure. In particular, a non-Hermitian SSH model with energy gain and the observation of topological edge states has been realized recently in cold atoms~\cite{Zhao:2025}. Dissipative non-Hermitian dynamics can also be engineered and controlled, as demonstrated in Refs.~\cite{Li:2020,Tsuno:2025}. Non-reciprocal hopping may be implemented using asymmetric ring configurations, where dissipation and non-Hermiticity coexist and skin modes have been directly observed~\cite{Gou:2020,Liang:2022}. Furthermore, effective pairing interactions can be induced in Fermi gases via Feshbach resonances, providing a natural route to realizing the pairing terms considered here~\cite{Chin:2004,Chin:2010}. 
The integration of the above elements allows for the realization of the physical setup that we analyze here. 

Compared to the model considered in Ref.~\cite{Okuma:2019}, here we introduce the sublattices, which are known to influence the symmetry properties. Equivalently, this replaces the Hatano-Nelson terms in the normal part with a non-Hermitian SSH model, where the non-Hermiticity arises from both the non-reciprocal hopping and dissipation terms.  
We will see that the incorporation of the sublattices can alter the symmetries and therefore influence the topological properties.

\begin{table*}[t]
\centering
\caption{Symmetry relations, matrix representations,
and the corresponding unitary operators for our system. 
Here, the operator $H$ represents the full Hamiltonian
$ {H}_{\rm nHSC}^{\rm pbc} (k) 
+ \delta h_x \eta^z \tau^0 \sigma^{x}$, with $ {H}_{\rm nHSC}^{\rm pbc} $ defined in Eq.~\eqref{Eq:H_pbc2}  and  $U$ denotes certain unitary operators fulfilling the corresponding relations.
}
\begin{tabular}{l c l l }
\hline
\hline
\textbf{Symmetry type (abbreviations)} & \textbf{Symbol} & \textbf{Relation~\cite{Kawabata:2019}} & \textbf{Unitary operator} \\ \hline
Time-reversal symmetry (TRS) & $T_+$ & $U_{T_+} H^{*}(k) U_{T_+}^{\dagger}= H(-k)$ & $U_{T_+} = \eta^{y} \tau^{z} \sigma^{y}$ \\ \hline
Particle-hole symmetry (PHS) & $C_{-}$ & $U_{C_-} H^{T}(k) U_{C_-}^{\dagger}= -H(-k)$ & $U_{C_-} = \eta^{x} \tau^{0} \sigma^{0}$ \\ \hline
Time-reversal dagger symmetry (TRS$^{\dagger}$) & 
$C_{+}$ & $U_{C_+} H^{T}(k) U_{C_+}^{\dagger} = H(-k)$ & $U_{C_+} = \eta^{z} \tau^{0} \sigma^{x}$ \\ \hline
Particle-hole dagger symmetry (PHS$^{\dagger}$) & 
 $T_{-}$ 
& $U_{T_-} H^{*}(k) U_{T_-}^{\dagger} = -H(-k)$ & $U_{T_-} = \eta^{0} \tau^{z} \sigma^{z}$ \\ \hline
Chiral symmetry (CS) & $\Gamma$ & $U_{\Gamma} H^{\dagger}(k) U_{\Gamma}^{\dagger} = -H(k)$ & $U_{\Gamma} = \eta^{z} \tau^{z} \sigma^{y}$ \\ \hline
Pseudo-Hermiticity (pH) & $\pH$ & $U_{\pH} H^{\dagger}(k) U_{\pH}^{\dagger} = H(k)$ & $U_{\pH} = \eta^{x} \tau^{z} \sigma^{z}$ \\ \hline
Sublattice symmetry (SLS) & $S$ & $U_{S} H(k) U_{S}^{\dagger} = -H(k)$ & $U_{S} = \eta^{y} \tau^{0} \sigma^{x}$ \\ 
\hline \hline
\end{tabular}
\label{Table:Symmetries_with_perturbation}
\end{table*}

In addition to the above, we incorporate a perturbation term in the form of an uniform transverse magnetic field along $x$ direction,
\begin{equation}
 \mathcal{H}_{\rm pt} =    \sum_{j} { \delta h_x \big( a^{\dagger}_{j,\uparrow}a_{j, \downarrow} + b^{\dagger}_{j,\uparrow} b_{j, \downarrow}  \big) + {\rm H.c.}} ,
 \label{Eq:H_pt}
\end{equation}
which is known to gap out the non-Hermitian skin modes but leave Majorana zero modes (when present) stabilized~\cite{Okuma:2019}. 
We will first discuss the properties of the model in the absence of the perturbation terms and we will consider the full Hamiltonian later on.

Under the PBC, we can express  $\mathcal{H}_{\rm nHSC} $ as
\begin{subequations}
\begin{eqnarray}
     \mathcal{H}_{\rm nHSC} &=& \frac{1}{2} \sum_{k}\Psi^{\dagger}_{k} {H}_{\rm nHSC}^{\rm pbc} (k)\Psi_{k}, 
\end{eqnarray}  
where we introduce the spinor $\Psi^{\dagger}_{k}= \left(a_{k,\uparrow}^{\dagger},a_{k,\downarrow}^{\dagger},b_{k,\uparrow}^{\dagger},b_{k,\downarrow}^{\dagger},
a_{-k,\uparrow},a_{-k,\downarrow},b_{-k,\uparrow},b_{-k,\downarrow} \right)$ in the momentum space,
and the 8-by-8 matrix, 
   \begin{eqnarray}
 {H}_{\rm nHSC}^{\rm pbc} (k) &= &  -i\frac{\Gamma_{+}}{2}\eta^{z}-i\frac{\Gamma_{-}}{2}\eta^{z}\tau^{z} 
 +t_{\inter}\sin\,(ka_{0}) \eta^{z} \tau^{y} 
 \nonumber \\
&    & - \Big[ t_{\inter}\cos\,(ka_{0})+t_{\intra} \Big] \eta^{z}  \tau^{x}
    \nonumber \\
 &   & -i \Big[ \frac{g_{\inter}}{4} 
    \cos   (ka_{0})-\frac{g_{\intra}}{4} \Big] \tau^{y} \sigma^{z} \nonumber \\
  &  & -i  \frac{g_{\inter}}{4}\sin\,(ka_{0}) \tau^{x} \sigma^{z}   -\Delta_{0}\eta^{y} \sigma^{y} ,
\label{Eq:H_pbc2}
\end{eqnarray}   
\end{subequations}
with   
$\Gamma_{\pm} \equiv (\Gamma_{a} \pm \Gamma_{b})/2$, the lattice constant $a_{0}$, and the components $\mu \in \{x,y,z\}$  of the Pauli matrices $\eta^{\mu}$, $\tau^{\mu}$ and $\sigma^{\mu}$ acting on the particle-hole, sublattice, and spin indices, respectively. 
Below we discuss the symmetry properties of the model and  its energy spectrum.

\subsection{Symmetry properties and classification }
\label{Sec:Symmetry}

We now discuss the symmetry class of our model.
Without the perturbation term, the Hamiltonian ${H}_{\rm nHSC}^{\rm pbc}$ defined in Eq.~\eqref{Eq:H_pbc2} commutes with unitary operators and can therefore be block-diagonalized (see  Appendix~\ref{Appendix:Symmetry} for the details).
With the perturbation terms,  the full Hamiltonian 
${H}_{\rm nHSC}^{\rm pbc}  
+ \delta h_x \eta^z \tau^0 \sigma^{x}$ 
is no longer block-diagonalizable, and the symmetry can be fully determined by antiunitary operators~\cite{Ryu:2010,Kawabata:2019}, which allows us to characterize the system in the symmetry classification. 

To proceed, we examine the symmetries of the full Hamiltonian and obtain Table~\ref{Table:Symmetries_with_perturbation}
(where we also list the corresponding abbreviations for various symmetries).
Notably, the model respects all the possible internal symmetries characterized by antiunitary operators.  
We find that the full system belongs to the BDI class in the real AZ classification, characterized by the TRS (squaring to +1) and PHS (squaring to +1)~\cite{Kawabata:2019}. 
It also fulfills BDI$^\dagger$ class in the real AZ$^\dagger$ classification, characterized by the TRS$^\dagger$ (squaring to +1) and PHS$^\dagger$ (squaring to +1), which is guaranteed due to the presence of the SLS, which gives TRS$^\dagger$ operator $U_{C_+} \propto U_{S} U_{C_-}$ and PHS$^\dagger$ operator $U_{T_-} \propto U_{S} U_{T_+}$, up to phase factors. 
As indicated in Table~\ref{Table:Symmetries_with_perturbation}, we have SLS with $U_{S} = \eta^{y} \tau^{0} \sigma^{x}$, which follows the relations with the TRS and PHS operators: $U_{S} U_{T_+} = + U_{T_+} U_{S}^{*}$ and $U_{S} U_{C_-} = + U_{C_-} U_{S}^{*}$, denoted as $S_{++}$ in Ref.~\cite{Kawabata:2019}. 
Along with BDI class and one dimension, this gives topological invariants\footnote{As a remark, we have also considered an example including longer-range hopping terms, which are known to stabilize multiple Majorana zero modes in transverse-field quantum spin chains within Hermitian systems~\cite{Niu:2012}. In the present case, such longer-range hoppings alter the symmetry of the system, placing it in class D of the real AZ classification (equivalent to class AI$^\dagger$ in the real AZ$^\dagger$ classification), characterized by $Z$ invariant for both point and line gaps. 
}
$Z$, $Z \oplus Z$ and $Z \oplus Z$ for point gap, real line gap and imaginary line gap, respectively~\cite{Kawabata:2019}.

Interestingly, the symmetry properties of our model significantly affect its energy spectrum. In non-Hermitian systems, the PHS implies that if $E_0  $ is an eigenvalue, then $-E_0^*  $ must also be an eigenvalue. This corresponds to a mirror symmetry about the imaginary axis in the complex energy plane.
Meanwhile, the pH ensures that $E_0^* $ is also an eigenvalue, reflecting the spectrum across the real axis. When both symmetries are present, the spectrum exhibits a higher degree of structure, with eigenvalues constrained to appear symmetrically in all four quadrants of the complex plane, as is seen throughout this work.

\subsection{PBC energy spectrum}

To proceed, we consider the PBC and $ {H}_{\rm nHSC}^{\rm pbc} $ in Eq.~\eqref{Eq:H_pbc2} and obtain the analytic expression of the energy spectrum, which forms eight energy bands,
\begin{subequations}
    \label{Eq:E_pbc} 
    \begin{eqnarray}
\label{Eq:general-conditions} 
E^{\pm}_{\lambda\epsilon}(k) 
&=& \pm \frac{1}{4} \left[ \sqrt{F_{r}(k) + i \lambda F_{i}(k) } + 2 i \epsilon D_{+} \right] ^{1/2} \nonumber \\
&& \times \left[ \sqrt{F_{r}(k) + i\lambda F_{i}(k) }- 2 i \epsilon D_{-} \right] ^{1/2} , \\  
    F_{r}(k) &=& f_{r 0}+ f_{r 1} \cos\,(ka_{0}),\\
    F_{i}(k) &=&  f_{i} \sin\,(ka_{0}) , 
\end{eqnarray}
\end{subequations} 
with the overall sign $\pm$ and the indices $\epsilon, \lambda \in \{ +, -\}$ labeling the bands.
In the above, we have introduced the following quantities for notational convenience,
\begin{subequations}
\label{Eq:converting-A} 
\begin{eqnarray}
 D_{\pm}
    &=& 2\Delta_{0} \pm \Gamma_{+}, \\
  f_{r 0} &=& - ( g^{2}_{\inter} + g^{2} _{\intra}) +16(t^{2}_{\inter}+t^{2}_{\intra}) - 4\Gamma^{2}_{-}, \\
   f_{r 1}  &=& 2(g_{\inter}g_{\intra}+16t_{\inter}t_{\intra})  , \\
    f_{i} &=&8(g_{\intra}t_{\inter}+g_{\inter}t_{\intra}) . 
\end{eqnarray}
Up to here, we keep general onsite dissipation terms with arbitrary $\Gamma_{a,b}$.  
Below we focus on uniform dissipation limit with $\Gamma_{a} = \Gamma_{b} \equiv \Gamma_{0}$ (unless otherwise stated), and will comment on general cases with nonuniform dissipation terms in Sec.~\ref{sec:non-uniform}. 
For the present parameter choice, we have 
\begin{eqnarray}
 D_{\pm}
    &\to & 2\Delta_{0} \pm \Gamma_{0}, \\
  f_{r 0} & \to & - ( g^{2}_{\inter} + g^{2} _{\intra}) +16(t^{2}_{\inter}+t^{2}_{\intra})  . 
\end{eqnarray}
\end{subequations} 

In the presence of a small onsite transverse field, $\delta h_x$, the PBC spectrum can be obtained numerically; however, it shows only minimal deviation from Eq.~\eqref{Eq:general-conditions} due to the small magnitude of the introduced field relevant for us.
 Therefore, for practical purpose, we will describe the PBC spectrum using Eq.~\eqref{Eq:E_pbc} throughout this work.

\section{Spectral properties
\label{Sec:PBC}
}

In this section, we discuss the properties of the PBC  spectrum $E^{\pm}_{\lambda\epsilon}(k) $ in Eq.~\eqref{Eq:E_pbc}. 
We discuss the conditions for the closure of the PBC energy gap, which is related to the topological phase transition.  

\subsection{Gap closing curves }
Since we have complex energy spectra, the gap closing curves correspond to the parameter sets at which the PBC energy contours  pass through the origin of the complex energy plane, i.e., $E^{\pm}_{\lambda\epsilon}(k)   =0$. 
Below we consider two cases separately, with gap closing when (i) $\sin (ka_{0}) = 0$ or (ii) $\sin (ka_{0}) \neq 0$.

\begin{figure}[t]
\centering
\includegraphics[width=0.23\textwidth]{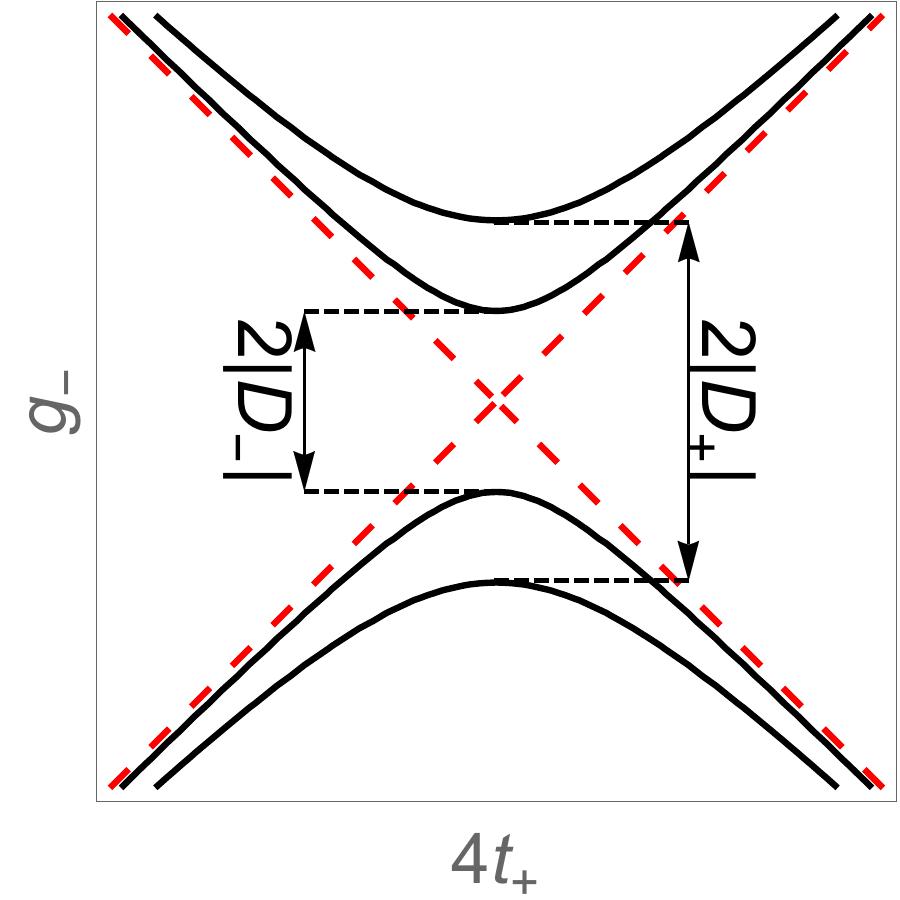} 
\includegraphics[width=0.23\textwidth]{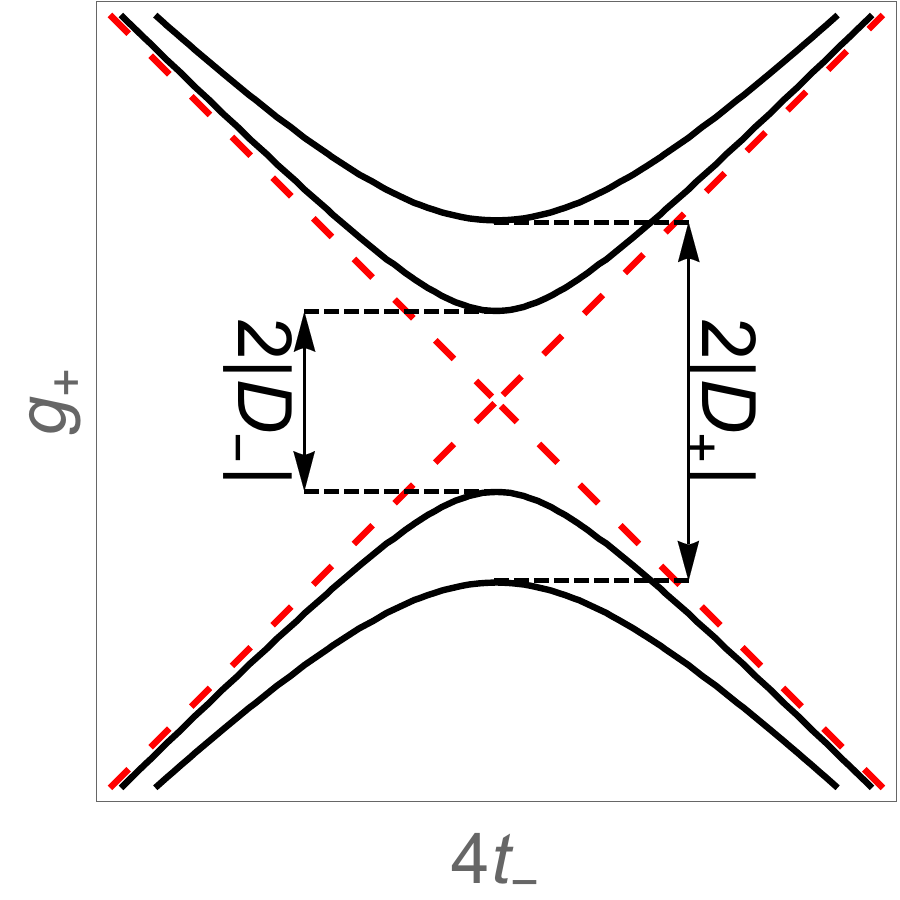}
\caption{Gap closing curves for Case~(i) in the $t_{+}$--$g_{-}$ (left) and $t_{-}$--$g_{+}$ (right) planes for a general  parameter set.
The solid curves are derived from Eq.~\eqref{Eq:Transition_H_nHSC(i)} and can be used to deduce phase diagrams.
The red dashed lines, $ |g_{\pm }| = 4 |t_{\mp}| $,  correspond to the asymptotic limit of $D_{\pm}  = 0$.
}
\label{Fig:GapClosing_H_nHSC}
\end{figure}

\subsubsection*{{\rm Case~(i):} Gap closing at $ka_{0} = 0$ or $\pm \pi$ }
\label{Sec:case(i)}

For Case~(i), we can separate the situation further into two conditions, where
the gap can close at
(ia)  $k = 0 $ 
or
(ib)  $k = \pm \pi / a_{0}$.
The above two cases lead to the following two relations respectively,
\begin{subequations}
\label{Eq:Transition_H_nHSC(i)}
\begin{eqnarray}
    \text{Case (ia)} & \quad &
    - g_{+} ^{2}+16 t_{-} ^{2} +D^{2}_{\pm}=0 ,
\label{Eq:Transition_H_nHSC(ia)}
\\
 \text{Case (ib)} & \quad & 
-g_{-}^{2}+16t_{+}^{2}+D^{2}_{\pm}=0 ,
\label{Eq:Transition_H_nHSC(ib)}
\end{eqnarray} 
\end{subequations}
where we introduce the following notations,
    \begin{subequations}
    \label{Eq:def_tg}
    \begin{eqnarray}
        g_{\pm} &=& (g_{\inter} \pm g_{\intra})/2 , \\
        t_{\pm} &=& (t_{\inter} \pm t_{\intra})/2 .
    \end{eqnarray} 
    \end{subequations}    
In general, Eq.~\eqref{Eq:Transition_H_nHSC(i)} forms four sets of curves, with two in the $g_{+}$-$t_{-}$ parameter space and the other two in the $g_{-}$-$t_{+}$ parameter space.

The above change of variables allows us to explore three different parameter regimes regarding the phase diagrams.
In the first regime,
we have both $D_{+}$ and $D_{-}$ non-vanishing, and the four equations in Eq.~\eqref{Eq:Transition_H_nHSC(i)}  form four pairs of hyperbolic curves in the parameter space of $t_+$-$g_-$ and $t_-$-$g_+$.
A general example for the gap closing curves is given in Fig.~\ref{Fig:GapClosing_H_nHSC},
which are plotted based on Eq.~\eqref{Eq:Transition_H_nHSC(ia)} and Eq.~\eqref{Eq:Transition_H_nHSC(ib)}.
In each panel, two pairs of hyperbolic curves are shown in the respective parameter space, distinguished by their intercepts on the $g_{\pm}$ axis, given by $|D_{\pm}|$.
In the second regime, $D_{-} = 0$ while $D_{+}$ remains nonzero. Two of the four hyperbolic curves in Eq.~\eqref{Eq:Transition_H_nHSC(i)} and Fig.~\ref{Fig:GapClosing_H_nHSC} reduce to straight lines passing through the origin. In the third regime, where $\Gamma_{0} = \Delta_{0} = 0$, all gap-closing curves become straight lines intersecting at the origin.
These three regimes allow us to systematically  explore the energy spectra under the OBC, as will be discussed  in Sec.~\ref{sec:OBC spectra}.

\subsubsection*{{\rm Case (ii):} Gap closing away from $ka_{0}  =  0$ or $\pm \pi$ }
\label{Sec:case(ii)}

We now discuss Case (ii), where the gap closes at momenta labeled by $k_g$, away from the high symmetry points $ka_{0}  =  0$ or $ \pm \pi$.
This can happen only when $f_i = 0$ and, equivalently,
\begin{eqnarray}
t_{+} g_{+} = t_{-} g_{-} .
        \label{Eq:Case(ii)_criterion} 
\end{eqnarray}
Interestingly, the above is also a condition related to the appearance of exceptional points in the non-Hermitian SSH model without superconductivity~\cite{Halder:2022,Perturbation_theory_for_linear_operators}.  
Using the notations in Eq.~\eqref{Eq:def_tg}, the following expressions in Eq.~\eqref{Eq:converting-A} can be rewritten as
\begin{subequations}
\label{Eq:converting-B}
    \begin{eqnarray}
         f_{r0} 
        & =& -2(g_+^{2}+g_-^{2})+32(t_+^{2}+t_-^{2}) ,
        \label{Eq:converting_2} \\
         f_{r1}
        & = &     2(g_+^{2}-g_-^{2})+32(t_+^{2}-t_-^{2}) . 
        \label{Eq:converting_4} 
    \end{eqnarray}
\end{subequations}
Using the relation $-1 \leqslant  \cos\,(k_{g} a_{0}) \leqslant 1$, we further simplify the equations by discussing two separate conditions: (iia) $f_{r1}>0$ and (iib) $f_{r1}<0$.

\begin{figure}[t]
    \centering
    \includegraphics[width=0.49\linewidth]{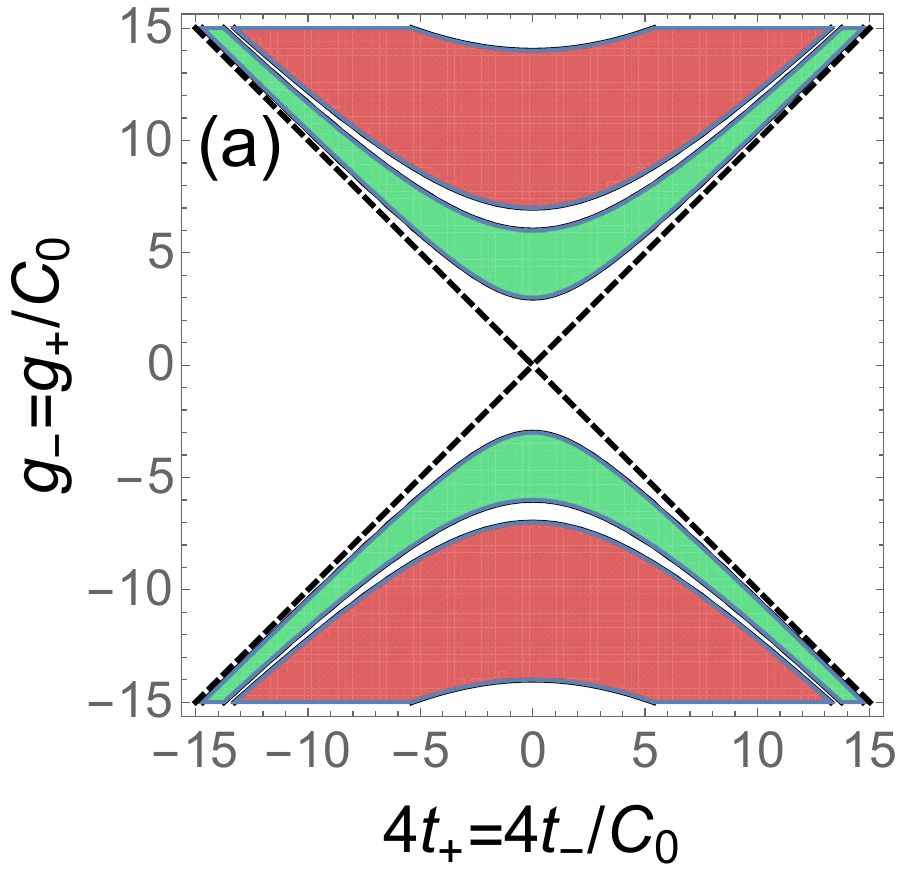}
    \includegraphics[width=0.468\linewidth]{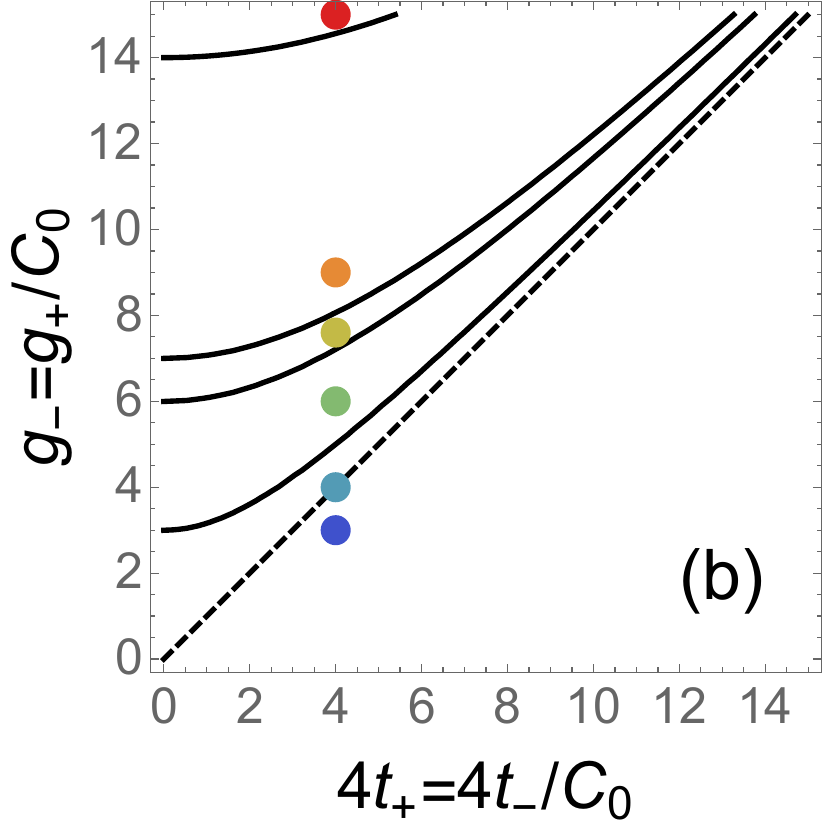}
    \\
    \includegraphics[width=0.49\linewidth]{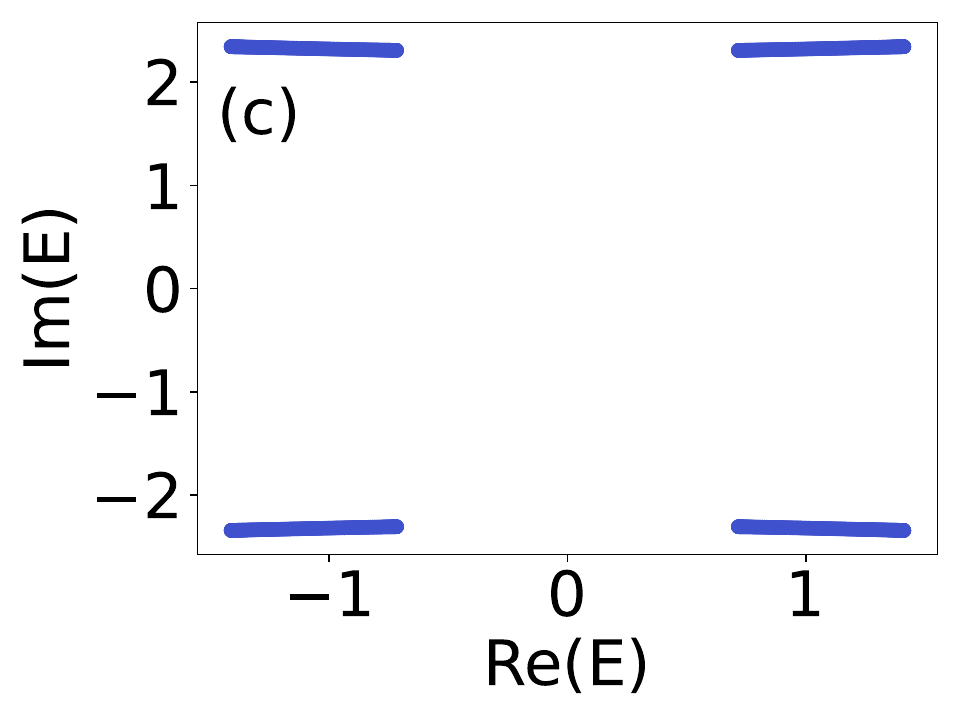}
    \includegraphics[width=0.49\linewidth]{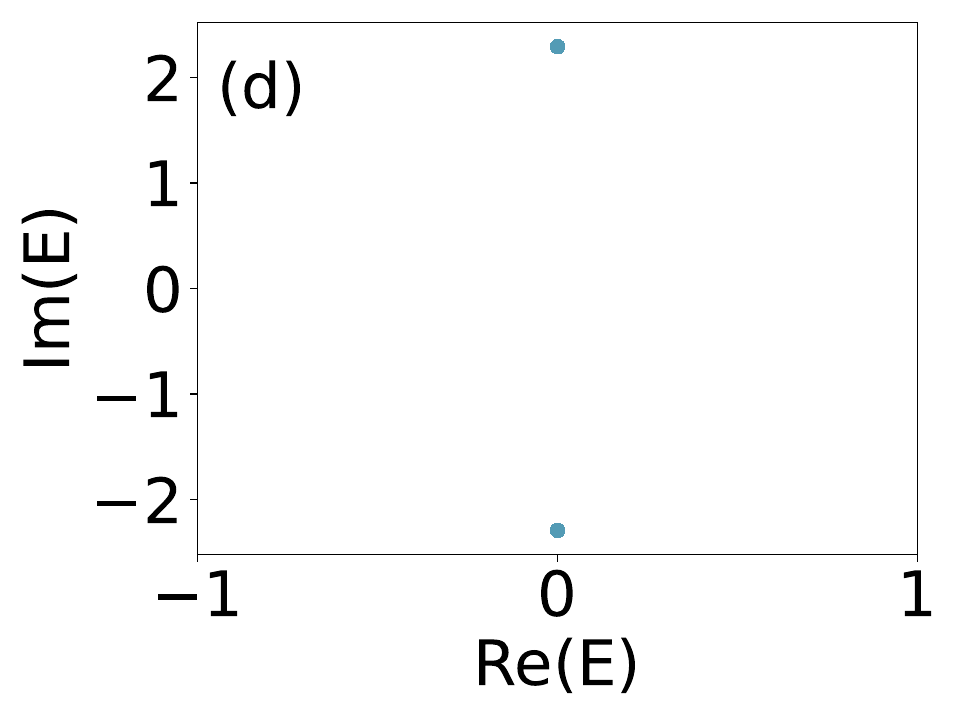}
    \label{fig:iso_point_(a)}
    \\
    \includegraphics[width=0.49\linewidth]{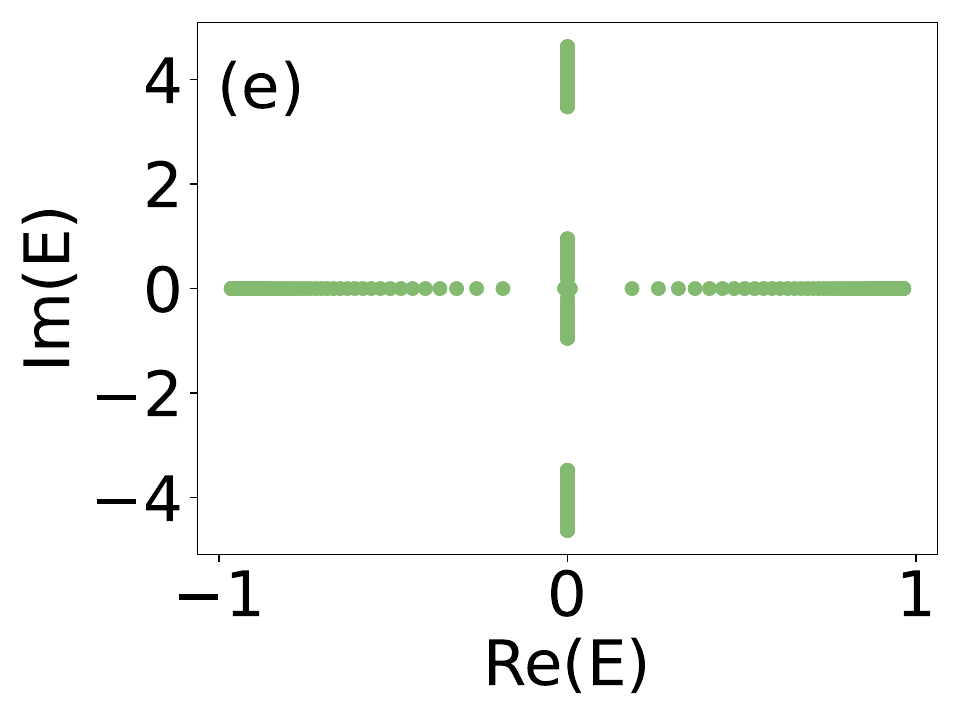}
    \includegraphics[width=0.49\linewidth]{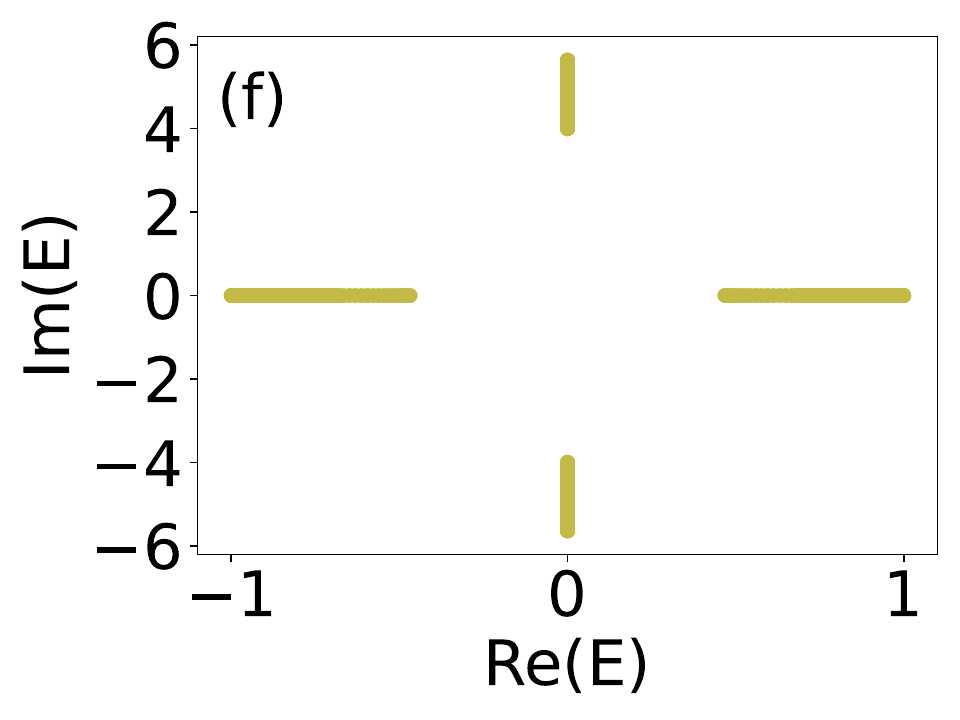}
    \\
    \includegraphics[width=0.49\linewidth]{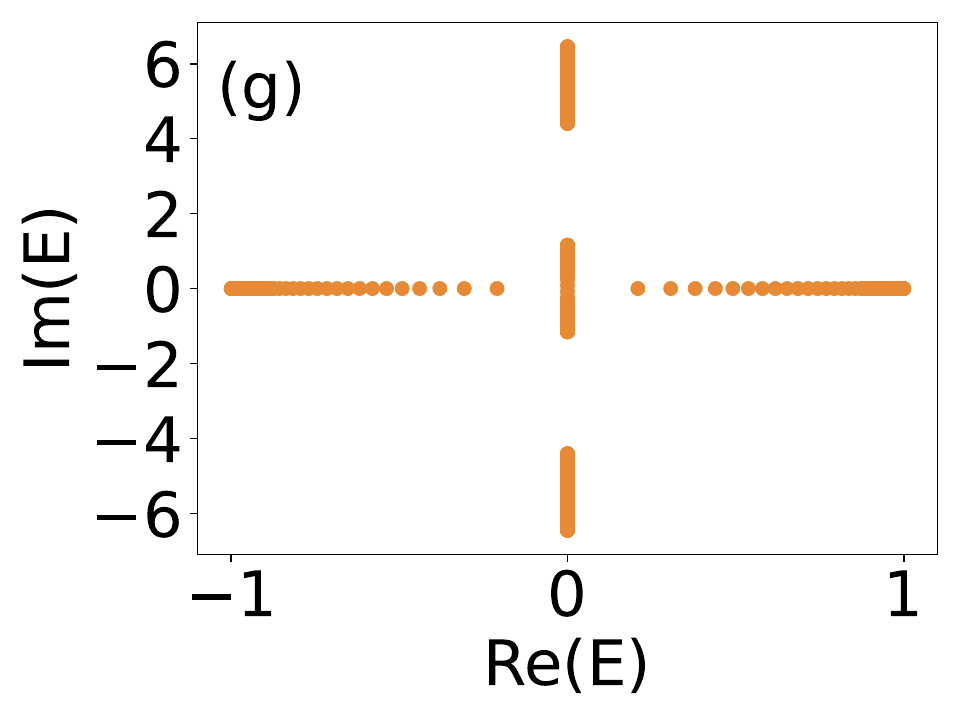}
    \includegraphics[width=0.49\linewidth]{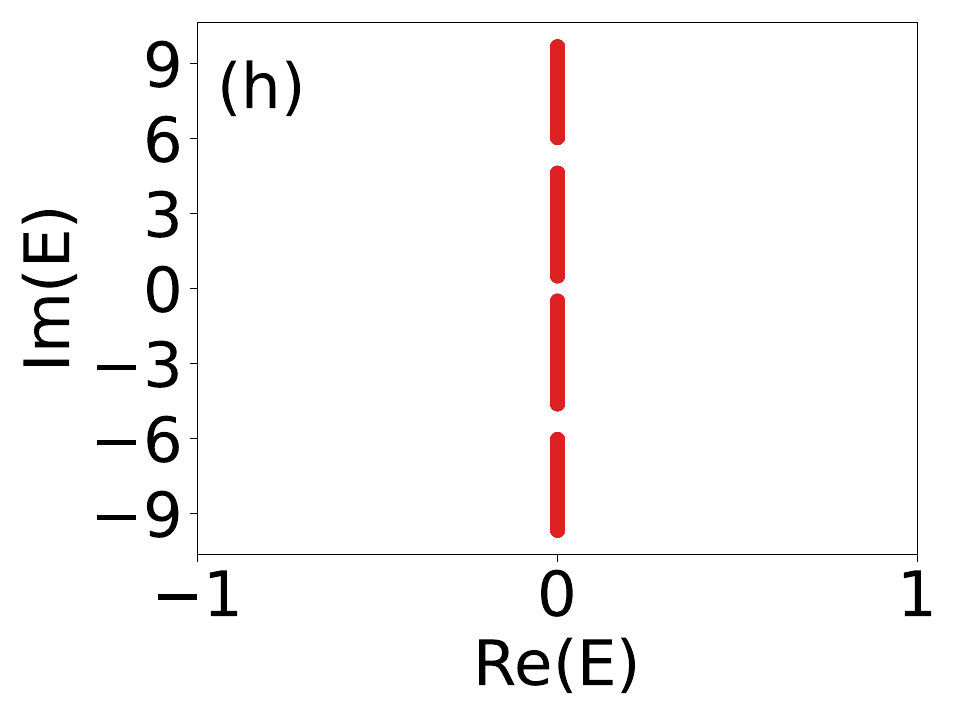}
    \caption{(a,b) Phase diagrams and (c--h) energy spectra for Case~(iib), where Eqs.~\eqref{eq:constraintconditions1}--\eqref{Eq:GaplessSC_condition} hold with $C_0=0.5$; in this case, $g_{\pm}$ and $t_{\pm}$ are mutually dependent. (a) Phase diagram with gapless superconducting phase in the shaded regions. (b) The first quadrant of Panel~(a), with dots indicating parameter sets of $t_+ = 1$ and $g_{-} = 3$, 4, 6, 7.6, 9, and $15$ at $\Gamma_0 = 5.0$ and $\Delta_0 = 1.0$, which are marked by the corresponding colors to the energy spectra in Panels~(c--h). See Table~\ref{Table:Parameters} for the adopted values of the full parameter set. Here, the PBC spectra coincide with the OBC ones.
    }
    \label{fig:energy_spectry_case(iib)}
\end{figure}

By defining a dimensionless quantity,  
\begin{equation} 
    C_{0} \equiv \frac{g_{+}}{g_{-}}=\frac{t_{-}}{t_{+}} , 
\label{eq:constraintconditions1}
\end{equation}
we find that $C^{2}_{0}>1$ and $C^{2}_{0}<1$ automatically establish from $f_{r1}>0$ and $f_{r1}<0$, respectively. In consequence, the condition for Case~(ii) can be simplified as 
\begin{subequations}
  \label{Eq:GaplessSC_condition}
\begin{eqnarray}
    \text{Case (iia)} \quad 
	 C_{0}^2 > 1 \; & \& \; 
    1 \leqslant \frac{D^{2}_{\pm}}{g^{2}_{-} - 16t^{2}_{+}} 
      \leqslant C_{0} ^2 , 
      \label{Eq:GaplessSC_condition(iia)} \\
 \text{Case (iib)} \quad 
  C_{0}^2 < 1  \; & \&  \; 
        C_{0} ^2 \leqslant \frac{D^{2}_{\pm}}{g^{2}_{-} - 16t^{2}_{+}} \leqslant 1.
        \label{Eq:GaplessSC_condition(iib)}
\end{eqnarray}
\end{subequations}
As presented in the above relation, the parameter space is divided into regions separated by 
the four sets of hyperbolic curves given in Eq.~\eqref{Eq:Transition_H_nHSC(i)}
and the two asymptotic lines, $4t_+  \pm g_- = 0$. 

In Fig.~\ref{fig:energy_spectry_case(iib)}, we show the correspondence between the system parameters and the energy spectra, where the shaded areas in Fig.~\ref{fig:energy_spectry_case(iib)}(a) mark the parameter regions where the inequalities in Eq.~\eqref{Eq:GaplessSC_condition} are fulfilled. 
Since we consider $C_{0}=0.5 $ and   $\Gamma_{0},~\Delta_{0}>0$ 
for the case shown in Fig.~\ref{fig:energy_spectry_case(iib)}, Case~(iib) is fulfilled in the shaded regions. Specifically, the red regions in Fig.~\ref{fig:energy_spectry_case(iib)} satisfy the inequalities in Eq.~\eqref{Eq:GaplessSC_condition(iib)} with $D_{+}$, while the green regions correspond to those with $D_{-}$.
For $16t^2_+ > g^2_-$, since Eq.~\eqref{Eq:GaplessSC_condition} cannot be fulfilled, there is no gapless phase in this regime; an example is shown in Fig.~\ref{fig:energy_spectry_case(iib)}(c).
In the above, we focus on $C_{0} < 1$. Alternatively, choosing $C_{0} > 1$ realizes Case~(iia), and one can obtain gapless superconducting phases when Eq.~\eqref{Eq:GaplessSC_condition(iia)} is fulfilled.  

From the spectra, we see that gapless superconducting phases emerge in certain regimes of the hopping parameters. We have explicitly checked that these gapless phases are unstable against an onsite transverse magnetic field, indicating that they are not generically protected. Notably, although the model contains only onsite $s$-wave pairing, gapless superconducting spectra can nevertheless arise due to the interplay of nonreciprocal hopping and non-Hermiticity. While the resulting spectra may resemble those of nodal superconductors, with gap closings occurring away from high-symmetry points, this does not imply an underlying nodal pairing symmetry. 

Before moving forward, we remark that the conditions in Eq.~\eqref{Eq:Transition_H_nHSC(i)}, Eq.~\eqref{Eq:Case(ii)_criterion} 
and Eq.~\eqref{Eq:GaplessSC_condition} could be simultaneously fulfilled. In addition, additional gap closing curves from the latter condition may appear in the $t_{+}$-$g_{-}$ or $t_{-}$-$g_{+}$ plots, leading to rich phase diagrams, as will be demonstrated below. 
Interestingly, in addition to the parameter regime where the system exhibits gapless superconductivity with nodal points $\pm k_{g}$, in certain regimes of Case~(ii), the system also exhibits unusual features in the energy spectrum, as we discuss in more detail below.

\subsection{Uncommon spectral features for Case~(ii) }
\label{Sec:feature}

In this section, we discuss various features in the  energy spectra when the system is within the parameter regime described by Eq.~\eqref{Eq:Case(ii)_criterion} 
and 
Eq.~\eqref{Eq:GaplessSC_condition}, as displayed in Fig.~\ref{fig:energy_spectry_case(iib)}. 
Here we focus on the regime of $4 |t_+| < |g_-|$ unless otherwise noted. 
While we mainly discuss the features under PBC, guided by the analytical form in Eq.~\eqref{Eq:E_pbc}, most of the features discussed here also applies to the OBC spectra, which are constrained to follow the  PBC bands~\cite{Gong:2018}, the latter having zero enclosing area for the parameter sets of interest here.

\subsubsection{Real or purely imaginary eigenvalues}

Below we first discuss the conditions for 
real  PBC spectrum,
${\rm Im} \big[ E^{\pm}_{\lambda\epsilon}(k) \big] =0$,
or 
purely imaginary PBC spectrum, ${\rm Re} \big[ E^{\pm}_{\lambda\epsilon}(k) \big] = 0$.
It can be seen from Eq.~\eqref{Eq:E_pbc} that  $ E^{\pm}_{\lambda\epsilon}(k)$ is real or purely imaginary when 
$F_{i}(k) = 0$ [equivalently, Eq.~\eqref{Eq:Case(ii)_criterion}] and $  F_{r}(k) <0 $, and whether it is  real or purely imaginary depends on the value of the square root in that expression. 
To be precise, we summarize the conditions as follows, 
\begin{subequations}
 \label{Eq:Re-Im-Conditions}
\begin{eqnarray}
\hspace{-20pt}
E^{\pm}_{\lambda\epsilon}(k)  \in \mathcal {R}:  
 & \quad &
 F_{i}(k) = 0 \; \& \; F_{r}(k) <0 \; \& \nonumber \\
 &\quad &    4 \Delta_{0} >   \sqrt{ | F_{r}(k) | } + 2 \Gamma_{0}     , \\
\hspace{-20pt}  E^{\pm}_{\lambda\epsilon}(k)  \in \mathcal {I} :
&\quad & F_{i}(k) = 0 \; \& \; F_{r}(k) <0 \; \& \nonumber \\
& \quad  &   4 \Delta_{0} < \Big|  \sqrt{ | F_{r}(k) | } - 2   \Gamma_{0} \Big|  ,
\end{eqnarray}
which are simplified using  $\Gamma_{0},\Delta_{0} >0$. 
Interestingly, there is another condition,
\begin{eqnarray}
 & &      F_{i}(k) = 0 \; \& \; F_{r}(k) <0 \; \& \nonumber \\
 & &    \Big|  \sqrt{ | F_{r}(k) | } - 2 \Gamma_{0} \Big| < 4 \Delta_{0} < \sqrt{ | F_{r}(k) | } + 2 \Gamma_{0},
\end{eqnarray}
\end{subequations}  
for which one obtains an energy spectrum in which half of the bands are real while the others are purely imaginary.

Since the evolution of the PBC spectrum must be continuous upon varying the parameters, for the energy bands to transit from real to purely imaginary (or the reverse), they must pass through a gap-closing point. 
Using the condition derived from Eqs.~\eqref{Eq:Case(ii)_criterion}, \eqref{eq:constraintconditions1} and \eqref{Eq:GaplessSC_condition}, 
we obtain
$F_r(k)=2( 16t_+^2 - g_{-}^2) \big[ (1 + C_0^2) + (1 - C_0^2) \cos\, (k a_0) \big]$,
from which we find that the gap starts to close at the momentum $ka_0 = 0$ or $\pm \pi$ when $ D^{2}_{\pm}= g^{2}_{-} - 16t^{2}_{+}$ or 
$ D^{2}_{\pm} =  C_{0} ^2 (g^{2}_{-} - 16t^{2}_{+}) $.
These conditions, together with $ g_+ t_+ = g_- t_-$ and $16t^2_+ < g^2_-$, guarantee that 
$F_{r}(k) <0$ for all $k \in [-\pi/a_0, \pi/a_0 ]$. 
Combining these with Eq.~\eqref{Eq:Re-Im-Conditions}, we get the following extrema of the energy bands, 
\begin{subequations}
\begin{eqnarray}
    E^{\pm}_{\lambda\epsilon} (0) 
    &=&
    \pm  
    \sqrt{ -\Big(  \sqrt{  g^2_- -  16t^2_+ } +  \epsilon \Gamma_{0} \Big)^2 / 4 +   \Delta_{0}^{2}} , \nonumber \\
    \\
   E^{\pm}_{\lambda\epsilon} \Big(\pm \frac{\pi}{a_0} \Big) 
   &=&
   \pm  
    \sqrt{ -\Big( |C_0 | \sqrt{  g^2_- -  16t^2_+ } +  \epsilon \Gamma_{0} \Big)^2 / 4 +   \Delta_{0}^{2}} . \nonumber \\
\end{eqnarray}
\end{subequations} 
Comparing this conclusion with Fig.~\ref{fig:energy_spectry_case(iib)}, we obtain that for the shaded areas in Fig.~\ref{fig:energy_spectry_case(iib)}, the energy spectrum consists of both real and purely imaginary bands with some of them passing through the origin [see Panels~(e,g)].
In the region sandwiched by these shaded areas, the system exhibits a spectrum consisting of both real and purely imaginary bands with a point gap, as shown in Panel~(f), 
whereas in the other white region the system shows either a real or purely imaginary spectrum [see Panels~(d,h)]. 

Notably, when the parameter set lies on the asymptotic lines [see Panel~(d)], isolated spectral points emerge in both the PBC and OBC spectra. In the following section, we look into this feature in more detail.

\subsubsection{Complex flat bands}

In Fig.~\ref{fig:energy_spectry_case(iib)}(d), we observe that the PBC spectra collapse onto isolated points, forming complex flat bands in which the energy bands remain constant for all momentum values.  
From Eq.~\eqref{Eq:general-conditions}, we derive the conditions for the emergence of such complex flat bands, and show that it is given by Eq.~\eqref{Eq:Case(ii)_criterion}  along with one of the following conditions,  
\begin{subequations}
\label{Eq:complex_flex_band_condition}
\begin{eqnarray}
&&
g^{2}_{+}=16t^{2}_{-}   \; \& \;
g^{2}_{-}=16t^{2}_{+} 
, 
\label{Eq:complex_flex_band_condition(iso-a)}
\\
&&
t^{2}_{-}=t^{2}_{+} \; \& \;
g_+ =  g_-.
\label{Eq:complex_flex_band_condition(iso-b)}
\end{eqnarray}
\end{subequations} 
Under these conditions  we discuss the PBC energy spectra. 

First, when Eq.~\eqref{Eq:Case(ii)_criterion} and Eq.~\eqref{Eq:complex_flex_band_condition(iso-a)} are fulfilled, the energy spectrum becomes highly degenerate, with the following spectral points, 
\begin{equation}
\label{Eq:Epbc_H_nHSC_(iso_a)}
\begin{split}
     E^{\pm}_{\lambda\epsilon}(k)= 
    \left\{
    \begin{array}{l}
    \pm \frac{i}{2} | D_{+}D_{-}
    |^{1/2} ,~~  {\rm for} ~ D_{-} < 0,  \\
    \pm \frac{1}{2} | D_{+}D_{-} |^{1/2} , ~~ {\rm for} ~ D_{-} > 0 ,
    \end{array}
    \right.
\end{split}
\end{equation}
which are independent of the momentum and the signs of $\lambda$ and $\epsilon$.
An example is given in Fig.~\ref{fig:energy_spectry_case(iib)}(d) discussed earlier, in which  we consider $D_{-} < 0$,
and the two spectral points are thus located on the imaginary axis.
It is also possible to have $D_{-} > 0$ and real flat bands (not shown).

\begin{figure}[t]
    \centering
    \includegraphics[width=0.49\linewidth]{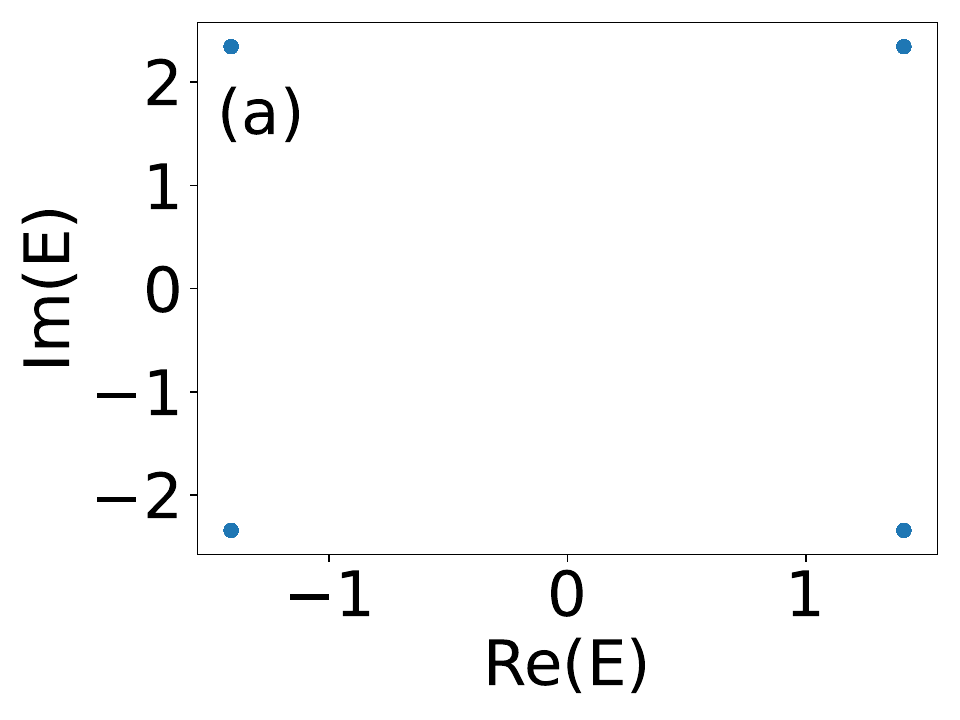}
    \includegraphics[width=0.49\linewidth]{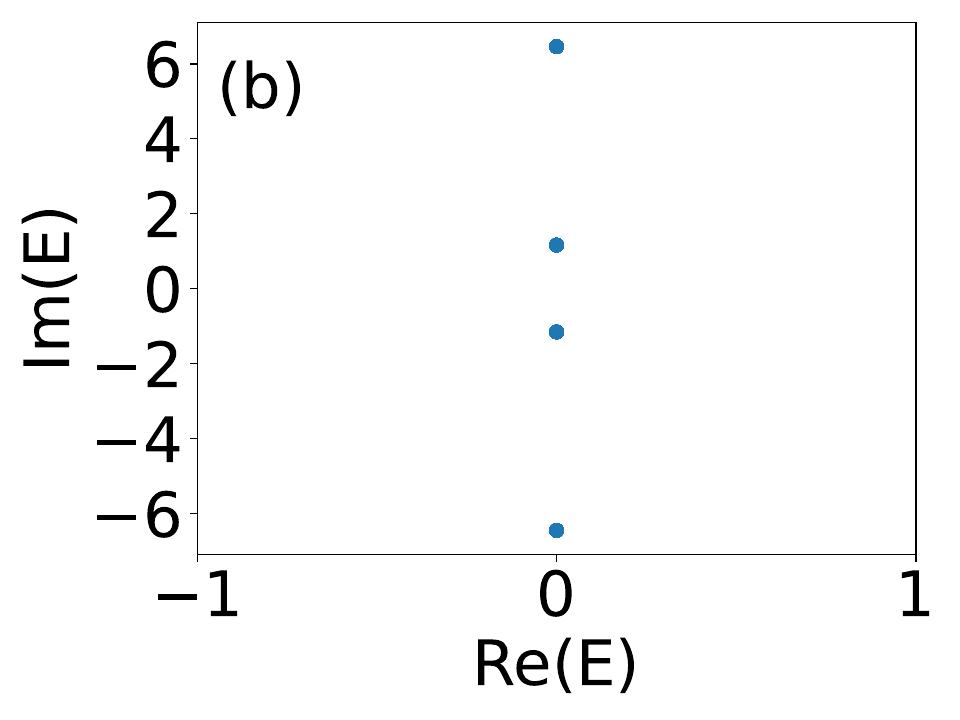}\\
    \includegraphics[width=0.49\linewidth]{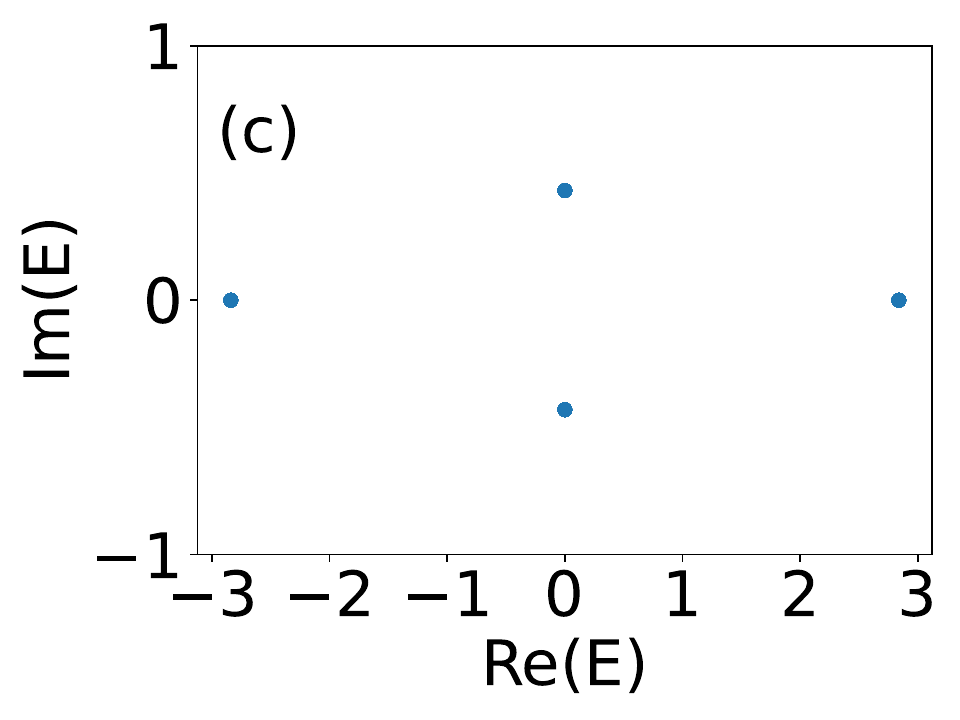}
    \includegraphics[width=0.49\linewidth]{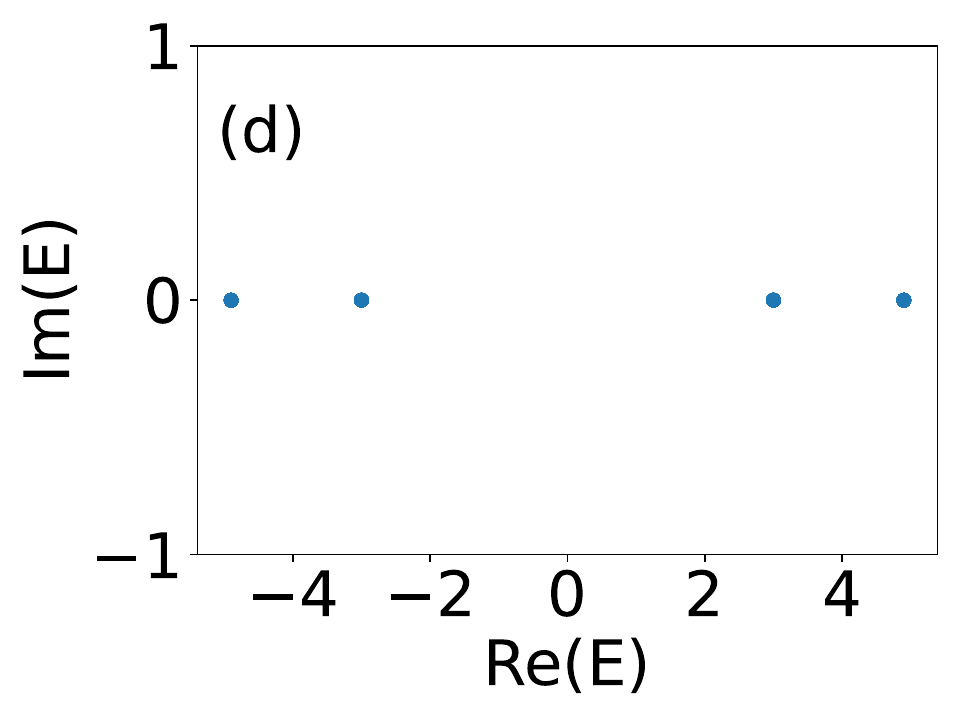}
    \caption{Energy spectra of the system when Eq.~\eqref{Eq:Case(ii)_criterion} and Eq.~\eqref{Eq:complex_flex_band_condition(iso-b)} are fulfilled  with $t_{+} = t_{-} = 1$.
    (a) $g_{+}=g_{-}=3$, $\Gamma_{0}=4$, and $\Delta_{0}=3$. (b) $g_{+}=g_{-}=9$, $\Gamma_{0}=5$, and $\Delta_{0}=1$. (c) $g_{+}=g_{-}=4.5$, $\Gamma_{0}=4$, and  $\Delta_{0}=3$. (d) $g_{+}=g_{-}=5$, $\Gamma_{0}=5$, and  $\Delta_{0}=5$. See Table~\ref{Table:Parameters} for the adopted values of the full parameter set. 
    Here, the PBC spectra   coincide with the OBC ones.
    }
    \label{fig:energy_spectrum_(iso_b)}
\end{figure}

Second, when Eq.~\eqref{Eq:Case(ii)_criterion} and Eq.~\eqref{Eq:complex_flex_band_condition(iso-b)} are fulfilled, the energy spectrum collapses onto the following spectral points, 
\begin{eqnarray}
    E^{\pm}_{\lambda\epsilon}(k)
     =
   \pm \frac{1}{2}
    \sqrt{
    -\left( \sqrt{g^{2}_{+}-16t^{2}_{+} } +\epsilon  \Gamma_{0}  \right)^2 + 4 \Delta_{0}^2 },
    \label{Eq:Epbc_H_nHSC_(iso_b)}
\end{eqnarray}
which are generally complex and depend on the overall $\pm$ sign and $\epsilon$ (but independent of $\lambda$).
In Fig.~\ref{fig:energy_spectrum_(iso_b)}, we discuss this regime. 
There are four types of spectral patterns, depending on the other parameters. 
In Fig.~\ref{fig:energy_spectrum_(iso_b)}(a), 
 we consider $4 t_{+} > g_{+} > 0$, and the spectrum consists of four points with conjugated complex pairs, as indicated by Eq.~\eqref{Eq:Epbc_H_nHSC_(iso_b)}.
In Fig.~\ref{fig:energy_spectrum_(iso_b)}(b)--(d), we consider the regime of  $g_{+} > 4 t_{+} >  0$.
For Panel~(b), we have $|\sqrt{g^{2}_{+}-16t^{2}_{+}} \pm \Gamma_{0}| > 2|\Delta_{0}|$, which leads to negative values in the square root of Eq.~\eqref{Eq:Epbc_H_nHSC_(iso_b)}  for both $\epsilon = \pm$ bands, and 
all the spectral points  stay on the imaginary axis.
For Fig.~\ref{fig:energy_spectrum_(iso_b)}(c), we consider an intermediate regime with 
\[ 
|\sqrt{g^{2}_{+}-16t^{2}_{-}} - \Gamma_{0}| < 2|\Delta_{0}| < |\sqrt{g^{2}_{+}-16t^{2}_{-}} + \Gamma_{0}|.
\]
In this case, we have purely imaginary spectral points for the $\epsilon=+$ band while real spectral points for the $\epsilon=-$ band.  
Finally, for Panel~(d), we consider  $|\sqrt{g^{2}_{+}-16t^{2}_{+}} \pm \Gamma_{0}| < 2|\Delta_{0}|$, which results in  real spectral points  for both $\epsilon = \pm$ bands in Eq.~\eqref{Eq:Epbc_H_nHSC_(iso_b)}, indicating true flat bands, as in the Hermitian regime.

\begin{figure*}[th]
    \centering
 \includegraphics[width=0.99\linewidth]{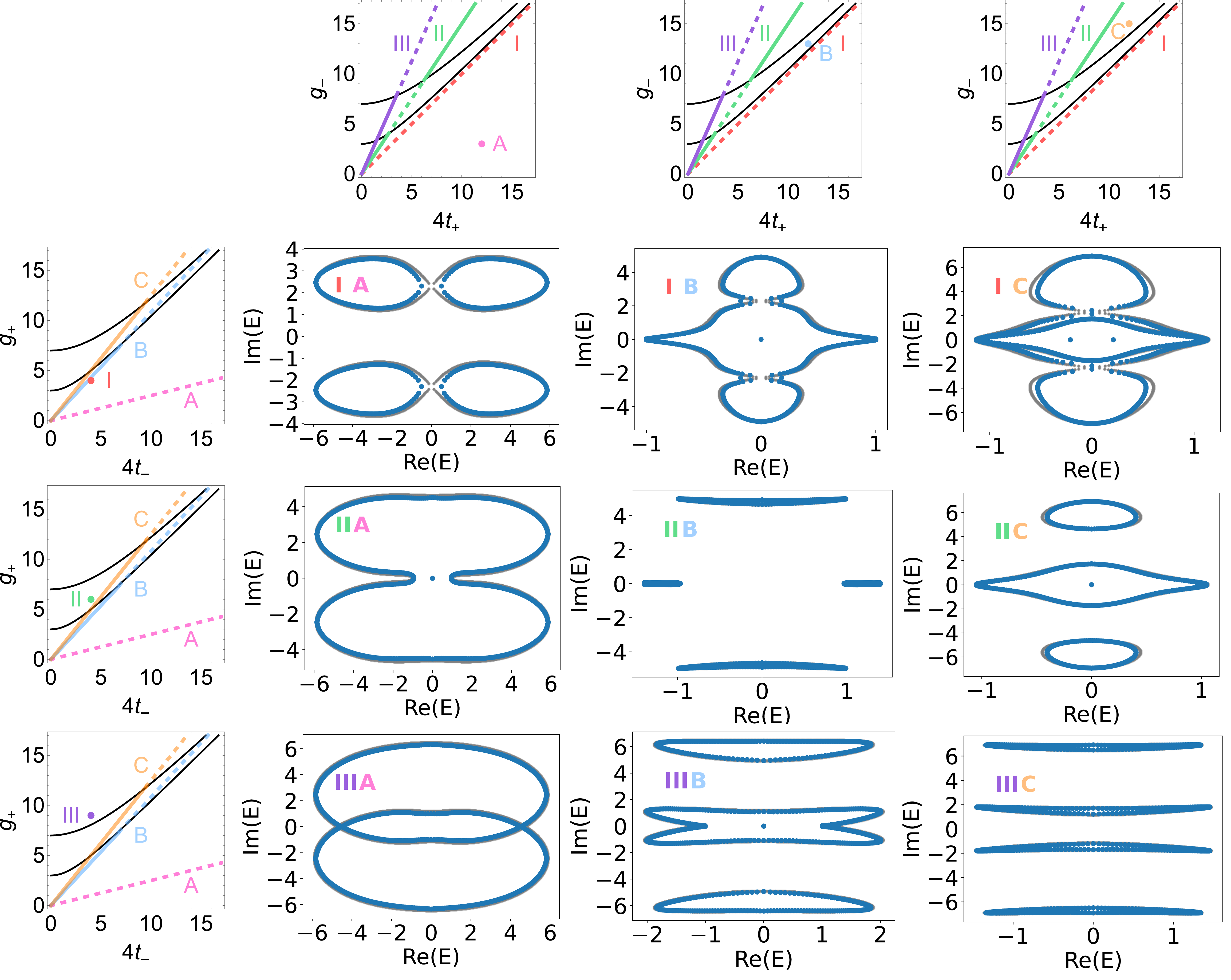}    
 \caption{Energy spectra under the PBC (gray curves) and OBC (blue dots) for $\Gamma_{0}=5$,  $\Delta_{0}=1$, $\delta h_x = 0.01$ and $N = 200$ (corresponding to 100 unit cells).
    The range of the parameter diagrams corresponds to the first quadrant of Fig.~\ref{Fig:GapClosing_H_nHSC}. 
    In each row, we adopt the values for the parameters  $(t_{-},g_{+})$ as labeled by the dots I, II and III in the leftmost panels, whereas in each column, we adopt the values for $(t_{+},g_{-})$ as labeled by the dots A, B and C in the topmost panels.
    In the leftmost panels, the relations in Eq.~\eqref{Eq:Case(ii)_criterion} are shown as dashed lines, with colors corresponding to sets A, B and C. Along these lines, the solid segments mark the additional gap closing curves satisfying Eq.~\eqref{Eq:GaplessSC_condition}. Similarly, in the topmost panels, the corresponding dashed lines and solid segments are shown for sets I, II and III. 
    See Table~\ref{Table:Parameters} for the adopted values of the full parameter sets. 
    }
    \label{Fig:phase_and_band_diagram_case1_perturbation}
\end{figure*}

It is of interest to examine the stability of the complex flat bands beyond the clean limit shown in Fig.~\ref{fig:energy_spectrum_(iso_b)}. To this end, we introduce random dissipation terms, with the results presented in Appendix~\ref{Sec:OBC-PBC-Spectra_Region2}. As shown in Figs.~\ref{fig:energy_spectrum_(iso_b)_disorder1}--\ref{fig:energy_spectrum_(iso_b)_disorder2}, the complex flat bands are generally unstable against such disorder and acquire a finite bandwidth when the disorder strength becomes sufficiently large. 
Having discussed the spectral features arising mainly from the PBC spectra, below we  discuss topological zero modes under OBC when an onsite transverse field is present.

\section{Energy spectra  in the presence of onsite transverse fields }
\label{sec:OBC spectra}

In this section we discuss the spectra when the onsite transverse field term in Eq.~\eqref{Eq:H_pt} is included. Here we discuss the OBC spectra and compare them to the PBC ones.
To better organize our results for the rather large parameter space, below we explore the energy spectra in two parameter regimes\footnote{In the third regime, where $D_{+} = D_{-} = 0$, there would be no Majorana zero modes, and therefore we do not discuss it here. 
} separately---one with both $D_{+}$ and $D_{-}$ nonzero and the other with $D_{+} \neq 0$ but $D_{-} =0$.

\begin{figure*}
\centering   
\includegraphics[width=0.99\textwidth]{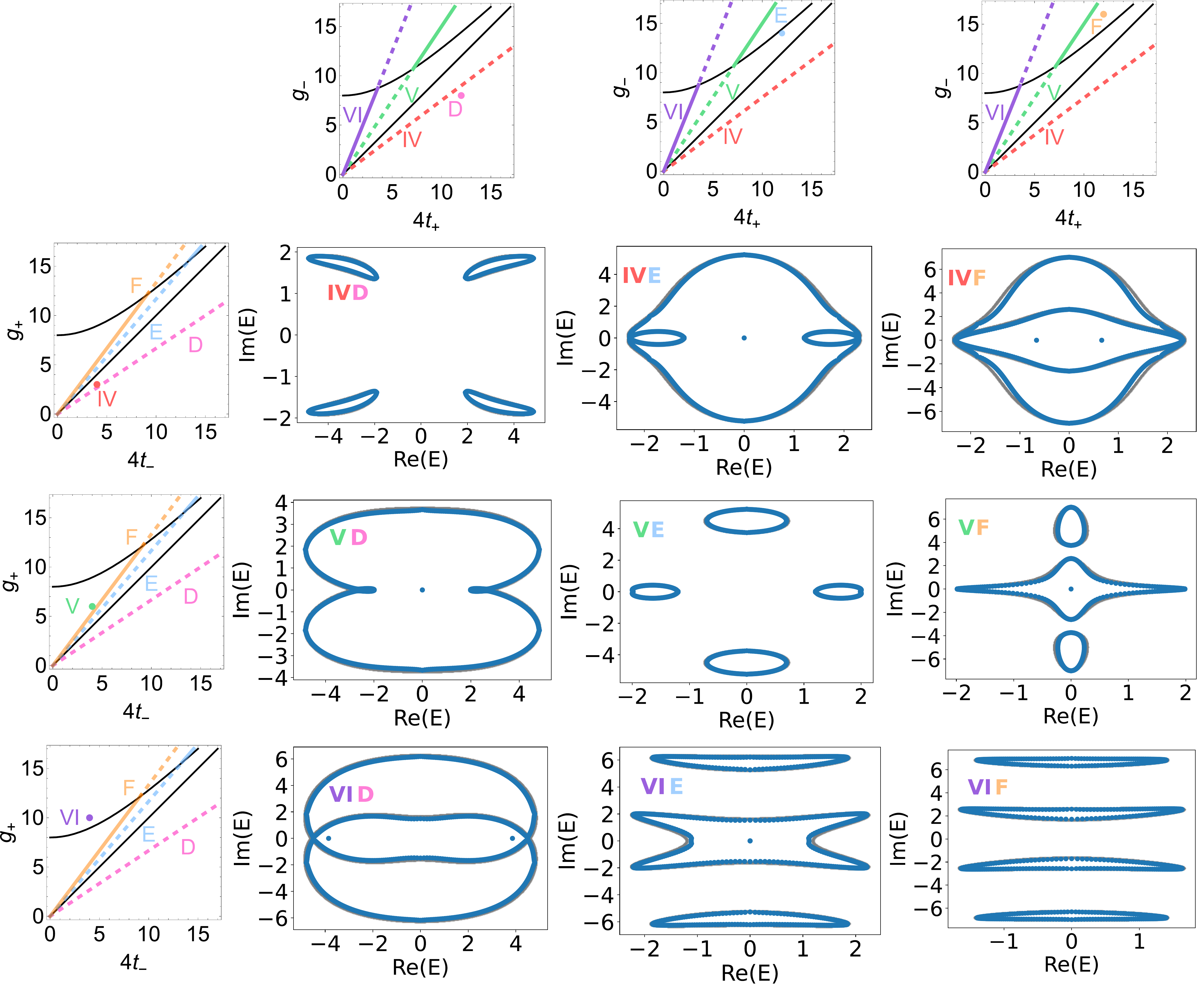}
    \caption{Similar plots as Fig.~\ref{Fig:phase_and_band_diagram_case1_perturbation}, but with $D_{-} = 0 $.
    The representative parameter sets in the $(t_{-},g_{+})$ space are labeled as IV, V and VI, and those in $(t_{+},g_{-})$ space are labeled as  D, E and F. 
    See Table~\ref{Table:Parameters} for the adopted values of the full parameter sets. 
    }
\label{Fig:phase_and_band_diagram_case2_perturbation}
\end{figure*}

\subsection{When both $D_{+}$ and $D_{-}$ non-vanishing}

In the first regime, one can derive the gap closing conditions in 
Eq.~\eqref{Eq:Transition_H_nHSC(i)}
 and obtain Fig.~\ref{Fig:GapClosing_H_nHSC} for a general set of $t_{\pm}$ and $g_{\pm}$. 
In Fig.~\ref{Fig:phase_and_band_diagram_case1_perturbation} we organize the corresponding OBC and PBC spectra for different parameter sets, along with the corresponding gap closing curves [see Eq.~\eqref{Eq:Transition_H_nHSC(i)} and Eq.~\eqref{Eq:Case(ii)_criterion}] in the parameter space. 
As indicated in Fig.~\ref{Fig:GapClosing_H_nHSC}, the parameters $t_{\pm}$ and $g_{\pm}$ are independent, which motivates us to label the entire parameter space using two sets of diagrams. The first set of  diagrams describes the $t_{-}$–$g_{+}$ plane, where we define three regions (I, II and III) separated by the hyperbolic gap-closing curves in Eq.~\eqref{Eq:Transition_H_nHSC(ia)}. Similarly, we introduce a second set of  diagrams for 
the $t_{+}$–$g_{-}$ plane and Eq.~\eqref{Eq:Transition_H_nHSC(ib)}, with regions labeled as A, B and C. 
Combining the two diagrams sets yields a total of nine parameter regions, which are  further complicated by the additional gap closing conditions given in Eq.~\eqref{Eq:GaplessSC_condition}, with the corresponding energy spectra shown in Fig.~\ref{Fig:phase_and_band_diagram_case1_perturbation}. 

From these spectra, several notable features emerge. 
First, in all the panels, both the OBC and PBC spectra exhibit symmetry with respect to the real and imaginary axes, consistent with the constraints imposed by pH and PHS, as discussed in Sec.~\ref{Sec:Symmetry}.
Second, the OBC spectra here asymptotically approach the PBC spectra. This behavior arises because the onsite transverse field mixes and gaps out the non-Hermitian skin modes, thereby suppressing the skin effect~\cite{Okuma:2019}. This behavior is in stark contrast to the case without the onsite transverse field; see Appendix~\ref{Appendix:no-h_x} for details. 
We also note that, when the onsite magnetic field term is absent, one can perform imaginary gauge transformation and derive the PBC and OBC spectra; see Appendix~\ref{Appendix:IGT} for details. 
Third, notably, isolated modes appear in the OBC spectra in certain panels. By examining their energy values, we identify  zero modes in Panels I~B, II~A, II~C, and III~B of Fig.~\ref{Fig:phase_and_band_diagram_case1_perturbation}.
We will further examine their topological origin in Sec.~\ref{Sec:winding}.

\subsection{When $D_{+} \neq 0$ but $D_{-} =0$ }

In the second regime, we set one of $D_{\pm}=0$, so that one family of hyperbolic curves reduces to two straight lines. Since we restrict to $\Gamma_{0}, \Delta_{0}>0$, this corresponds to $D_{-}=0$.
As in the previous case, we summarize the parameter diagrams together with the OBC and PBC spectra in Fig.~\ref{Fig:phase_and_band_diagram_case2_perturbation}. In this regime, three representative parameter sets in the $t_{+}$–$g_{-}$ parameter space are labeled D, E, and F, while those in the $t_{-}$–$g_{+}$ space are labeled IV, V, and VI.

The spectral features closely parallel those in Fig.~\ref{Fig:phase_and_band_diagram_case1_perturbation}. In particular, with pH and PHS, the spectra remain symmetric with respect to both the real and imaginary axes. As before, the onsite transverse field hybridizes the skin modes and suppresses the skin effect, causing the OBC spectra to asymptotically coincide with the PBC spectra. Consistent with Fig.~\ref{Fig:phase_and_band_diagram_case1_perturbation}, Majorana zero modes emerge in Panels IV~E, V~D, V~F, and VI~E of Fig.~\ref{Fig:phase_and_band_diagram_case2_perturbation}.
 
In the above, we only focus on the energy spectra with finite $\delta h_x$ in the regime $g_{+} t_{+} \neq g_{-} t_{-}$. 
As discussed in Sec.~\ref{Sec:feature} and Appendix~\ref{Sec:OBC-PBC-Spectra_Region2}, in the regime of $g_{+} t_{+} = g_{-} t_{-}$, all PBC spectra enclose zero area, leading to identical PBC and OBC spectra and the absence of non-Hermitian skin effect. It is therefore unnecessary to introduce the onsite magnetic field in this case. 
As already analyzed in Fig.~\ref{fig:energy_spectry_case(iib)}, there is no appearance of Majorana zero modes in this regime. 
For completeness, we have also numerically confirmed the absence of zero modes even when a finite $\delta h_x$ is introduced (not shown for brevity).

\subsection{Winding number  }
\label{Sec:winding}

To characterize the phases in each panels of Fig.~\ref{Fig:phase_and_band_diagram_case1_perturbation}--Fig.~\ref{Fig:phase_and_band_diagram_case2_perturbation}, we introduce a topological invariant. 
Motivated by Refs.~\cite{Kawabata:2019,Okuma:2019}, we begin with constructing the following Hermitianized form of the Hamiltonian,
\begin{equation}
   \tilde{H}(k)= \left(
   \begin{array}{cc}
       0 &  {H}_{\rm nHSC}^{\rm pbc} (k) \\
       \big[ {H}_{\rm nHSC}^{\text{pbc }}(k) \big]^{\dagger} & 0
   \end{array} \right) .
\end{equation}
with $ {H}_{\rm nHSC}^{\rm pbc} $ defined in Eq.~\eqref{Eq:H_pbc2}, which now includes general $\Gamma_{\pm}$ terms. 

While the Hermitianization procedure is primarily a mathematical construction, from a physical perspective the Hamiltonianized Hamiltonian may be viewed as an enlarged system consisting of the original system and its Hermitian-conjugate counterpart. We note that this procedure preserves locality, as it does not introduce nonlocal couplings or string operators.

By introducing the Pauli matrix $\xi^{\mu}$ acting on the ``Hermitianization space,'' we obtain a unitary operator $\tilde{U}_{} = \xi^{z} \otimes U_{S}$ with $U_{S}=\eta^{y} \tau^{0} \sigma^{x}$ as listed in Table~\ref{Table:Symmetries_with_perturbation}, which commutes with $ \tilde{H} $. To proceed, we introduce another unitary matrix $\tilde{V}$ that diagonalizes $\tilde{U}$.
It can be shown that the matrix $\tilde{V}$ also block-diagonalizes the Hermitianized Hamiltonian, resulting in 
\begin{eqnarray}
    \tilde{V} \tilde{H}(k) \tilde{V}^{-1}=\left(
    \begin{array}{cccc}
       0  & H_{-,-} (k) & 0 & 0 \\
       H_{+,-} (k)  & 0 & 0 & 0 \\
       0  & 0 & 0 & H_{-,+} (k) \\
       0  & 0 & H_{+,+} (k) & 0
    \end{array}\right). \nonumber \\ 
\end{eqnarray}
With this procedure, the eigenstates of the upper block in the above expression are also eigenstates of $U_S$ with eigenvalue $+1$, while those of the lower block correspond to eigenstates with eigenvalue $-1$.

In the above, we have expressed the blocks in terms of the 4-by-4 matrix, 
\begin{equation}
\begin{split}
    &H_{\delta,\pm}=
    \left(
    \begin{array}{cccc}
       i\delta P_{-,  \pm}  & 0 & h_{\delta,-} (k) & 0 \\
       0  & i\delta P_{-,  \mp} & 0 & h_{-\delta,-} (k) \\
       h_{-\delta,+} (k)  & 0 & i\delta P_{+,  \pm} & 0 \\
       0  & h_{\delta,+} (k) & 0 & i\delta P_{+,  \mp}
    \end{array}\right) , 
    \label{Eq:H_blocks-for-winding}
\end{split}
\end{equation}
where we have defined
\begin{equation}
\begin{split} 
    &P_{\zeta,  \pm} = \frac{1}{2}(\zeta \Gamma_{-} \pm D_{\pm}),\\
   & h_{\delta,\pm} (k)= \Big( t_{\intra} +  \delta\frac{g_{\intra}}{4} \Big) +
   \Big( t_{\inter} -\delta\frac{g_{\inter}}{4} \Big) e^{\pm i ka_{0}} ,
\end{split}
\label{Eq:def_P-fn}
\end{equation}
with the indices $\zeta$, $\delta \in \{+,-\}$. 
 
For each of $H_{\delta,\pm}$, we can define a winding number as 
\begin{equation}
    W_{\delta,\pm} = \int_0^{2\pi} \frac{dk}{2\pi i} \frac{d}{dk} \ln \left\{ \det \left[ H_{\delta,\pm}(k)  \right] \right\}.
    \label{Eq:winding}
\end{equation}
One can verify that, for general $\Gamma_{\pm}$, the conditions $\det \left[H_{\delta,\pm}(k)\right]=0$ at $k=0$ or $k=\pm \pi/a_0$ coincide with the corresponding gap-closing conditions; see Eq.~\eqref{Eq:Transition_H_nHSC(i)_general} below. We remark, however, that when $t_{+} g_{+} = t_{-} g_{-}$, the gap can also close at  momenta away from these points, enabling additional transitions between phases with different winding numbers, as will be discussed below.
Importantly, for each of $H_{\delta,\pm}$ in Eq.~\eqref{Eq:H_blocks-for-winding}, one can plot the PBC spectrum in the complex energy plane and find that the value of Eq.~\eqref{Eq:winding} consistently matches the number of the spectral trajectory winds around the origin; see Appendix~\ref{Appendix:Winding} for more details.

The above formula allows us to characterize the phases in  different parameter regimes.
For $\Gamma_{-}=0$, corresponding to Figs.~\ref{Fig:phase_and_band_diagram_case1_perturbation}--\ref{Fig:phase_and_band_diagram_case2_perturbation},
we numerically confirm that the winding number follows the relation of $W_{+,+}=W_{-,-}=-W_{+,-}=-W_{-,+}$ for a fixed parameter set.
In addition, the winding number  $W_{\delta,\pm}  $ vanishes for all $\delta \in \{+,-\}$  and the $ \pm$ sign in all the regimes  without Majorana zero modes, whereas the winding number is nonzero with $W_{+,+} =-1 $ for the regimes of I~B, II~A, II~C, VI~E, V~D and V~F, and $W_{+,+} =1 $ for III~B and VI~E, consistent with the appearance of the topological zero modes in Figs.~\ref{Fig:phase_and_band_diagram_case1_perturbation}--\ref{Fig:phase_and_band_diagram_case2_perturbation}.  
We will discuss the winding number for more general cases below. 

Having established the connection between the emergence of zero modes and the topological invariant, below we discuss the eigenstate profiles under the OBC, which reveal additional symmetry-enriched properties.

\section{Symmetry-enriched density profile features for the Majorana zero modes}
\label{Sec:density}

For a thorough understanding of this system, we plot the density profiles of the zero-energy eigenstate
by focusing on the parameter set II~A in Fig.~\ref{Fig:phase_and_band_diagram_case1_perturbation}.
For a general energy eigenstate in our non-Hermitian system, one can distinguish the right and left eigenstates as  
\begin{equation}
\begin{split}
    & (H^{\rm obc}_{\rm nHsc}+H_{\rm pt}) \ket{ {E}_{}} =E_{ }\ket{ {E}_{}} , \\
    & \langle\!\langle  {E}_{} | (H^{\rm obc}_{\rm nHsc} +H_{\rm pt}) = \langle\!\langle  {E}_{} | E  ,  
\end{split}
\end{equation}
for the right ($\ket{ {E}_{}} $) and left ($\langle\!\langle  {E}_{ } |$) eigenstates with the energy $E$. 
Based on this, we define the sector-resolved density profiles of the right and left eigenstates, separating the particle/hole and up-/down-spin components.  
To be precise, we have
\begin{subequations}
\label{Eq:density-profiles}
\begin{eqnarray}
\rho_{RR,\eta,\sigma}^{E}  (r_j) &\equiv& \langle  {E} |r_j,\eta,\sigma \rangle \langle r_j,\eta,\sigma  |  {E}  \rangle= \left|  \psi^{E}_{\eta,\sigma} (r_j)  \right|^2 ,  
\hspace{14pt} ~ \\
\rho_{LL,\eta,\sigma}^{E}  (r_j) &\equiv& \langle\!\langle  {E}_{}  |r_j,\eta,\sigma\rangle \!\rangle \langle\!\langle r_j,\eta,\sigma |  {E}_{}  \rangle\!\rangle 
= \left| \bar{ \psi}^{E}_{\eta,\sigma} (r_j)  \right|^2 , \nonumber \\
\end{eqnarray}
\end{subequations}
with the spin index $\sigma$, the particle/hole index $\eta \in \{ {e \equiv +, h \equiv -} \}$,  and $\psi$ ($\bar{\psi}$) denoting the wavefunctions of the right (left) eigenstates.
We also introduce the site index $r_j$, with $r_j=2j+1$ for sublattice A and $r_j=2j$ for sublattice B. 
From the above we see that the density profiles $\rho_{RR}^{E} $ and $\rho_{LL}^{E} $ are real.
 
Projecting the eigenstates onto a specific particle/hole and spin-up/down sector, we show the spatial profiles of the zero-energy modes in  Fig.~\ref{Fig:B,I,per,state,n=200}. 
As shown, the density profiles remain localized at the boundaries across all particle/hole and spin components.
Interestingly, for each of the right and left eigenstates, we observe correlations between the particle component with spin $\sigma$ and the hole component with opposite spin $-\sigma$. 
Furthermore, between the right and left eigenstates, additional correlations emerge, where we find that  the particle component of the right eigenstate is identical to the hole component of the left eigenstate with the same spin (and vice versa).

\begin{figure}[t]
\centering    
\includegraphics[width=0.99\linewidth]{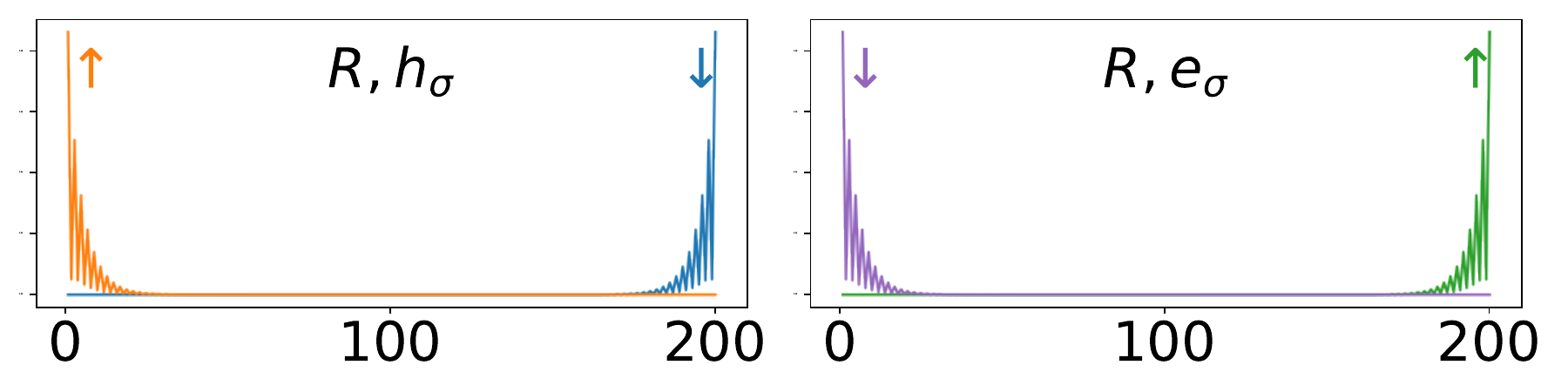} 
\includegraphics[width=0.99\linewidth]{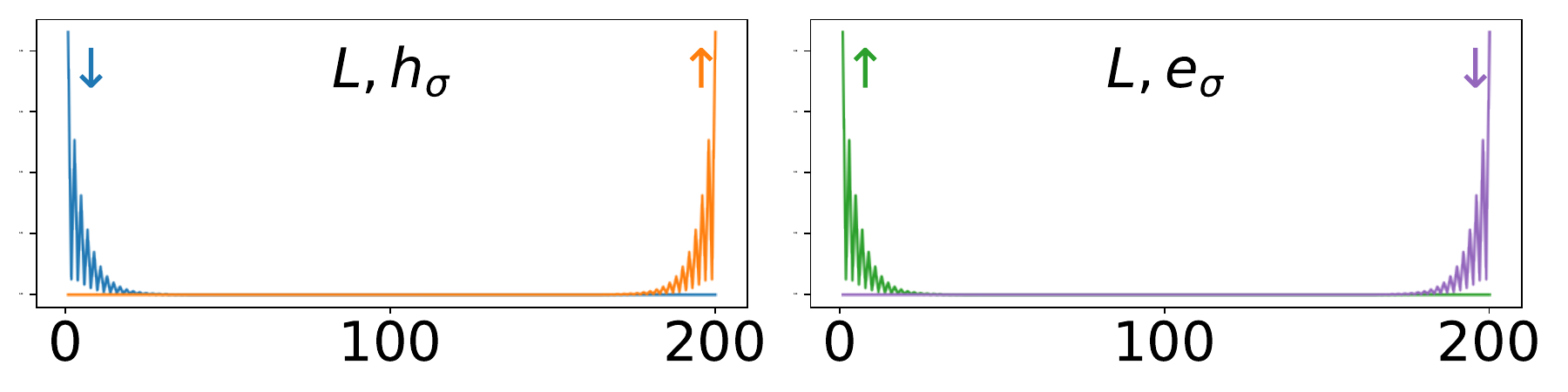}   
\caption{Density profiles of the zero-energy eigenstate under OBC. 
    We separate the particle/hole (labeled by $\eta \in \{e,h\}$) and up/down-spin ($\sigma \in \{ \uparrow, \downarrow \}$) components of the $E=0$ mode for  
    $\rho_{RR,\eta,\sigma}^{0}  $ (top row) and $ \rho_{LL,\eta,\sigma}^{0} $ (bottom row) 
    defined in Eq.~\eqref{Eq:density-profiles}. The adopted values of the parameters are given by $t_{+}=3$, $t_{-}=1$, $g_{+}=6$, $g_{-}=3$, $\Gamma_{0}=5$, and $\Delta_{0}=1$ with $N =200$ (that is, 100 unit cells),  corresponding to the parameter set II~A in Fig.~\ref{Fig:phase_and_band_diagram_case1_perturbation}.
}
\label{Fig:B,I,per,state,n=200}
\end{figure}

To further explore this phenomenon, we note that our model respects the CS and pH, as discussed in Sec.~\ref{Sec:Symmetry}. With these symmetries, it is straightforward to show that the right and left eigenstates are connected through the following relations,
\begin{subequations}
\begin{eqnarray}
    \ket{ {E}_{ }} &=& U_{\Gamma}  |  {-E^{*}} \rangle\!\rangle, 
    \label{Eq:CS-relation} \\    
     \ket{ {E}_{ }} &=& U_{\pH} |  {E^{*}} \rangle \!\rangle,
     \label{Eq:pH-relation}
\end{eqnarray}    
\end{subequations}
for a general energy $E$.  
As given in Table~\ref{Table:Symmetries_with_perturbation}, in our system one can find $U_{\Gamma}=\eta^{z} \tau^{z} \sigma^{y}$ and $U_{\pH}=\eta^{x}\tau^{z}\sigma^{z}$ for the CS and pH, respectively.

To proceed, we first use the relation from the CS.
Projecting the eigenstates to the basis labeling the particle-hole $\eta$ and spin $\sigma$ states, the left-hand side of Eq.~\eqref{Eq:CS-relation} is given by  $\psi^{E}_{\eta,\sigma} (r_j)$ and the right hand side becomes
\begin{equation}
\begin{split}
    & \langle \! \langle \eta| \otimes \langle \!\langle  r_j| \otimes \langle \!\langle \sigma| \eta^{z} \tau^{z} \sigma^{y} | {-E^{*}}\rangle\!\rangle\\
    & =  \eta (-1)^{r_j+1} (-i \sigma) \bar{\psi}^{-E^{*}}_{\eta,-\sigma} (r_j),
\end{split}
\end{equation}
thereby establishing the relation between the density profile of the right eigenstate $\rho^{0}_{RR, \eta,\sigma} $ and  
the left eigenstate $\rho^{0}_{LL,\eta,-\sigma} $ with the opposite spins  for the zero modes, consistent with Fig.~\ref{Fig:B,I,per,state,n=200}.
Similarly, using the pH operator, one gets $  \psi^{E}_{\eta,\sigma} $ and $ (-1)^{r_j+1} \sigma \bar{\psi}^{ E^{*} }_{ -\eta,\sigma} $ on the two sides of Eq.~\eqref{Eq:pH-relation}.
As a consequence, it guarantees the relation between the density profile of the right eigenstate and 
that of the left  eigenstate with the opposite  $\eta$ and the same $\sigma$ (that is, switching the particle and hole components), as demonstrated in Fig.~\ref{Fig:B,I,per,state,n=200}. 
Furthermore, combining these relations derived from the CS and pH, one can further connect the particle and hole components with the opposite spin for each of the right or left eigenstates, again consistent with Fig.~\ref{Fig:B,I,per,state,n=200}.

We conclude this section by remarking on the symmetry-enriched features of the wavefunctions in the model, revealed by the relations between the particle/hole and up-/down-spin components of the left and right eigenstates, as imposed by CS, pH, and their combination.

\begin{figure}[t]
    \centering
\includegraphics[width=0.49\linewidth]{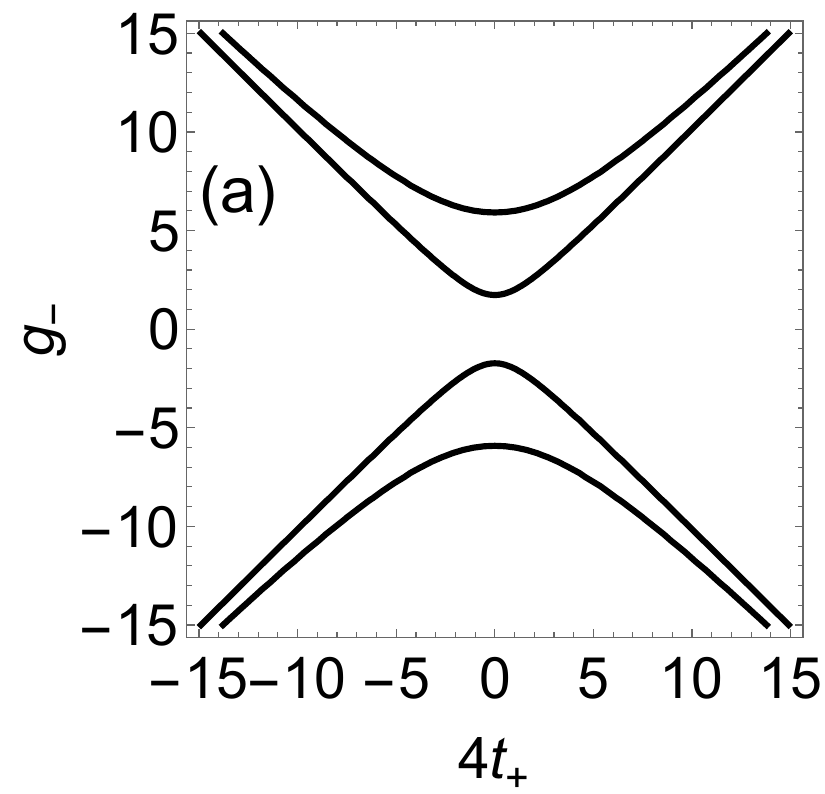}
\includegraphics[width=0.49\linewidth]{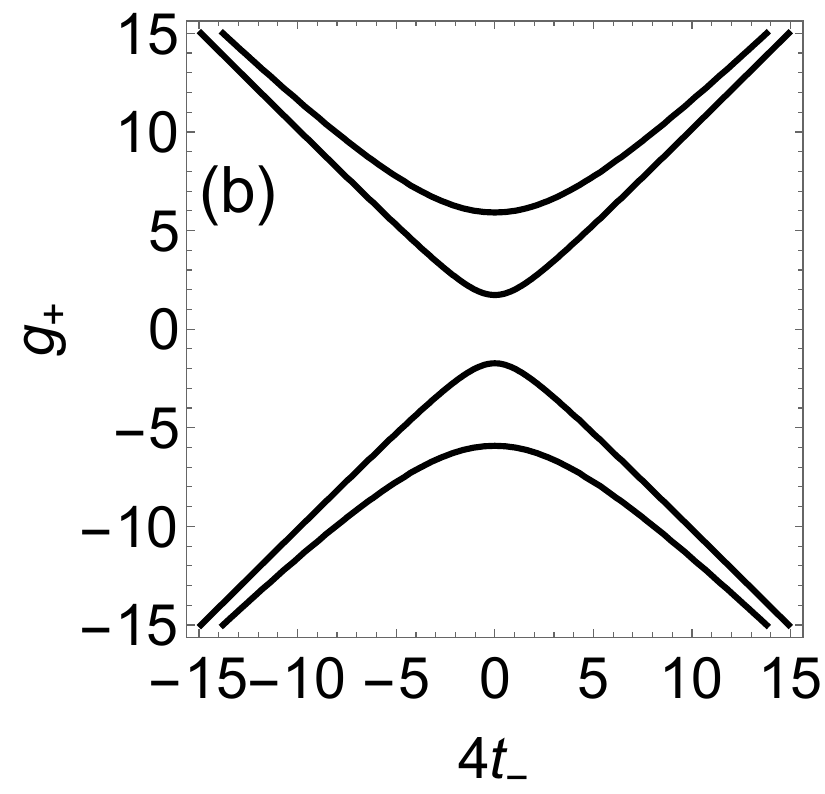}\\    \includegraphics[width=0.49\linewidth]{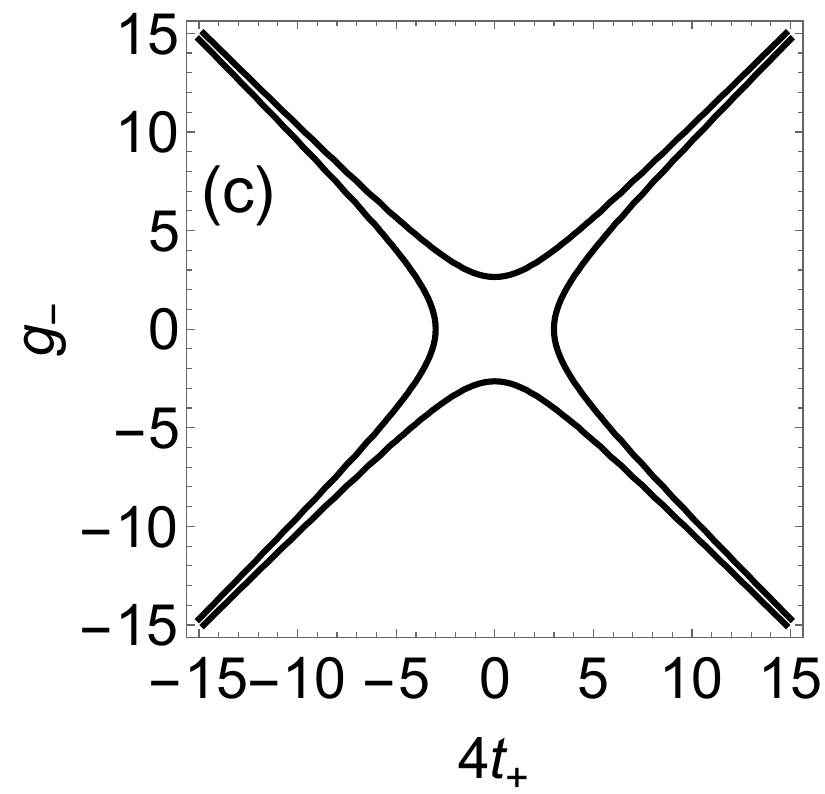}
\includegraphics[width=0.49\linewidth]{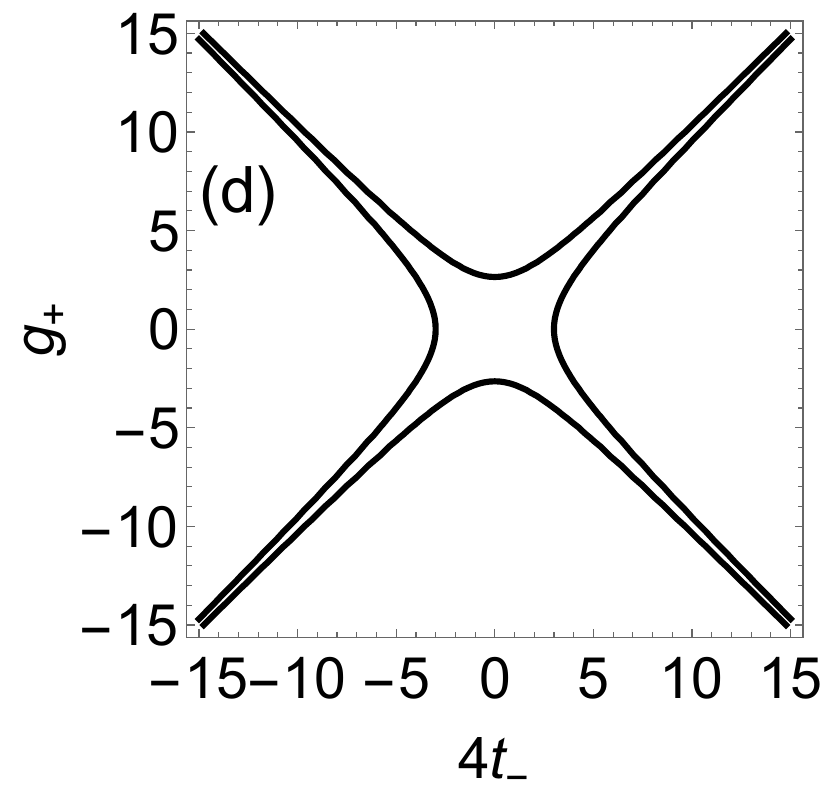}\\
\includegraphics[width=0.49\linewidth]{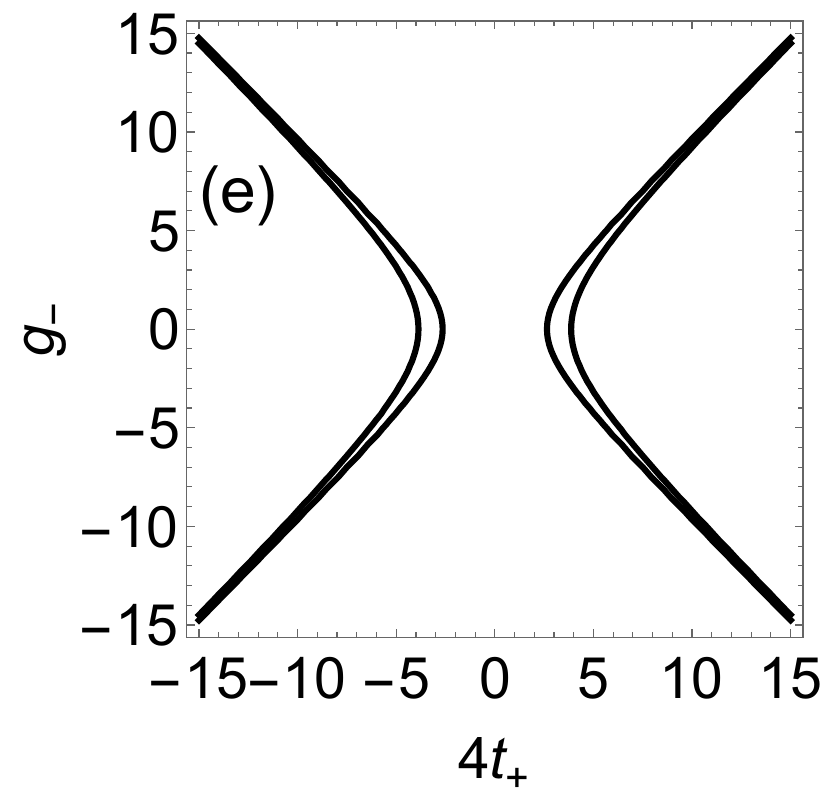}
\includegraphics[width=0.49\linewidth]{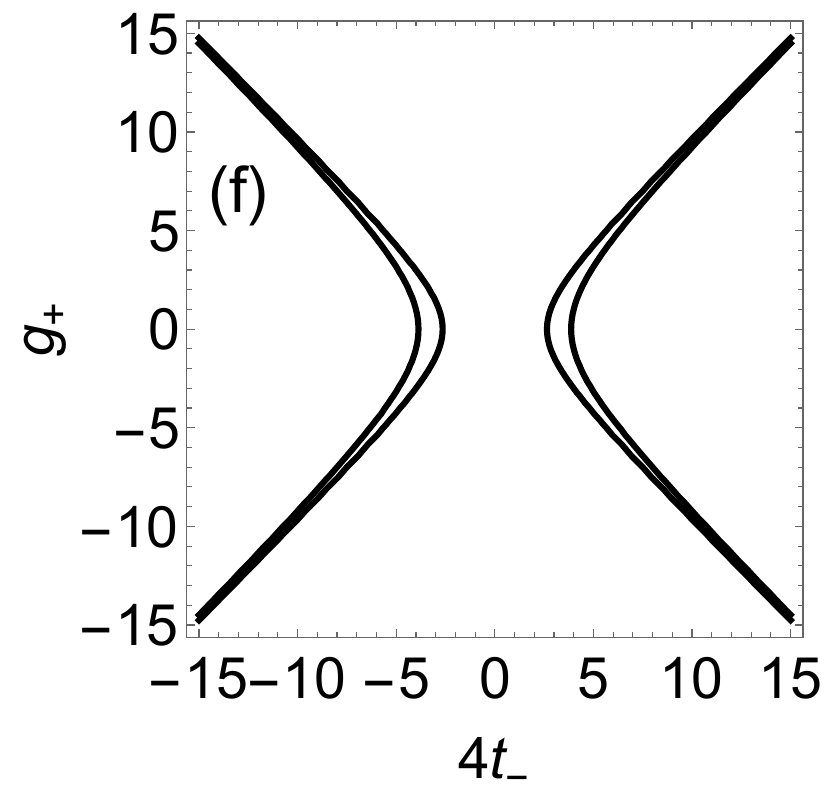}
\caption{Gap closing curves in the  $t_+$-$g_-$ (left column) and $t_-$-$g_+$ (right column) parameter space derived from Eq.~\eqref{Eq:Transition_H_nHSC(i)_general} with $\Delta_{0}=1$ for nonuniform onsite dissipation terms. The adopted values for the dissipation parameters are given by (a,b) $(\Gamma_{+}, \Gamma_{-}) =  (4, 1)$, (c,d) $(\Gamma_{+}, \Gamma_{-}) = (2, 3)$, and (e,f) $(\Gamma_{+}, \Gamma_{-}) = (1, 4)$.
    }
\label{fig:phase_change_with_gammaA_and_gammaB}
\end{figure}

\section{Nonuniform dissipation terms}
\label{sec:non-uniform}

So far we have mainly considered uniform dissipation terms. Below we discuss a more general case with $\Gamma_{a} \neq \Gamma_{b}$, which also includes the staggered limit with $\Gamma_{a} = -\Gamma_{b}$.

\subsection{Gap closing curves for general $\Gamma_a$ and $\Gamma_b$}

For general $\Gamma_a$ and $\Gamma_b$,
we once again have the PBC energy spectrum given by Eq.~\eqref{Eq:E_pbc}.
As before, we consider two cases, where the gap closes (i) at  $k=0,\pm \pi/a_0$ or (ii) away from these special points. 
The gap closing conditions for Case (i) now become
\begin{subequations}
\label{Eq:Transition_H_nHSC(i)_general}
\begin{eqnarray}
    \text{Case (ia$^\prime$)} & \quad &
    - g_{+} ^{2}+16 t_{-} ^{2}-\Gamma^{2}_{-}+D^{2}_{\pm}=0 ,
\label{Eq:Transition_H_nHSC(ia)_general}
\\
 \text{Case (ib$^\prime$)} & \quad & 
-g_{-}^{2}+16t_{+}^{2}-\Gamma^{2}_{-}+D^{2}_{\pm}=0 ,
\label{Eq:Transition_H_nHSC(ib)_general}
\end{eqnarray} 
\end{subequations} 
where we use the prime to denote the more general case.
It can be seen that these conditions differ from Eq.~\eqref{Eq:Transition_H_nHSC(i)} by an extra term proportional to $\Gamma_{-}^2$, but can still be represented in the 
 $t_{+}$–$g_{-}$ and $t_{-}$–$g_{+}$ planes.

In Fig.~\ref{fig:phase_change_with_gammaA_and_gammaB}, we show how these curves 
evolve when varying the parameters $\Gamma_{\pm} = (\Gamma_a \pm \Gamma_b)/2$.  From Fig.~\ref{fig:phase_change_with_gammaA_and_gammaB}(a) to Fig.~\ref{fig:phase_change_with_gammaA_and_gammaB}(f), 
we fix $\Gamma_a$ and gradually decrease $\Gamma_b$, thereby tuning the system from the nearly uniform limit (where $\Gamma_{+}$ is maximized and $\Gamma_{-}$ vanishes) toward the nearly staggered limit (characterized by maximal $\Gamma_{-}$ and vanishing $\Gamma_{+}$)\footnote{As a side remark, here we do not include the set $(\Gamma_{+},\Gamma_{-})=(3,2)$, since the results closely resemble Fig.~\ref{fig:phase_change_with_gammaA_and_gammaB}(c,d) and provide no additional information. For completeness, however, Fig.~\ref{Fig:phase_and_band_diagram_general_perturbation} below shows the corresponding parameter diagrams when we discuss the energy spectra. }.
As $\Gamma_{+}$ decreases and $\Gamma_{-}$ increases, the associated hyperbolic gap-closing curves rotate from a vertical (up-down) to a horizontal (left-right) orientation.
Initially, both pairs of hyperbolae intersect the vertical axis [see Panel~(a,b)]; during the transition [see Panel~(c,d)], one pair shifts and begins to intersect the horizontal axis; eventually, in Fig.~\ref{fig:phase_change_with_gammaA_and_gammaB}(e) and (f), both pairs intersect the horizontal axis.
Accompanying this evolution, the phase regions sandwiched between the two pairs of hyperbolic curves first merge near the origin and then become separated again.

\begin{figure}
    \centering
    \includegraphics[width=0.49\linewidth]{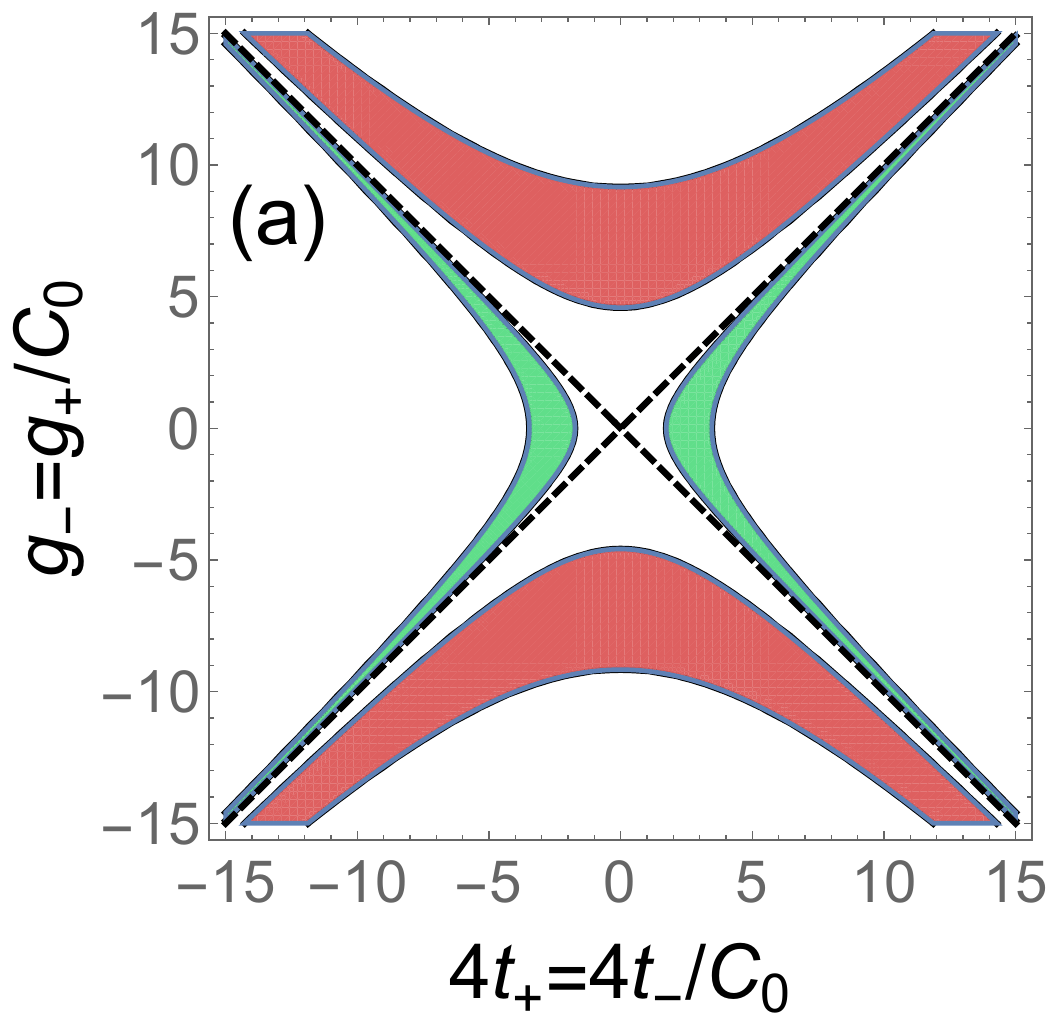}
    \includegraphics[width=0.468\linewidth]{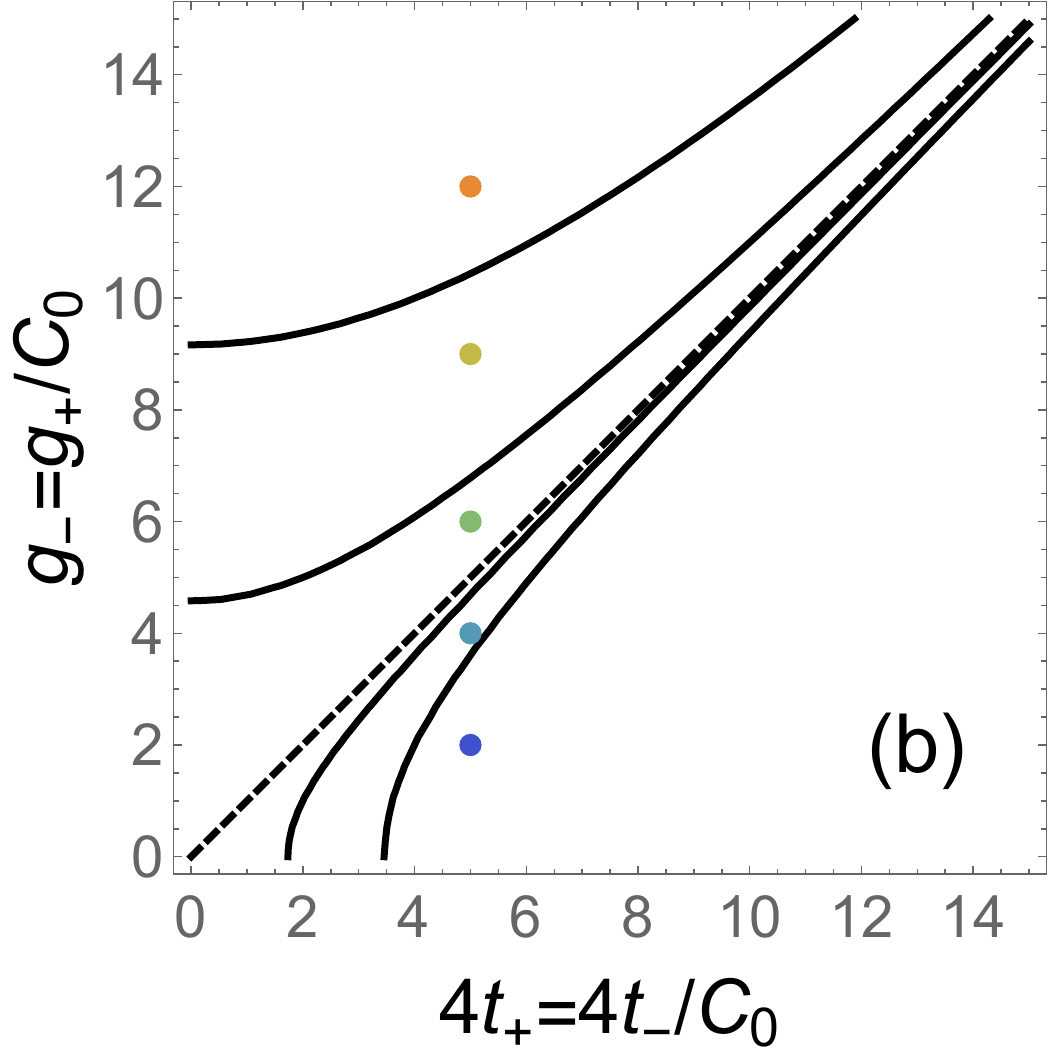} 
    \caption{Phase diagrams and energy spectra for Case (iib$^\prime$) with $\Gamma_{+} = 3$, $\Gamma_{-} = 2$, $\Delta_0 = 1$, and $C_0=0.5$. (a) Phase diagram. The shaded regions show the gapless superconducting phases given by Eq.~\eqref{Eq:GaplessSC_condition(iia_prime)}. (b) A magnified view of the first quadrant of Panel~(a), with the dot colors corresponding to the spectra in Fig.~\ref{fig:general_gamma_gap_close_case(ii)_spectra} (see below). 
    } 
    \label{fig:general_gamma_gap_close_case(ii)}
\end{figure}

We now turn to Case~(ii), where Eq.~\eqref{Eq:Case(ii)_criterion}  is fulfilled. Unlike in the uniform limit, here the sign of $f_{r1}$ is no longer fixed solely by whether $C_0$ exceeds unity due to the presence of a nonzero $\Gamma_{-}$.
However, the conditions under which the gap closes away from $k=0,\pm \pi/a_0$ can still be determined by the value of $C_0$, as given by
\begin{subequations}
\label{Eq:GaplessSC_general_gamma_condition}
\begin{eqnarray}
    \text{Case (iia$^{\prime}$)} 
    \quad 
	 C_{0}^2 > 1 \; & \& \; 
    1 \leqslant \frac{D^{2}_{\pm}-\Gamma^{2}_{-}}{g^{2}_{-} - 16t^{2}_{+}} 
      \leqslant C_{0} ^2 , 
      \label{Eq:GaplessSC_condition(iia_prime)} \\
 \text{Case (iib$^{\prime}$)}
  \quad 
  C_{0}^2 < 1  \; & \&  \; 
        C_{0} ^2 \leqslant \frac{D^{2}_{\pm}-\Gamma^{2}_{-}}{g^{2}_{-} - 16t^{2}_{+}} \leqslant 1 ,
        \label{Eq:GaplessSC_condition(iib_prime)}
\end{eqnarray}
\end{subequations}
according to which we obtain the parameter regimes where the gapless superconductivity appears. 
An example for $C_0^2 < 1$ is presented in Fig.~\ref{fig:general_gamma_gap_close_case(ii)}, with the gapless superconducting phases marked in the shaded areas.

In contrast to the uniform $\Gamma_{a,b}$ limit in 
Eq.~\eqref{Eq:GaplessSC_condition} and Fig.~\ref{fig:energy_spectry_case(iib)}, here the quantities, $ D^{2}_{\pm}-\Gamma^{2}_{-}$, 
in Eq.~\eqref{Eq:GaplessSC_general_gamma_condition} can take negative value(s), allowing for the hyperbolic curves to have intercepts with the horizontal axis. Indeed, as shown in Fig.~\ref{fig:general_gamma_gap_close_case(ii)}, we identify two distinct pairs of gapless regions: one (red) bounded by two sets of hyperbolic curves with intercepts on the vertical axis, and the other (green) bounded by hyperbolic curves with intercepts on the horizontal axis.

Before analyzing the OBC spectra in the nonuniform dissipation limit, we first examine the gap closing condition in the staggered dissipation limit and show that in this case Majorana zero modes do not appear.

\subsection{Gap closing curves in the staggered limit }
\label{Sec:Staggered dissipation term}

We now turn to the gap-closing condition in the limit opposite to uniform dissipation. In the literature on the non-Hermitian SSH model without pairing~\cite{Zhang:2018,Halder:2022}, a staggered dissipation term ($\Gamma_{a}=-\Gamma_{b}\equiv \Gamma_{0}^{\prime}$) is often considered, which has been shown to yield interesting spectral features. However, we demonstrate below that this limit does not support Majorana zero modes in our system. In our notation, this limit corresponds to $\Gamma_{+}\to 0$ and $\Gamma_{-}\to \Gamma_{0}^{\prime}\neq 0$.

\begin{figure}[t]
    \centering    
\includegraphics[width=0.49\linewidth]{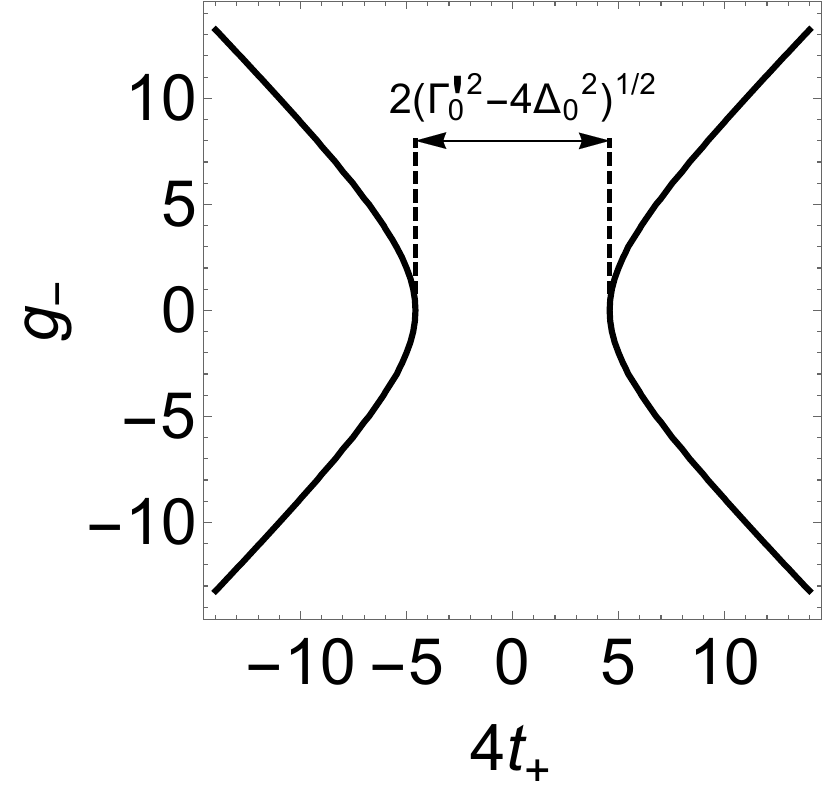}
\includegraphics[width=0.49\linewidth]{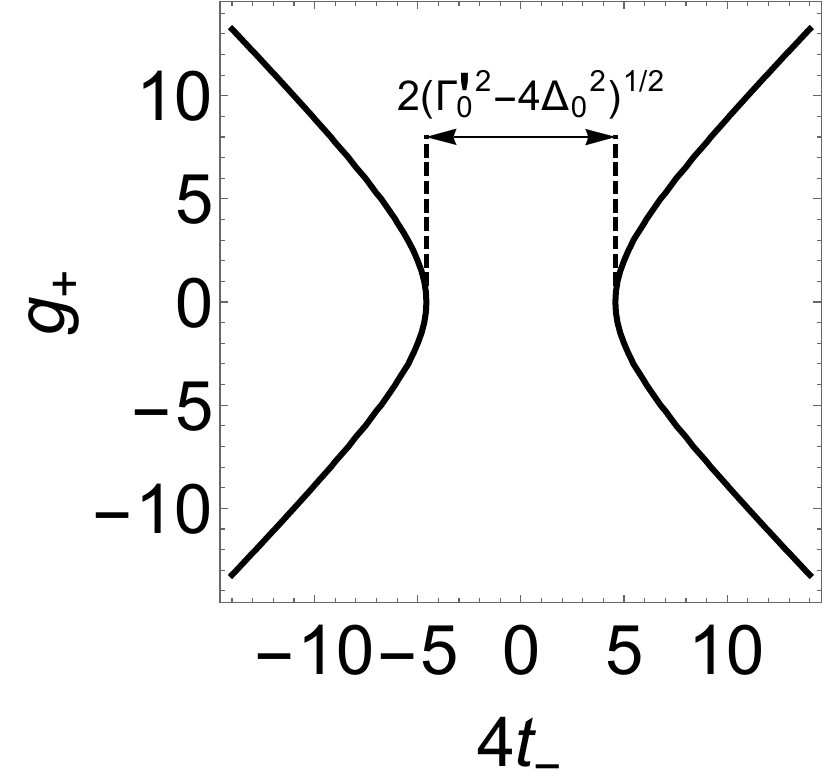}
    \caption{Gap closing curves in the $t_+$-$g_-$ (left)  and $t_-$-$g_+$ (right) space in the staggered limit. The adopted parameter values are given by  $\Gamma^{\prime}_0 = 5$ and $\Delta_0 = 1$. 
    }
\label{fig:phase_diagram_gammaA=-gammaB}
\end{figure}

In Fig.~\ref{fig:phase_diagram_gammaA=-gammaB}, we show the corresponding phase diagram and the gap-closing curves, as described by 
\begin{subequations}
\label{Eq:Transition_H_nHSC(i)_staggere}
\begin{eqnarray}
    & \quad &
    - g_{+} ^{2}+16 t_{-} ^{2}-(\Gamma^{\prime}_{0})^2 +4\Delta^{2}_{0}=0 ,
\label{Eq:Transition_H_nHSC(ia)_staggere}
\\
  & \quad & 
-g_{-}^{2}+16t_{+}^{2}-(\Gamma^{\prime}_{0})^2 + 4\Delta^{2}_{0}=0 ,
\label{Eq:Transition_H_nHSC(ib)_staggere}
\end{eqnarray} 
\end{subequations} 
obtained from Case (ia$^\prime$) and (ib$^\prime$) in Eq.~\eqref{Eq:Transition_H_nHSC(i)_general}.   
Since we choose 
$ |\Gamma^{\prime}_0| > |\Delta_0| $, 
it can be seen that these curves are hyperbolic with intercepts on the $t_{\pm}$ axes given by $\pm \sqrt{(\Gamma_{0}^{\prime})^2 - 4\Delta_{0}^{2}}$, in contrast to the vertical orientation in the uniform dissipation limit shown in Fig.~\ref{Fig:GapClosing_H_nHSC}.
Alternatively, if we choose 
$ |\Gamma^{\prime}_0| < |\Delta_0| $,
then we will instead obtain hyperbolic curves with intercepts on the $g_{\pm}$ axes.

\begin{figure*}[th]
    \centering
    \includegraphics[width=0.99\linewidth]{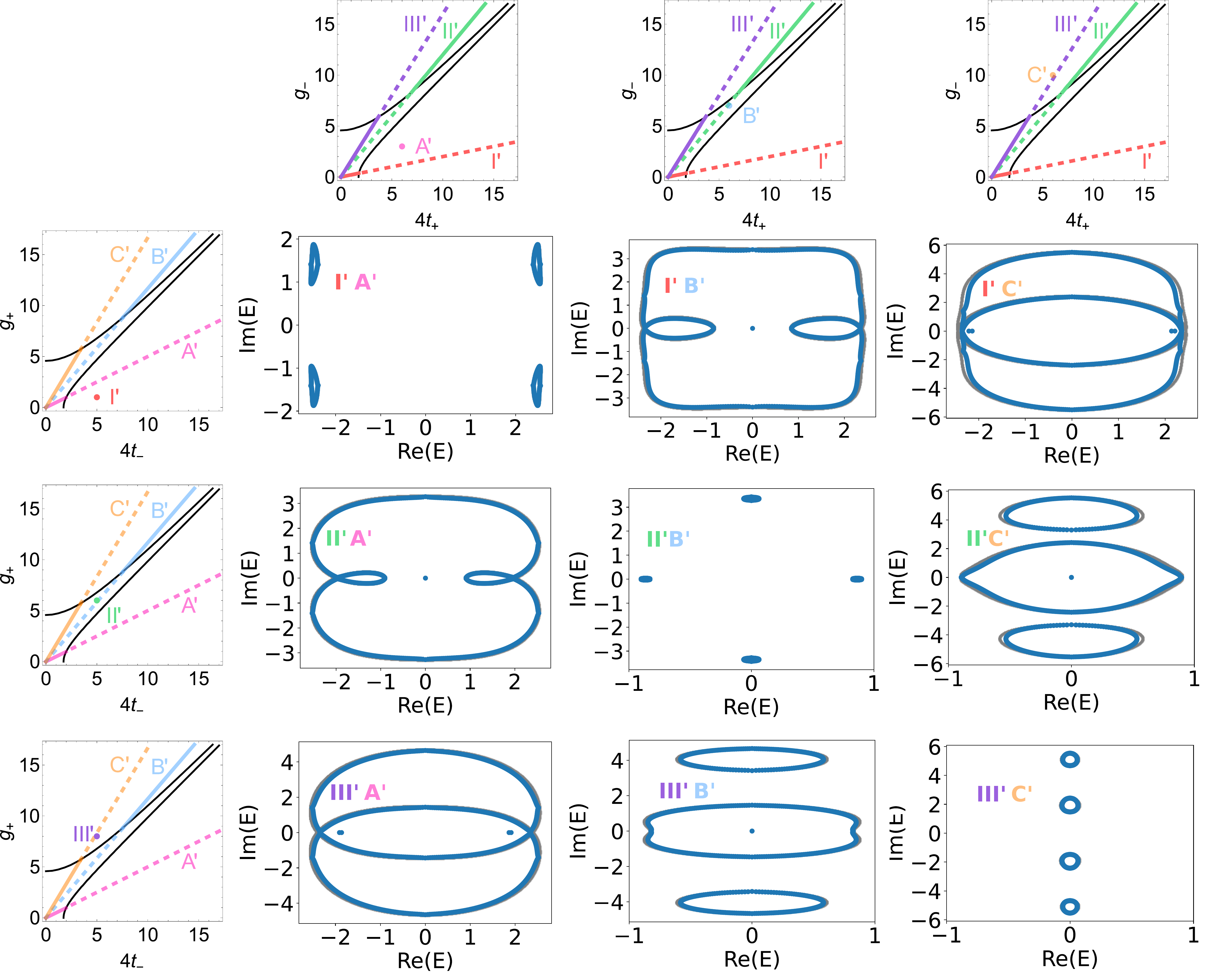}
    \caption{Similar plots as Fig.~\ref{Fig:phase_and_band_diagram_case1_perturbation}, but with nonuniform dissipation terms, $(\Gamma_{+},\Gamma_{-}) =(3,2)$, corresponding to $\Gamma_a = 5$ and $\Gamma_b = 1$. See Table~\ref{Table:Parameters} for the adopted values of the full parameter sets. 
    }
\label{Fig:phase_and_band_diagram_general_perturbation}
\end{figure*}

Importantly for the topological zero modes, when $\Gamma_{+}=0$ we have $D_{+}=D_{-}$, so that the two hyperbolic curves overlap in Fig.~\ref{fig:phase_diagram_gammaA=-gammaB}. Together with the evolution of the gap-closing curves shown in Fig.~\ref{fig:phase_change_with_gammaA_and_gammaB}, this implies that finite regions bounded by the hyperbolic curves (as required in at least one of the $t_{+}$-$g_{-}$ or $t_{-}$-$g_{+}$ planes for zero modes) are absent. As a result, the regions hosting Majorana zero modes are eliminated in the absence of uniform dissipation. Therefore, we expect that no Majorana zero modes appear in the staggered limit.

\subsection{Spectral features and Majorana zero modes  }
\label{Sec:spectra_general-gamma}

\begin{figure}[h]
    \centering
    \includegraphics[width=0.5\linewidth]{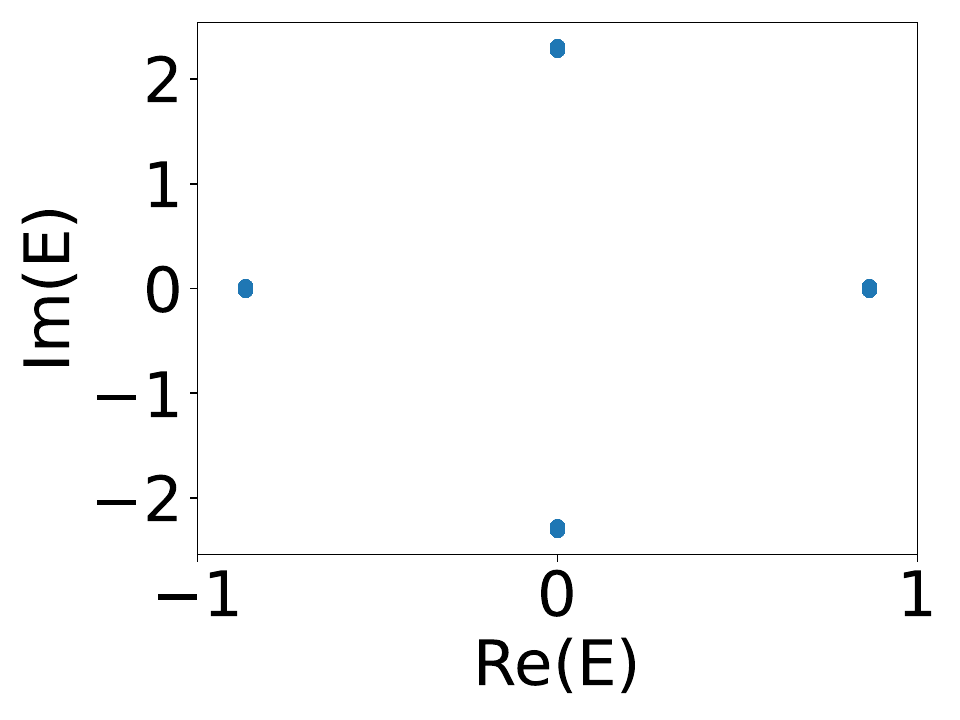}
    \caption{OBC (blue) and PBC (gray) spectra for  
    $t_{+}=6$, $t_{-}=5$, $g_{+}=5$, $g_{-}= 6$, $\Delta_{0}=1$,  and $\delta h_x=0.01$. The plot corresponds to the parameter set II~B in Fig.~\ref{Fig:phase_and_band_diagram_case1_perturbation}, 
    but with $\Gamma_{+}=2$ and $\Gamma_{-} =3$. 
    }
    \label{fig:energy_spectrum_general_(iso_a)}
\end{figure}

\begin{figure}[t]
    \centering
    \includegraphics[width=0.49\linewidth]{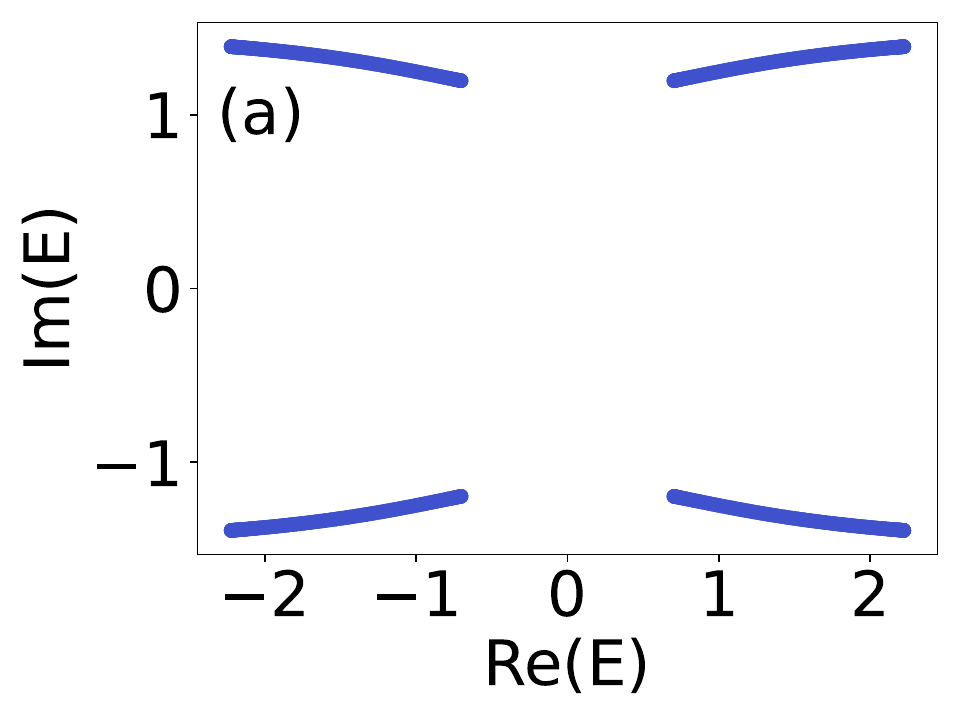}
    \includegraphics[width=0.49\linewidth]{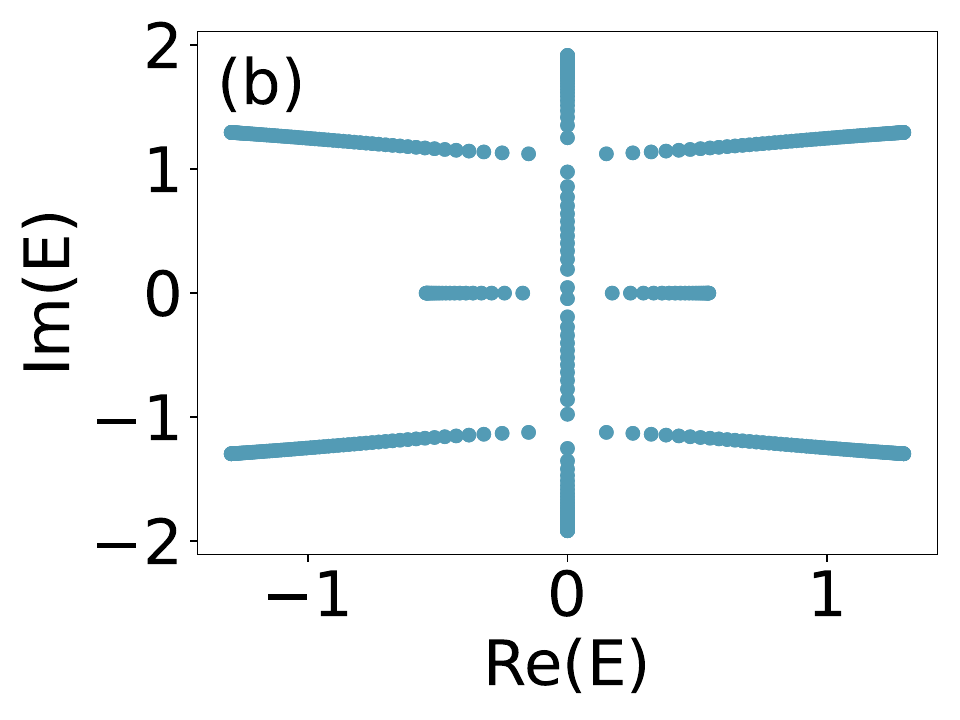}
    \\
    \includegraphics[width=0.49\linewidth]{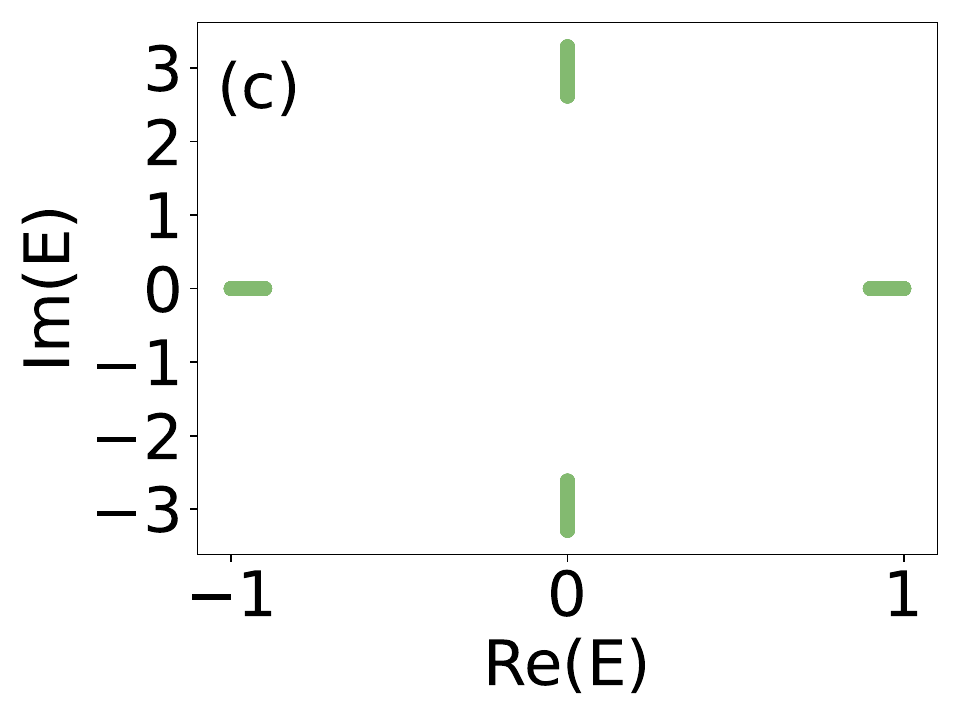}
    \includegraphics[width=0.49\linewidth]{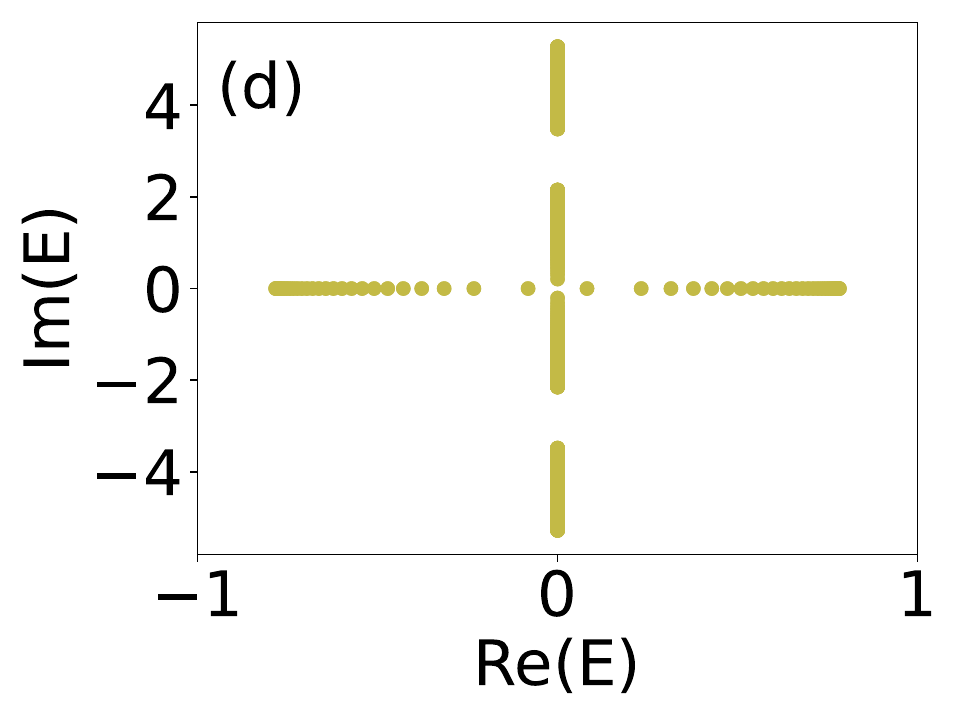}
    \\
    \includegraphics[width=0.49\linewidth]{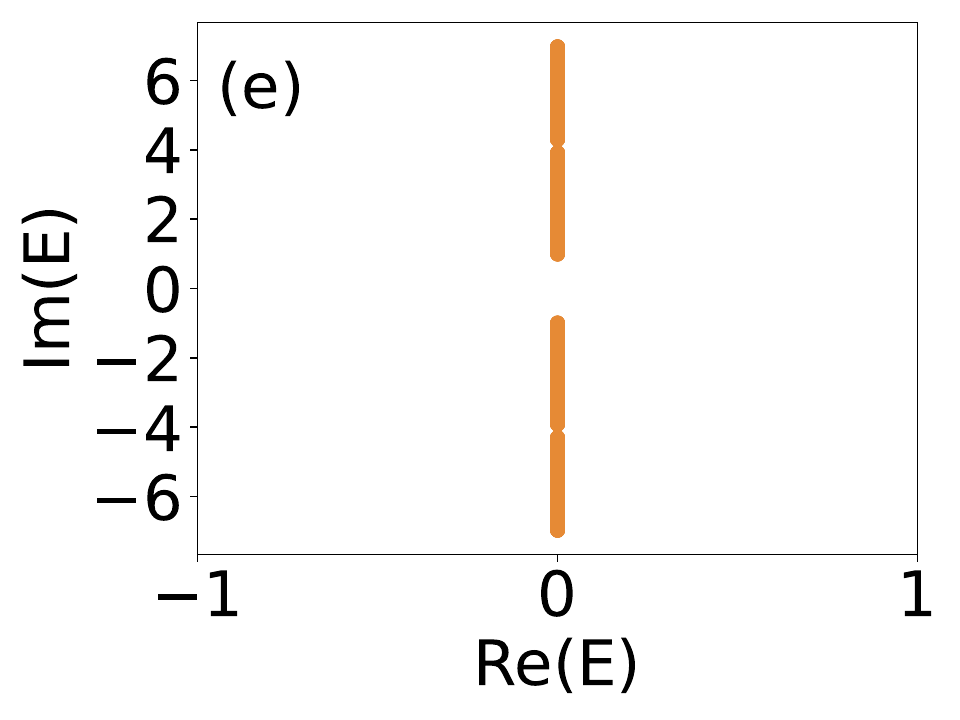}
    \caption{Energy spectra for Case~(iib$^\prime$) with 
    $\Gamma_{+}=3$, $\Gamma_{-}=2$, $\Delta_{0}=1$, and $C_{0}=0.5$. The spectra are computed at the parameter sets marked by the colored dots in Fig.~\ref{fig:general_gamma_gap_close_case(ii)}(b), with $t_{+}=5/4$ and $g_{-}=2$, $4$, $6$, $9$ and $12$. 
    See Table~\ref{Table:Parameters} for the adopted values of the full parameter sets. 
    Here, the PBC spectra   coincide with the OBC ones.
}
    \label{fig:general_gamma_gap_close_case(ii)_spectra}
\end{figure}

\begin{figure}[h]
    \centering
\includegraphics[width=0.99\linewidth]{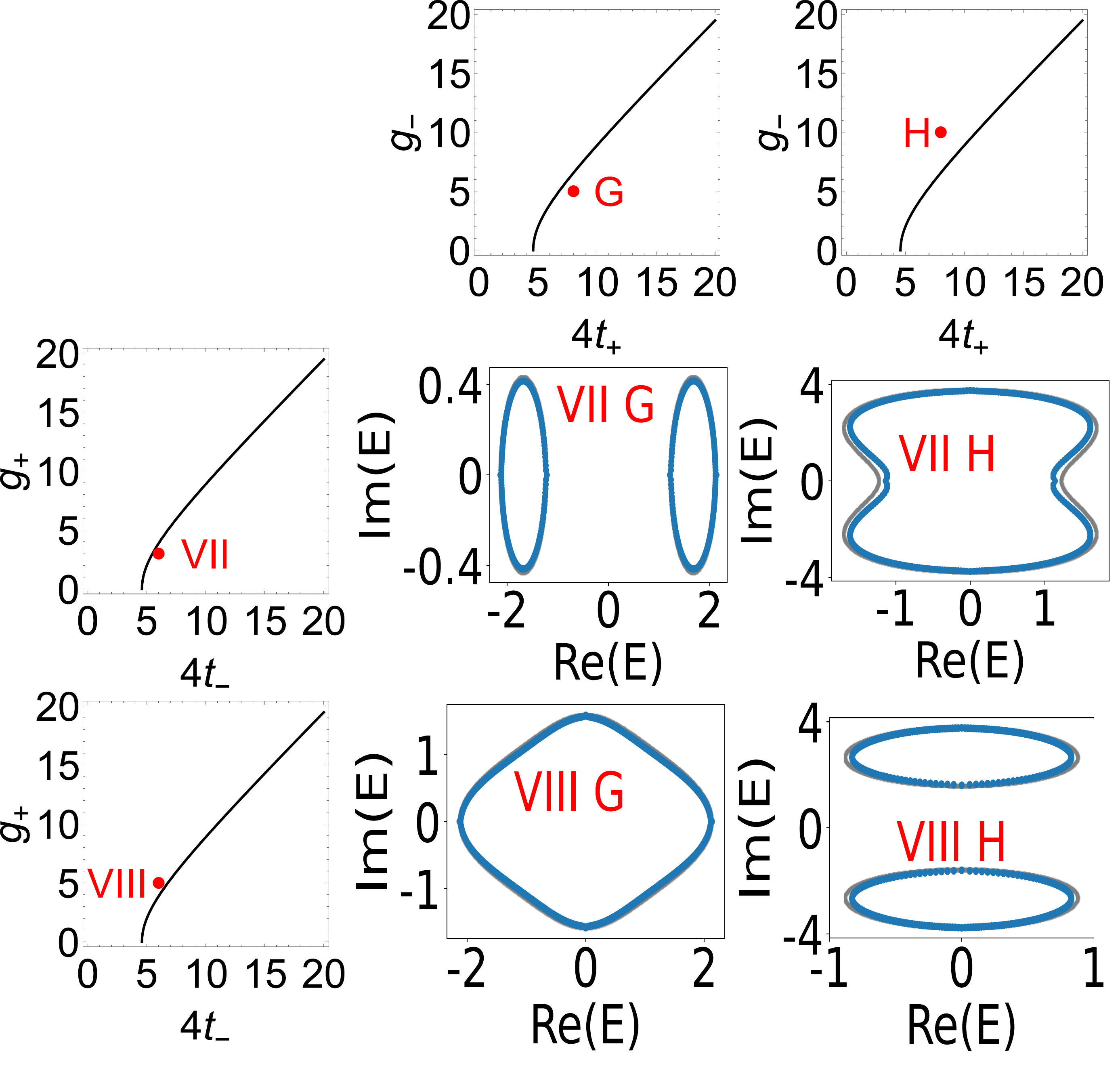}
    \caption{Parameter diagrams and energy spectra; similar plot as Fig.~\ref{Fig:phase_and_band_diagram_case1_perturbation}, but with $\Gamma^{\prime}_{0}= \Gamma_{a} = -\Gamma_{b} =5$,  $\Delta_{0}=1$, and $\delta h_x = 0.01$.
    See Table~\ref{Table:Parameters} for the adopted values of the full parameter sets. }
\label{fig:energy_spectra_GammaA=-GammaB}
\end{figure}

To support the above observation about the evolution of the gap closing curves for $\Gamma_a \neq \Gamma_b$, we compute the OBC and PBC energy spectra under small onsite transverse magnetic fields
as before.
The results are summarized 
in Figs.~\ref{Fig:phase_and_band_diagram_general_perturbation}
[corresponding to Fig.~\ref{fig:phase_change_with_gammaA_and_gammaB}(c,d)], 
Fig.~\ref{fig:energy_spectrum_general_(iso_a)}, 
Fig.~\ref{fig:general_gamma_gap_close_case(ii)_spectra}
(corresponding to Fig.~\ref{fig:general_gamma_gap_close_case(ii)}), 
and Fig.~\ref{fig:energy_spectra_GammaA=-GammaB} (corresponding to the staggered  dissipation limit in Fig.~\ref{fig:phase_diagram_gammaA=-gammaB}).

In Figs.~\ref{Fig:phase_and_band_diagram_general_perturbation}--\ref{fig:energy_spectrum_general_(iso_a)}, we present the spectra and gap-closing conditions for Case~(i). To explore a broader parameter space, in Fig.~\ref{Fig:phase_and_band_diagram_general_perturbation} we define three representative parameter sets, A$^\prime$, B$^\prime$ and C$^\prime$, in distinct regions determined by the hyperbolic gap-closing curves in the $t_{+}$–$g_{-}$ plane, and sets I$^\prime$, II$^\prime$ and III$^\prime$ in the $t_{-}$–$g_{+}$ plane. Remarkably, for general $\Gamma_{a,b}$, Majorana zero modes also appear in the parameter regimes I$^\prime$~B$^\prime$, II$^\prime$~A$^\prime$, II$^\prime$~C$^\prime$ and III$^\prime$~B$^\prime$, matching those in Fig.~\ref{Fig:phase_and_band_diagram_case1_perturbation}, though with boundaries shifted by the finite $\Gamma_{-}$. We have further verified that the corresponding zero-mode wavefunctions are localized at the system boundaries (not shown for brevity).

Notably, since the $\Gamma_{-}$ term preserves all existing symmetries of the system, most of the symmetry-protected features are also preserved. Specifically, the energy spectra remain symmetric with respect to both the real and imaginary axes, as guaranteed by the PHS and pH.
For nonuniform $\Gamma_{a,b}$,
complex flat bands also appear, since the conditions in Eq.~\eqref{Eq:Case(ii)_criterion} and Eq.~\eqref{Eq:complex_flex_band_condition(iso-a)} can also be satisfied in this regime. An example is illustrated in  Fig.~\ref{fig:energy_spectrum_general_(iso_a)}, which corresponds to the set II$^\prime$~B$^\prime$ of Fig.~\ref{Fig:phase_and_band_diagram_general_perturbation}, but with different $\Gamma_{+}$ and $\Gamma_{-}$ values.

Additionally, we employ the topological invariant defined in Eq.~\eqref{Eq:winding} to characterize the parameter regimes with $\Gamma_{-}\neq 0$. The winding number satisfies $W_{\delta,\pm}=0$ in regions without Majorana zero modes, while in the remaining regions it obeys
$ W_{+,+}=W_{-,-}=-W_{+,-}=-W_{-,+}$.
In particular, $W_{+,+}$ takes the values $(-1,-1,1,1)$ for the parameter sets I$^\prime$~B$^\prime$, II$^\prime$~A$^\prime$, II$^\prime$~C$^\prime$ and III$^\prime$~B$^\prime$, respectively, in agreement with the appearance of zero modes.

We now move on the spectra in  Case~(ii), which can host gapless superconducting phases, and present the results in 
Fig.~\ref{fig:general_gamma_gap_close_case(ii)_spectra}.  
As in the uniform limit, all spectra remain symmetric with respect to both the real and imaginary axes, with the OBC spectrum coinciding with the PBC spectrum due to the suppression of the skin effect in this regime.
With the inclusion of the $\Gamma_{-}$ term, a new type of spectrum emerges in the gapless non-Hermitian superconducting phase; see Fig.~\ref{fig:general_gamma_gap_close_case(ii)_spectra}(b). 
Furthermore, the spectra become either real or purely imaginary when the parameters lie above the asymptotic lines, as illustrated in Fig.~\ref{fig:general_gamma_gap_close_case(ii)_spectra}(c)--(e).

Finally, we examine the staggered dissipation limit, shown in Fig.~\ref{fig:energy_spectra_GammaA=-GammaB}. In the first row, two parameter sets (G and H) are defined in the $t_{+}$-$g_{-}$ plane, separated by a hyperbolic gap-closing curve, while in the first column, two sets (VII and VIII) are similarly defined in the $t_{-}$-$g_{+}$ plane. As anticipated from the evolution of the gap-closing curves (see the discussions in Sec.~\ref{Sec:Staggered dissipation term}), no Majorana zero modes appear in any of these regions. Consistently, the topological invariant vanishes throughout these regimes. In particular, we demonstrate that the commonly assumed staggered dissipation in non-Hermitian SSH model does not give rise to Majorana zero modes in this system.

\begin{figure}[h]
    \centering
    \includegraphics[width=0.49\linewidth]{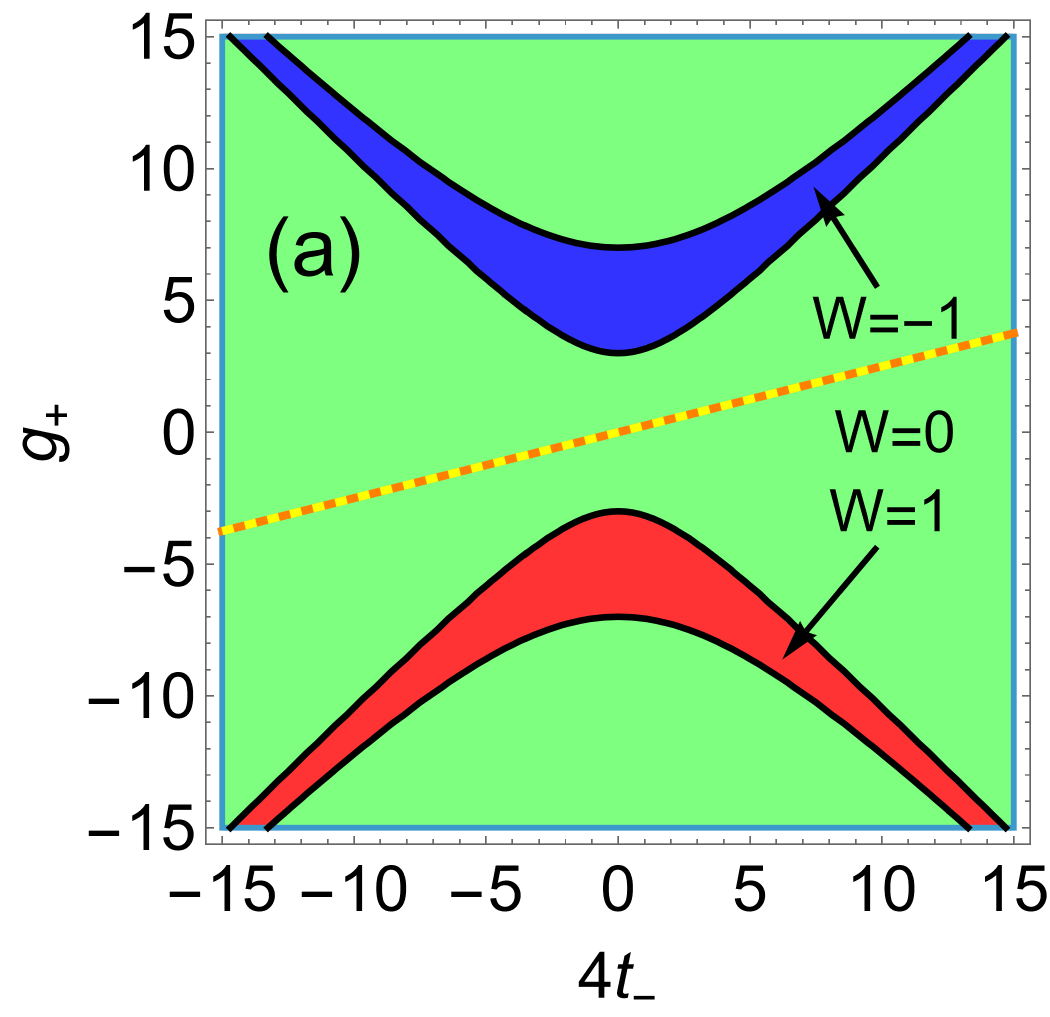}
    \includegraphics[width=0.49\linewidth]{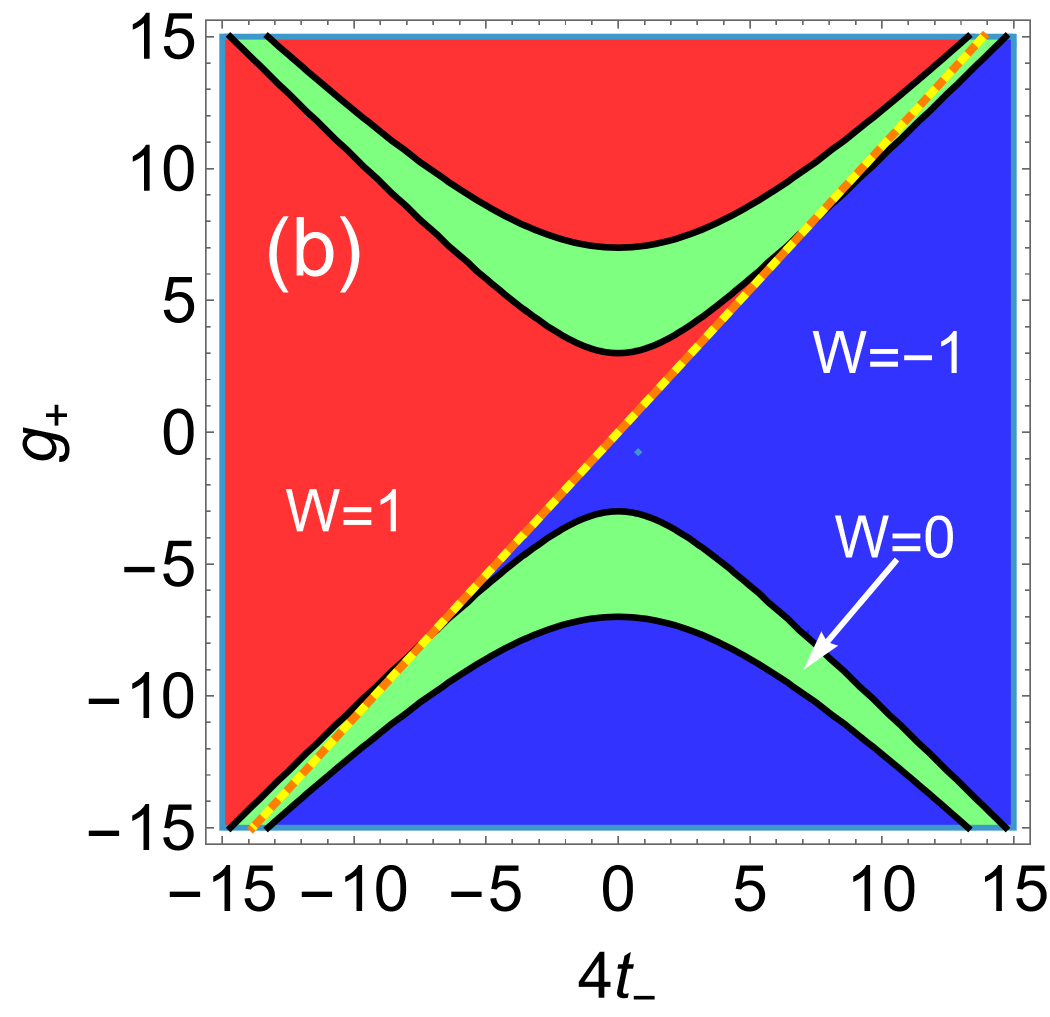}\\
    \includegraphics[width=0.49\linewidth]{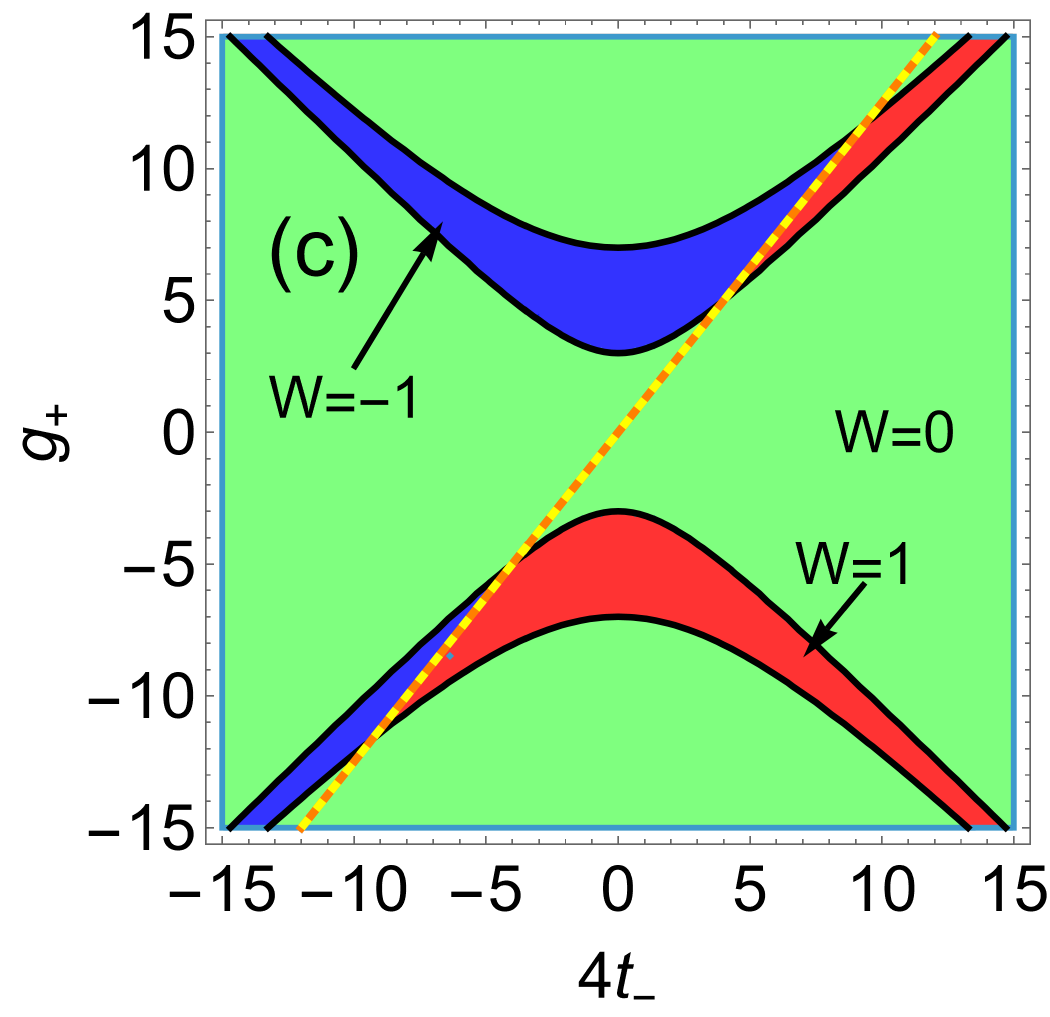} 
    \includegraphics[width=0.49\linewidth]{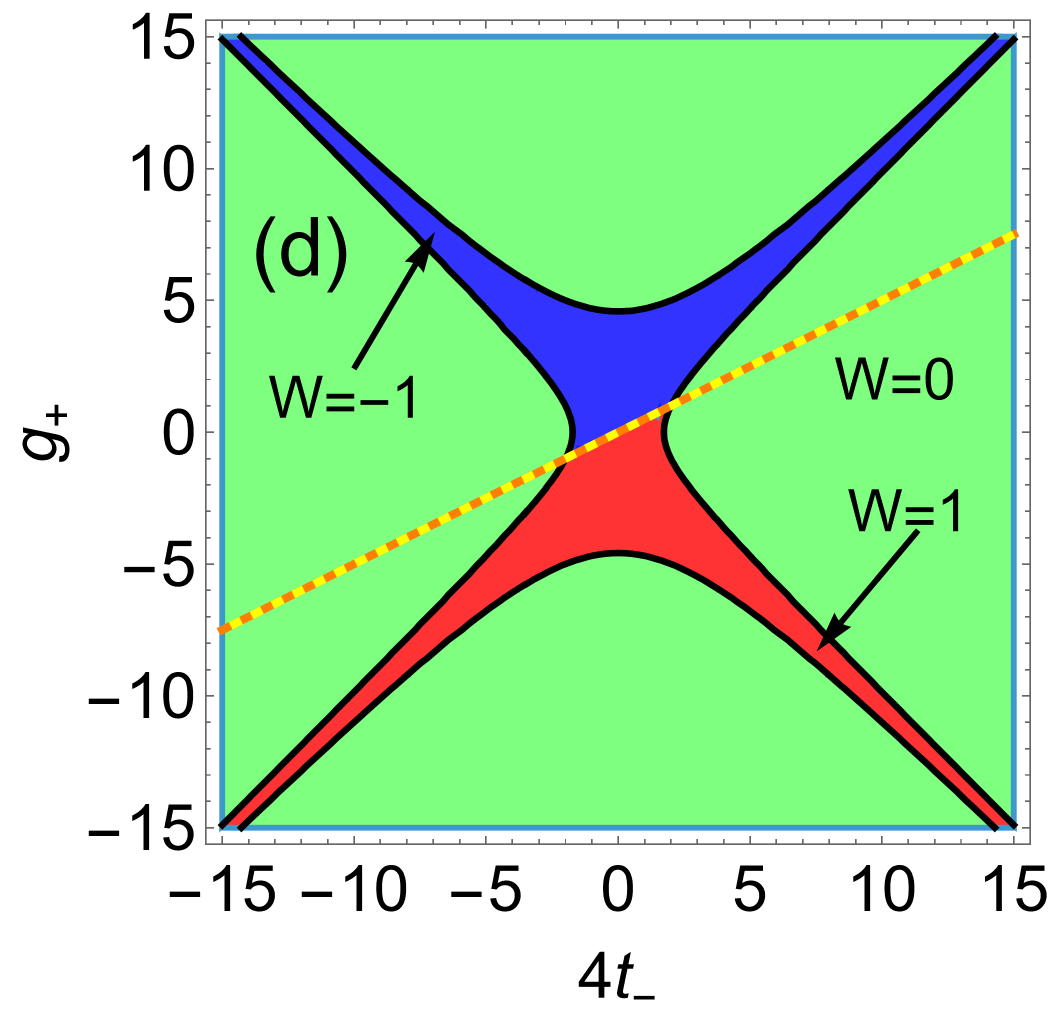} 
    \caption{Computed winding number, $W_{+,+}$, in the $t_{-}$–$g_{+}$ plane with $\Delta_{0}=1$. Panels~(a--c) correspond to $\Gamma_{+}=5$ and $\Gamma_{-}=0$, matching parameter sets A, B and C in Fig.~\ref{Fig:phase_and_band_diagram_case1_perturbation}, respectively, while Panel~(d) corresponds to $\Gamma_{+}=3$ and $\Gamma_{-}=2$, matching parameter set A$^\prime$ in Fig.~\ref{fig:energy_spectrum_general_(iso_a)}. In each panel, the winding number takes values ${0,\pm1}$, as indicated. The solid curves mark Eq.~\eqref{Eq:Transition_H_nHSC(i)_general}, and the dashed lines mark the condition for a gapless superconducting phase in Eq.~\eqref{Eq:Case(ii)_criterion}.
    }
    \label{Fig:Winding_number}
\end{figure}

\begin{figure}[h]
    \centering
\includegraphics[width=0.9\linewidth]{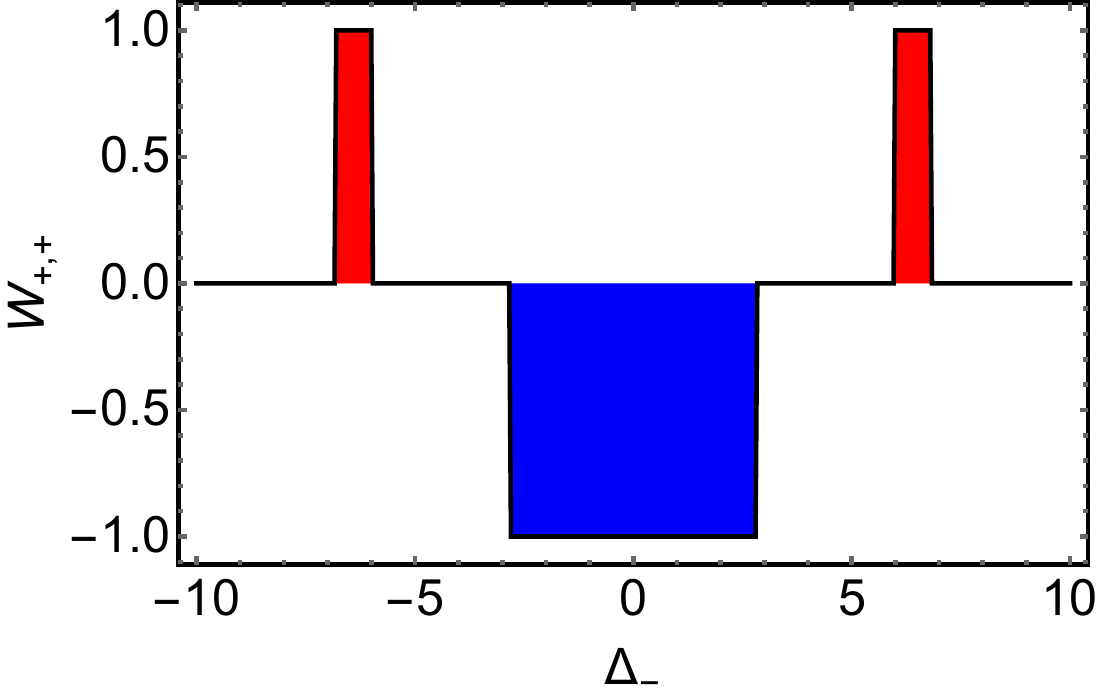}
    \caption{Computed winding number, $W_{+,+}$, as a function of the staggered component $\Delta_{-}$ of the pairings. The winding number is computed using Eq.~\eqref{Eq:winding} with the function $P_{\zeta, \pm}$ replaced with Eq.~\eqref{Eq:P-fn-nonuniform-pairing}. The shading colors correspond to those in Fig.~\ref{Fig:Winding_number}.
    The other parameters corresponding to the set II~A in Fig.~\ref{Fig:phase_and_band_diagram_case1_perturbation}. 
    } 
\label{Fig:Winding_IIA_DeltaMinus}
\end{figure}

\section{More general systems  }
\label{Sec:winding_general}

In this section, we consider more general settings with the sublattice structure and compute the corresponding winding number in the parameter space.  

\subsection{Topological phase diagram from the winding number }
\label{Sec:winding_general-gamma}

To further establish the connection between the topological phase transitions and the winding number in Eq.~\eqref{Eq:winding}, we compute the winding number across a broader parameter space. Since the winding numbers of different blocks are mutually dependent, satisfying
\begin{equation}
W_{+,+}=W_{-,-}=-W_{+,-}=-W_{-,+},
\end{equation}
we present the calculated $W_{+,+}$ in Fig.~\ref{Fig:Winding_number}.

For direct comparison with the spectra in Fig.~\ref{Fig:phase_and_band_diagram_case1_perturbation}, we choose three sets of $(t_{+},g_{-})$ corresponding to parameter sets A, B and C, and scan across the $t_{-}$-$g_{+}$ plane, as displayed in Fig.~\ref{Fig:Winding_number}(a)--(c). The winding number is evaluated in all regions separated by the gap-closing curves, allowing us to establish its correspondence with the appearance of Majorana zero modes: clearly, Majorana zero modes emerge when the winding number is $\pm 1$ and are absent when it is $0$.
Notably, the topological invariant changes not only at the gap-closing curves derived from Eq.~\eqref{Eq:Transition_H_nHSC(i)_general}, but also under the conditions of Eq.~\eqref{Eq:Case(ii)_criterion}, which characterize the gapless superconducting phases. 
Finally, Panel~(d) extends the analysis to the case of nonuniform dissipation, $\Gamma_{a} \neq \Gamma_{b}$, corresponding to parameter set A$^\prime$ in Fig.~\ref{Fig:phase_and_band_diagram_general_perturbation}.

We thereby complete the topological phase diagram deduced from the winding number.
Remarkably, while the winding number changes by $\pm 1$ when crossing the gap-closing curves determined by Eq.~\eqref{Eq:Transition_H_nHSC(i)_general}, it changes by $\pm 2$ when crossing those of Eq.~\eqref{Eq:Case(ii)_criterion}, and thus has no direct impact on the presence of Majorana zero modes in the present analysis.

\subsection{Nonuniform pairing}
\label{Sec:winding_general_pairing}

In the above analysis, we have considered uniform onsite pairing  in different  sublattices. In this section, we explore the topological properties in the presence of  nonuniform pairing terms on the sublattice A and B. To this end, we generalize the pairing terms in  Eq.~\eqref{Eq:H_nHSC} into the following form,
\begin{eqnarray}
    \sum_{j}\left(\Delta_{a} a_{j,\uparrow}^{\dagger}a_{j,\downarrow}^{\dagger}
    + \Delta_{b} b_{j,\uparrow}^{\dagger}b_{j,\downarrow}^{\dagger}+\mathrm{H. c.}\right) .
\end{eqnarray}
This amounts to include an additional term
$\Delta_- \eta^{y}\tau^{z}\sigma^{y}$, in the PBC Hamiltonian in Eq.~\eqref{Eq:H_pbc2}, representing the staggered component $\Delta_-= (\Delta_a - \Delta_b)/2 $ of the pairings on the sublattices. 
We verify that the system symmetry remains unchanged with this nonuniform onsite pairing. 

Following the same Hermitianization and block-diagonalization procedure as in Sec.~\ref{Sec:winding}, we obtain the winding number as Eq.~\eqref{Eq:winding} with the function $P_{\zeta, \pm}$ in Eqs.~\eqref{Eq:H_blocks-for-winding}--\eqref{Eq:def_P-fn} replaced with 
\begin{equation}
    P_{\zeta, \pm} (\Gamma_{-}) \rightarrow P_{\zeta, \pm} (\Gamma_{-} \pm \Delta_-). 
    \label{Eq:P-fn-nonuniform-pairing} 
\end{equation}

In Fig.~\ref{Fig:Winding_IIA_DeltaMinus}, we show how the generalized winding number evolves with the staggered pairing strength $\Delta_-$, while keeping the other parameters same as the parameter set II~A. 
For $\Delta_{-}=0$, the winding number $W_{+,+} = -1$ is consistent with the energy spectrum of II~A in Fig.~\ref{Fig:phase_and_band_diagram_case1_perturbation} and the winding number in Fig.~\ref{Fig:Winding_number}. As discussed above, topological zero modes appear when the winding number is nonzero. 
From the plot one observes the winding number remains unchanged for small $|\Delta_{-}|$, indicating the robustness of the topological phase and zero modes against a small stagger pairing component. As $|\Delta_-|$ further increases, the winding number changes, signifying additional topological phase transitions.

\section{discussion }
\label{Sec:discussion}

In this work, we analyze the energy spectra and density profiles of a one-dimensional non-Hermitian superconducting lattice with sublattices and onsite dissipation. Remarkably, the system accommodates all possible internal symmetries characteristic of non-Hermitian systems, including TRS, PHS, their dagger counterparts, CS, SLS, and pH, with all the relations and their corresponding unitary operators listed in Table~\ref{Table:Symmetries_with_perturbation}. These symmetries enforce stringent constraints, giving rise to rich spectral and wavefunction features. 
The symmetry-enforced relations between different components of the density profiles, across the right and left eigenstates, might be understood within the broader framework of  quantum mechanics and non-Hermitian symmetries~\cite{Sato:2012,Kohei:2018,Kohei:2019,Kawabata:2019,Okuma:2023}. In Hermitian systems, the TRS operator connects two eigenstates forming time-reversal partners~\cite{Hasan:2010,Hsu:2021}; in the non-Hermitian setting, this notion can be generalized to relations between right and left eigenstates through TRS and pH~\cite{Sato:2012}. In our case, the combined action of CS and pH further extends this scheme, leading to correlations between right and left eigenstates with particle-hole and opposite spin components that are absent in the Hermitian limit. Our results thus demonstrate how the interplay between conventional and dagger-type symmetries enriches the structure of non-Hermitian Majorana zero modes.

Furthermore, we have identified the conditions for the emergence of Majorana zero modes. In particular, a uniform component of onsite dissipation on the two sublattice sites is essential for stabilizing the zero modes, whereas a purely staggered dissipation suppress them. This finding underscores the crucial role of dissipation engineering, raising the broader question of how controlled gain and loss mechanisms, 
such as engineered dissipation in cold atoms using optical Feshbach resonances~\cite{Yoshida:2019}, 
may be exploited to drive topological phase transitions in non-Hermitian superconductivity.

With the construction of the winding number, we have shown that its changes coincide with the gap-closing conditions in the complex spectrum, thereby establishing a direct correspondence between the invariant and the emergence of Majorana zero modes. By adding longer-range hopping terms, the system leaves the original class and enters class D of the real AZ classification, characterized by an integer ($\mathcal{Z}$) invariant, while the winding number remains well defined. Since the case without longer-range hopping is a special limit of this more general model,  the stability of Majorana zero modes appears to be tied to the underlying PHS and SLS while TRS plays a less essential role. 

One open question concerns the effects of many-body interactions on the topological phases and the associated zero modes analyzed here. In the absence of pairing, interactions are known to induce unconventional non-Hermitian skin effects residing in the spin sector rather than the charge sector~\cite{Yoshida:2024}. In addition, interactions can reduce point-gap topological classifications and destabilize exceptional points in gapped non-Hermitian systems~\cite{Yoshida:2022,Yoshida:2023}. 
Moreover, a non-Hermitian many-body polarization was proposed as a topological invariant~\cite{Lee:2020} when the non-Hermitian skin modes are suppressed in a many-body system.  
Although a systematic analysis is beyond the scope of this work, it would be interesting to investigate whether a similar interaction-induced reduction occurs for the Majorana zero modes in a superconducting system such as the present model, possibly as a non-Hermitian generalization of the interacting classification discussed in Ref.~\cite{Fidkowski:2010}.

Finally, while its full realization might still be challenging, our model connects to several experimental platforms where both nonreciprocal hopping and dissipation are tunable.
First, as mentioned earlier, non-Hermitian Hamiltonians can be engineered in ultracold atomic systems through controlled loss mechanisms, such as light-induced atom loss or interactions with external reservoirs~\cite{Gou:2020,Liang:2022}.
Second, optical waveguides provide a well-established platform to realize non-Hermitian topological transitions, with gain/loss engineering enabling direct observation of bulk-edge correspondence and spectral winding~\cite{Zeuner:2015}.
Third, electrical circuits offer controllable non-Hermitian parameters and allow topological features to be directly probed via impedance measurements~\cite{Lee:2018,Helbig:2020}.
Finally, mechanical and acoustic metamaterials can reproduce non-Hermitian skin phenomena via asymmetric couplings, and have been used to realize higher-order topological phases~\cite{Kawabata:2018,Zhang:2021,Wang:2022}.

Together, these diverse platforms offer routes to experimentally access and manipulate non-Hermitian lattices, in analogy to the ``poor man’s Majorana modes'' realized in Hermitian setups~\cite{Leijnse:2012,Dvir:2023}. If realized in solid-state systems, additional opportunities arise, as the density profiles of the zero modes could be probed directly by scanning tunneling microscopy (STM) or other spatially resolved techniques, as demonstrated in the Hermitian regime~\cite{Machida:2019},
while the correlations between opposite spin components of the density profile could be detected using spin-resolved STM. These possibilities highlight not only the versatility of non-Hermitian platforms but also the potential to uncover symmetry-enriched Majorana physics in engineered  quantum platforms.

\begin{acknowledgments}
We thank Y.-Y.~Chang, C.~Chin, S.-J.~Cheng, Y.~Kato, S.~Li, R.~Okugawa, Y.-P.~Wang, and T.~Yoshida for interesting discussions. 
We thank H.-C.~Wang for inventing the symbol of $\intra$, which we adopt here to label the intra-unit-cell  terms.
This work was financially supported by the National Science and Technology Council (NSTC), Taiwan, through Grant No.~NSTC-112-2112-M-001-025-MY3 and Grant No.~NSTC-114-2112-M-001-057, and Academia Sinica (AS), Taiwan through Grant No.~AS-iMATE-114-12. We acknowledge the technical support from the Academia Sinica Grid Computing Center (ASGC), Taiwan through Grant No.~AS-CFII-112-103. 
K.S. acknowledges the financial support from JST SPRING (Grant No.~JPMJSP2151). 
N.O. acknowledges support from the Japan Society for the Promotion of Science (JSPS) KAKENHI Grant No.~JP23K03243. 

\end{acknowledgments}

\section*{Data Availability}
The data that support the findings of this study are available at Zenodo~\cite{data}.

\appendix

\section{More details about the system in the absence of the onsite transverse fields }
\label{Appendix:no_hx}

In the main text, we have discussed symmetry relations when the system is subject to a small onsite transverse field. For completeness, here we discuss the case when this onsite transverse field is absent and we have Eq.~\eqref{Eq:H_nHSC} alone.

\subsection{Symmetries }
\label{Appendix:Symmetry}

The model ${H}_{\rm nHSC}^{\rm pbc}$  defined in Eq.~\eqref{Eq:H_pbc2} commutes with the unitary operator,  $\eta^z \tau^0 \sigma^z$, and can therefore be block-diagonalized.
To be specific, we find  
\begin{align} 
 \left(
\begin{array}{cc} 
 {h}_{\rm nHSC}^{+} (k) & 0 \\
 0  & {h}_{\rm nHSC}^{-} (k)
\end{array}
\right),
\end{align}
with  the two 4-by-4 block matrices,
\begin{align} 
    {h}_{\rm nHSC}^{\pm} (k) &=
    i\frac{\Gamma_{0}}{2}\rho^{z}\omega^{0} \pm \Delta_{0}\rho^{x}\omega^{0}
    \nonumber \\
    &~+\Big[ t_{\intra}+t_{\inter}\cos(ka_{0}) \pm i\frac{g_{\inter}}{4}\sin\, (ka_{0}) \Big] \rho^{z}\omega^{x}
    \nonumber \\
    &~+\Bigg\{ t_{\inter}\sin\,(ka_{0}) \pm i\Big[ \frac{g_{\intra}}{4}-\frac{g_{\inter}}{4}\cos(ka_{0}) \Big] \Bigg\} \rho^{z}\omega^{y},
    \label{Eq:h_nHSC subblock}
\end{align}
and new sets of Pauli matrices,  $\rho^{\mu}$ and $\omega^{\mu}$.

The symmetry relations preserved by these blocks are listed in Table~\ref{Table:Symmetries2}, obtained similarly to Table~\ref{Table:Symmetries_with_perturbation}. Both the blocks belong to the AI class in the real AZ classification, characterized by the TRS (squaring to +1) and the absence of PHS (squaring to zero),
as well as the D$^\dagger$ class in the real AZ$^\dagger$ classification, characterized by the absence of TRS$^\dagger$ (squaring to 0) and the presence of PHS$^\dagger$ (squaring to +1). We have SLS with  $U_{S} = \rho^{y} \omega^{0}$, which commutes with the TRS operator: $U_{S} U_{T_+} = + U_{T_+} U_{S}^{*}$, giving rise to topological invariants $Z \oplus Z$, $Z$, and $Z$ for point gap, real line gap, and imaginary line gap, respectively~\cite{Kawabata:2019}.
While we summarize the block properties here for completeness, determining the topological invariant is more efficiently carried out using the Hermitianized Hamiltonian in the presence of the onsite magnetic field, as demonstrated in the main text.

\begin{table}[th]
\centering
\caption{Symmetry and their matrix representations for our system in the absence of the perturbation.
Here, we consider the block ${h}_{\rm nHSC}^{\pm}$  defined in Eq.~\eqref{Eq:h_nHSC subblock}.
}
\renewcommand{\arraystretch}{1.5} 
\begin{tabular}{c c l}
\hline
\hline
\textbf{Symmetry} & \textbf{Symbol} & \textbf{Unitary operators} \\ \hline
TRS & $T_+$ & $U_{T_+} = \rho^{x} \omega^{z}$ \\ \hline
PHS$^{\dagger}$ & $T_{-}$ & $U_{T_-} = \rho^{z} \omega^{z}$ \\ \hline
SLS & $S$ & $U_{S} = \rho^{y} \omega^{0}$ \\  
\hline \hline
\end{tabular}
\label{Table:Symmetries2}
\end{table}

\subsection{Energy spectra  }
\label{Appendix:no-h_x}

We now discuss the energy spectra of Eq.~\eqref{Eq:H_nHSC} without adding the transverse field. 
We remark that, in addition to the spectra from exact diagonalization as presented below, one can also perform imaginary gauge transformation and compute the corresponding spectra; see Appendix~\ref{Appendix:IGT} for details.

To proceed, we consider the following two regimes. 
In Appendix~\ref{Sec:OBC-PBC-Spectra_Region1}, we discuss the regime where Eq.~\eqref{Eq:Case(ii)_criterion}  does not hold, where the gap can close only at $k=0$ or $k= \pm \pi/a_0$, described by the hyperbolic curves in Eq.~\eqref{Eq:Transition_H_nHSC(i)}.
In Appendix~\ref{Sec:OBC-PBC-Spectra_Region2}, on the other hand, we explore the regime in which  Eq.~\eqref{Eq:Case(ii)_criterion}   holds and 
the system exhibits a nodal superconducting phase with the condition given in Eq.~\eqref{Eq:GaplessSC_condition}.

\begin{figure*}[]
    \centering 
\includegraphics[width=0.99\textwidth]{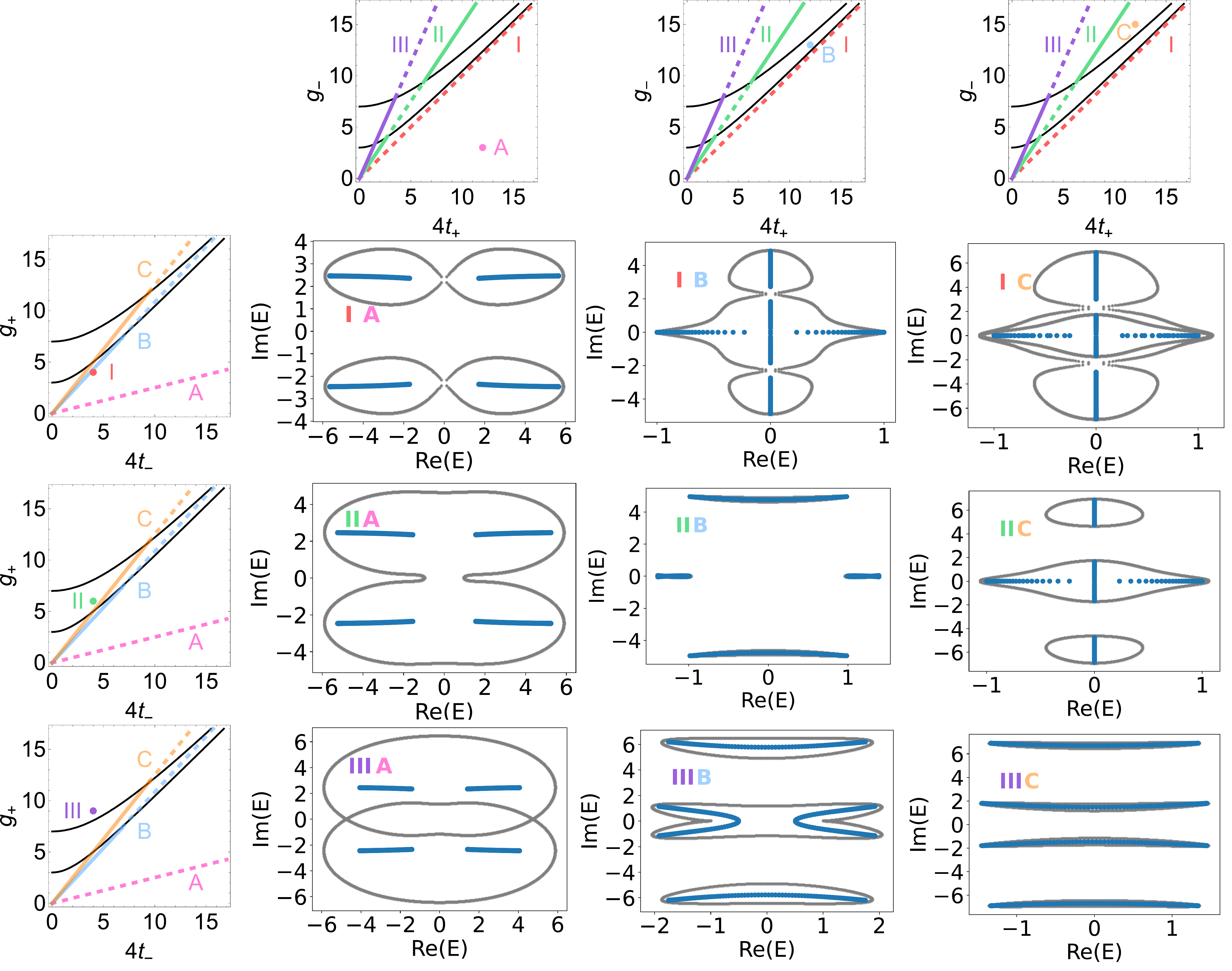}
    \caption{Similar plots to Fig.~\ref{Fig:phase_and_band_diagram_case1_perturbation}, except that we have $\delta h_{x} = 0$ here.
    See Table~\ref{Table:Parameters} for the adopted values of the full parameter sets. 
    }
\label{Fig:phase_and_band_diagram_case1}
    \end{figure*}

\begin{figure*}[]
\centering
\includegraphics[width=0.99\textwidth]{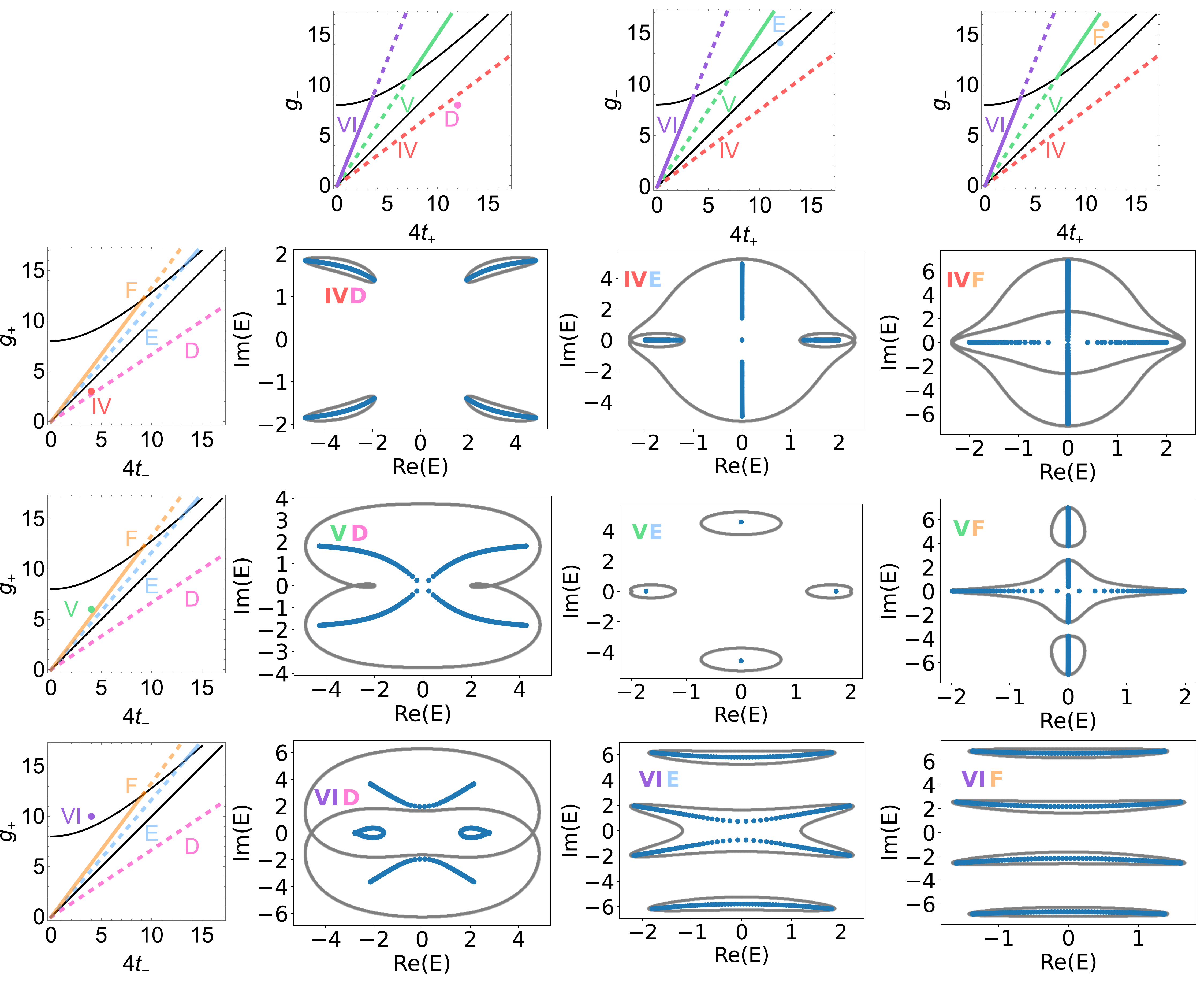}
    \caption{Similar plots to Fig.~\ref{Fig:phase_and_band_diagram_case2_perturbation}, except that we have $\delta h_{x} = 0$ here.
    See Table~\ref{Table:Parameters} for the adopted values of the full parameter sets. 
    }
\label{Fig:phase_and_band_diagram_case2}
\end{figure*}

\subsubsection{In the regime of $ g_+ t_+ \neq g_- t_-$ 
}
\label{Sec:OBC-PBC-Spectra_Region1}

In Fig.~\ref{Fig:phase_and_band_diagram_case1} and Fig.~\ref{Fig:phase_and_band_diagram_case2}, we present the  spectra corresponding to Fig.~\ref{Fig:phase_and_band_diagram_case1_perturbation} and Fig.~\ref{Fig:phase_and_band_diagram_case2_perturbation}, respectively. As discussed in the main text, all the spectra exhibit mirror symmetry with respect to the real and imaginary axes, owing to the pH and SLS.
A key feature is the pronounced difference between the OBC and PBC spectra, which is characteristic of non-Hermitian systems~\cite{Okuma:2020,Okuma:2023}. While the PBC spectra appear similar in both cases, the OBC spectra differ. 

In parallel to the rich PBC spectral features discussed in the main text, we now highlight some features of the OBC spectra. Specifically, we observe that the OBC spectra become either real or purely imaginary for certain parameter sets.
Following Ref.~\cite{Kawabata:2019}, we derive the OBC spectra in the absence of the onsite transverse magnetic field by employing the concept of the generalized Brillouin zone. This approach allows us to determine the conditions under which the spectra are real or purely imaginary, 
\begin{subequations}
\label{Eq:OBC_condition}
\begin{eqnarray}
    \text{~} & \quad &(g_{+}+g_{-})^{2} \ge (t_{+}+t_{-})^{2} \nonumber\\
   & \quad & \& ~~  (g_{+}-g_{-})^{2} \ge  (t_{+}-t_{-})^{2}, \label{Eq:OBC_condition1} \\
    \text{~ }
    & \quad &
    g_{+}g_{-} \ge t_{+}t_{-},  \label{Eq:OBC_condition2}
\end{eqnarray}
\end{subequations}
which are consistent with Fig.~\ref{Fig:phase_and_band_diagram_case1} (see the parameter sets I~B, I~C, and II~C) and Fig.~\ref{Fig:phase_and_band_diagram_case2} (see the sets IV~E, IV~F, V~E, and V~F).
Interestingly, the set V~E of Fig.~\ref{Fig:phase_and_band_diagram_case2} corresponds to
the equality in Eq.~\eqref{Eq:OBC_condition1}, where the OBC spectrum exhibits isolated spectral points. 
We note that this is not captured by the complex flat bands discussed in Eq.~\eqref{Eq:complex_flex_band_condition}, as the latter is derived for the system under the PBC.

\subsubsection{In the regime of $ g_+ t_+ = g_- t_-$
}
\label{Sec:OBC-PBC-Spectra_Region2}

As discussed in the main text,  we have a regime where the gap can close at  momenta other than $k=0$ or $k= \pm \pi/a_0$. 
The precise relation between the gap closure condition and the energy spectrum is presented in Fig.~\ref{fig:energy_spectry_case(iib)} in the main text. 
Here, when the parameters are in the shaded region in Fig.~\ref{fig:energy_spectry_case(iib)}, the conditions in Eq.~\eqref{Eq:GaplessSC_condition} are satisfied and the PBC energy gap closes. In contrast to the regime of $ g_+ t_+ \neq g_- t_-$, however, the PBC spectra here do not form loops with finite area. Without $\delta h_{x}$, we do not find any regimes with zero modes.

\begin{figure}[h]
    \centering
    \includegraphics[width=0.45\linewidth]{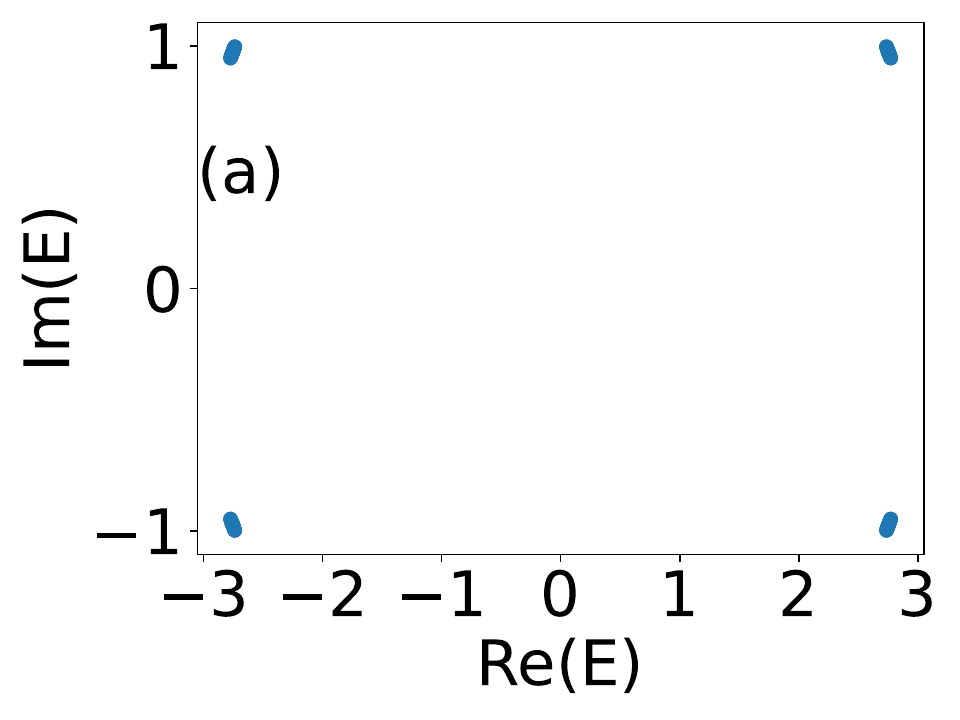}
    \includegraphics[width=0.45\linewidth]{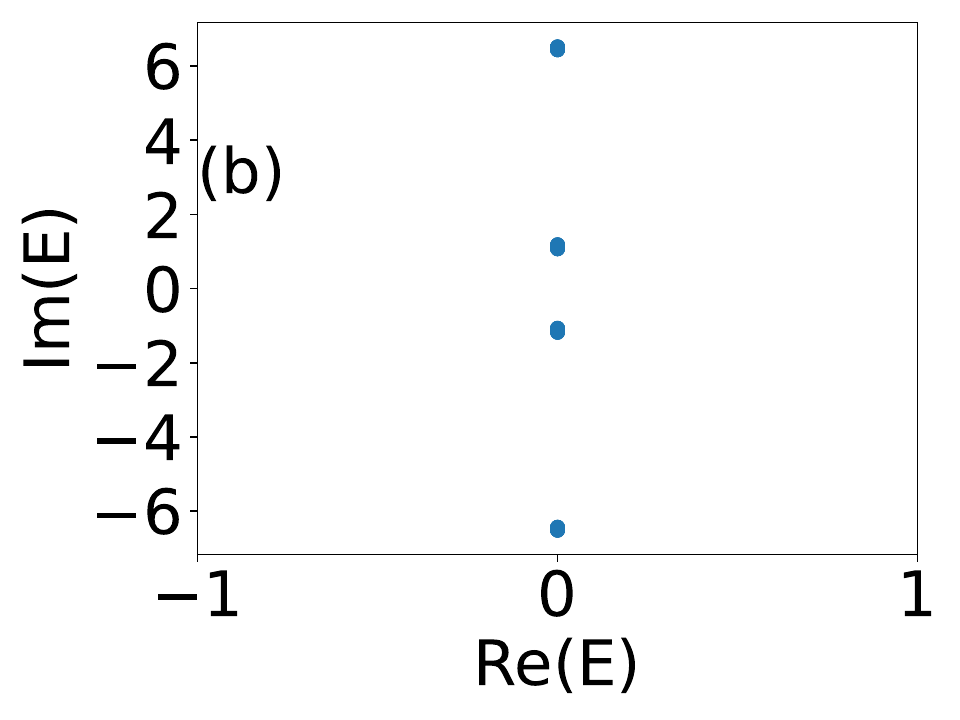}\\
    \includegraphics[width=0.45\linewidth]{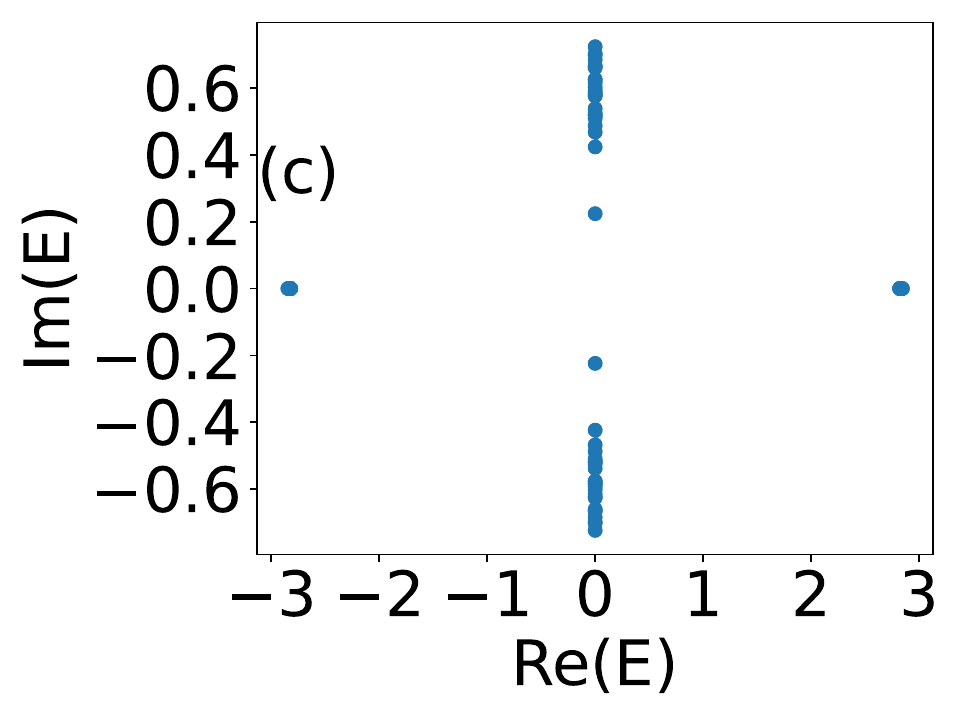}
    \includegraphics[width=0.45\linewidth]{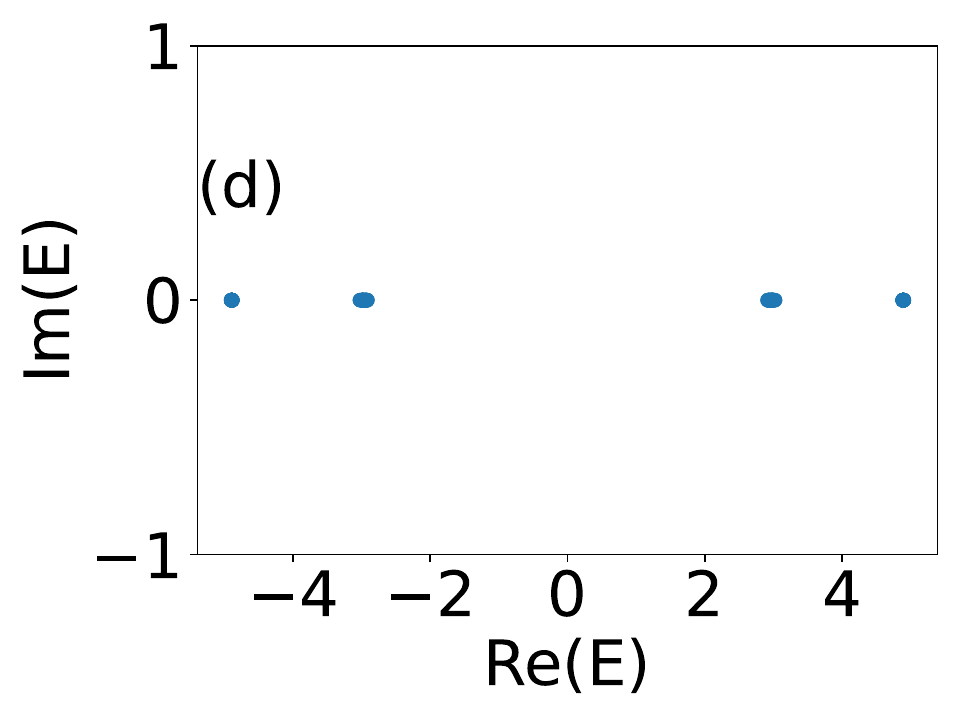}
    \caption{Similar to Fig.~\ref{fig:energy_spectrum_(iso_b)}, except that additional random dissipation terms in Eq.~\eqref{Eq:onsite_dissipation_disorder}, with $2$--$2.5~\%$ randomness. See Table~\ref{Table:Parameters} for the adopted values of the full parameter sets.}
    \label{fig:energy_spectrum_(iso_b)_disorder1}
\end{figure}

\begin{figure}[h]
    \centering
    \includegraphics[width=0.45\linewidth]{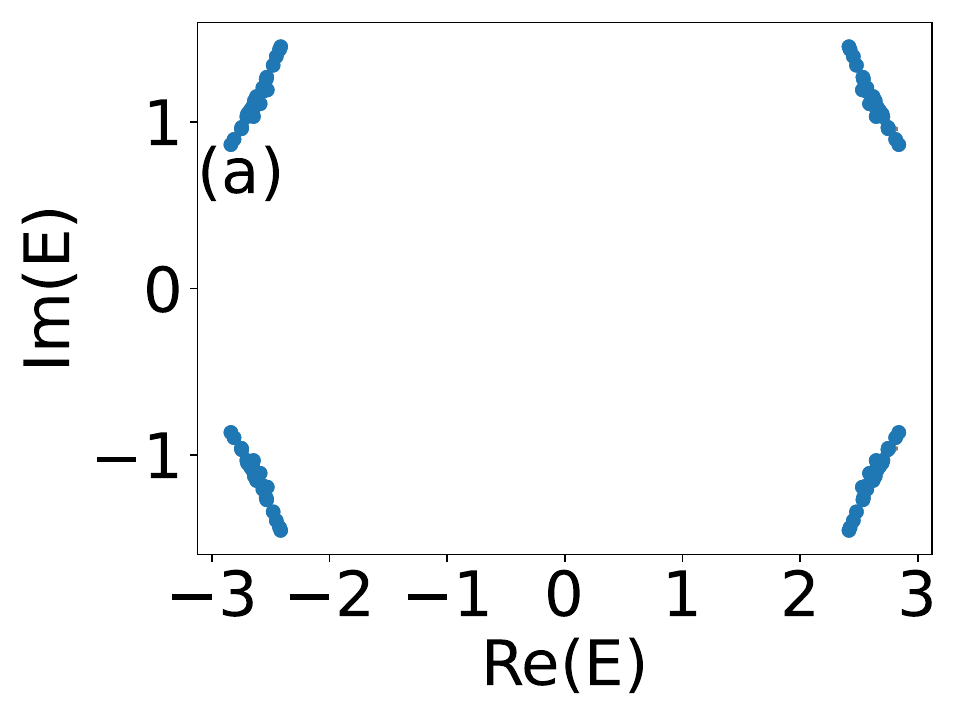}
    \includegraphics[width=0.45\linewidth]{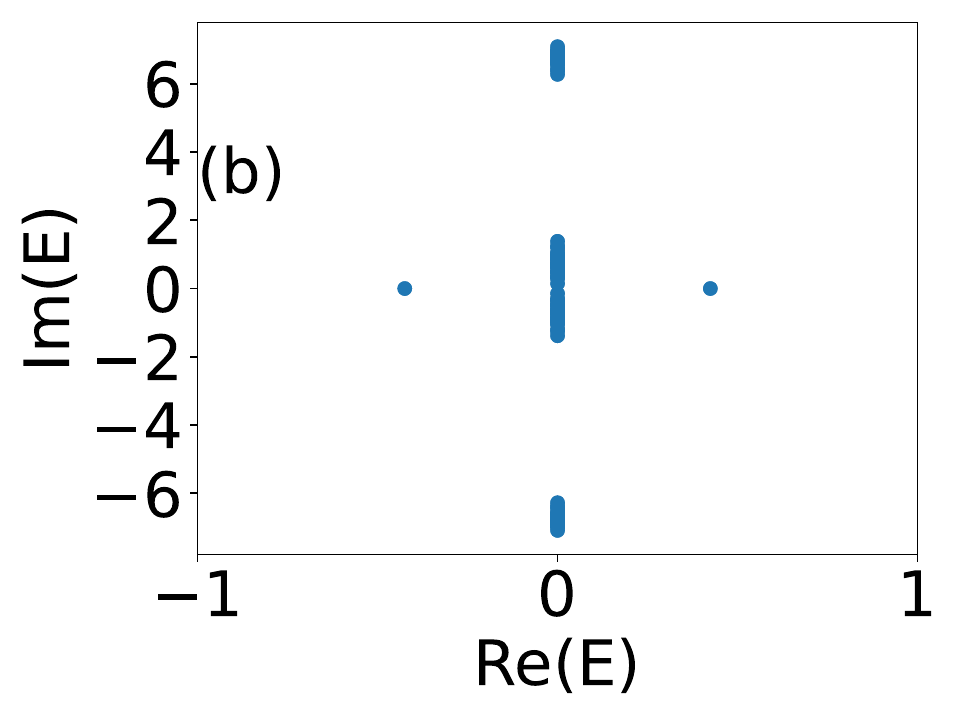}\\
    \includegraphics[width=0.45\linewidth]{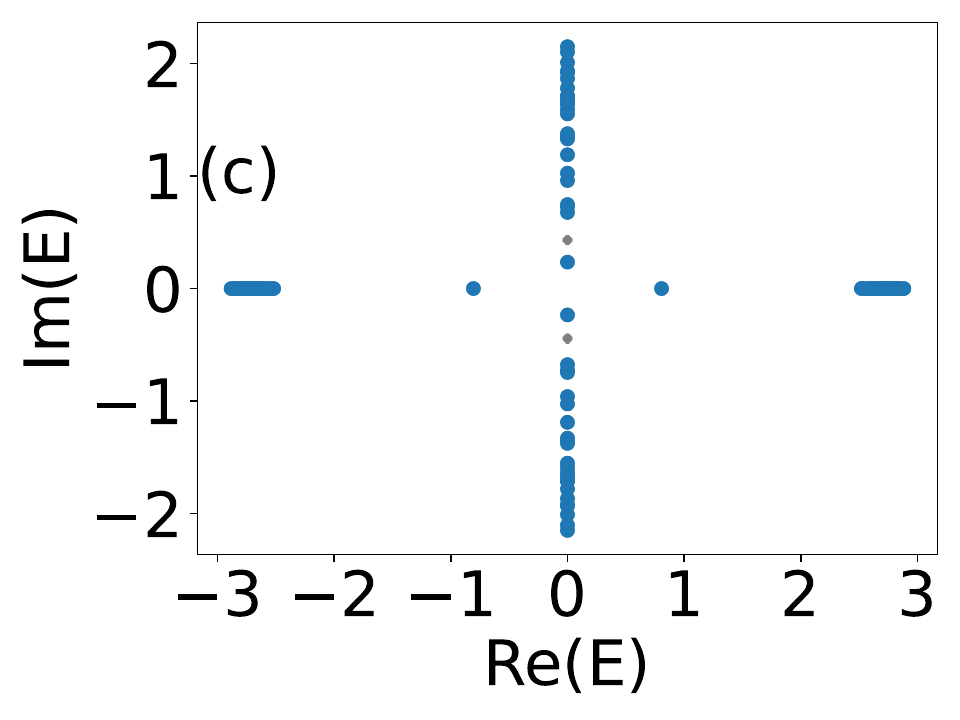}
    \includegraphics[width=0.45\linewidth]{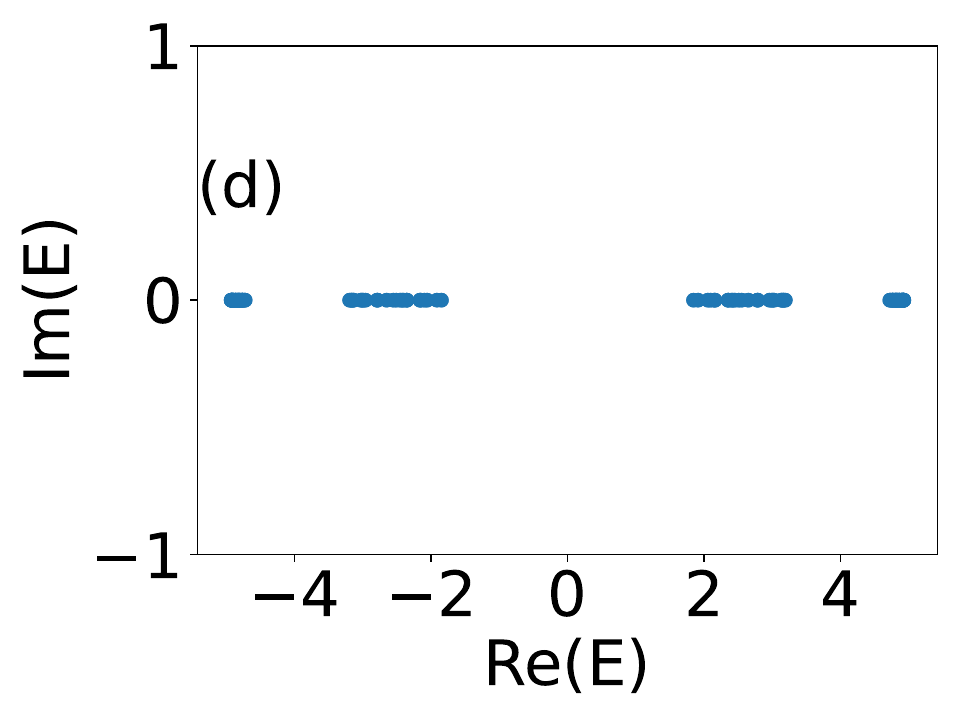}
    \caption{Similar to Fig.~\ref{fig:energy_spectrum_(iso_b)_disorder1}, except for stronger disorder with $20$--$25~\%$ randomness. See Table~\ref{Table:Parameters} for the adopted values of the full parameter sets. 
    }
    \label{fig:energy_spectrum_(iso_b)_disorder2}
\end{figure}

In addition, complex flat bands emerge in this regime, as shown in the main text. Here, we additionally discuss the stability of the complex flat bands emerging in Fig.~\ref{fig:energy_spectrum_(iso_b)}.  
In order to keep the symmetry class unchanged, we incorporate disorder in the form of onsite dissipation terms. 
To this end, we consider the real-space dissipation terms in following form, 
\begin{equation}
\label{Eq:onsite_dissipation_disorder}
\begin{split}
    \Gamma_{j,a} \rightarrow \Gamma_+ + \Gamma_-  + \delta \Gamma_{j,a}\\
    \Gamma_{j,b} \rightarrow \Gamma_+ - \Gamma_- +\delta \Gamma_{j,b}
\end{split}
\end{equation}
where $\delta \Gamma_{j,a}$ and $\delta \Gamma_{j,b}$ denotes random real numbers. 
Taking the spectra in Fig.~\ref{fig:energy_spectrum_(iso_b)} as the primary examples, we include the above disorder and compute the OBC spectra, as shown in Fig.~\ref{fig:energy_spectrum_(iso_b)_disorder1} and Fig.~\ref{fig:energy_spectrum_(iso_b)_disorder2}, representing different disorder strengths.

In Fig.~\ref{fig:energy_spectrum_(iso_b)_disorder1}, we consider $\delta \Gamma_{j,a}$ and $\delta \Gamma_{j,b}$ ranging from $-0.1$ to $0.1$,
which represents a maximum of $2$--$2.5~\%$ randomness as compared to the adopted values of $\Gamma_+=4$ or $\Gamma_+= 5$ in these panels. 
Most of the bands have little changes, except for Fig.~\ref{fig:energy_spectrum_(iso_b)_disorder1}(c), where half of the bands are stretched along the imaginary axis. 
We further increase the disorder strength to a maximum of $20$--$25~\%$ randomness; that is, 
$\delta \Gamma_{j,a}$ and $\delta \Gamma_{j,b}$ $\in [ -1 , 1$]. 
The results are presented in Fig.~\ref{fig:energy_spectrum_(iso_b)_disorder2}, where most of the bands now clearly have finite bandwidths. Again, among the energy spectra and bands, the imaginary bands in Fig.~\ref{fig:energy_spectrum_(iso_b)_disorder2}(c) evolve the most from their clean limit.

\section{Imaginary gauge transformation  }
\label{Appendix:IGT}

In this section we generalize the algebra in Ref.~\cite{Okuma:2019} to perform imaginary gauge transformation after block-diagonalizing our model and use it to compute the spectra.
We start with the Hamiltonian Eq.~\eqref{Eq:H_nHSC} in real space and derive the equation of motion for the fermion fields $a_{j,\sigma}$ and $b_{j,\sigma}$, and recast the differential equation into the following eigenvalue problem,
\begin{align}
&\begin{pmatrix}
H_+ & 0 & 0 & \Delta_{0} I_{2\times 2} \\
0 & H_- & -\Delta_{0} I_{2\times 2} & 0 \\
0 & -\Delta_{0} I_{2\times 2} & -H_- & 0 \\
\Delta_{0} I_{2\times 2} & 0 & 0 & -H_+
\end{pmatrix}
\begin{pmatrix}
\Psi_{j,\uparrow} \\
\Psi_{j,\downarrow}\\
\chi_{j,\uparrow}\\
\chi_{j,\downarrow}
\end{pmatrix} \nonumber \\
&=E
\begin{pmatrix}
\Psi_{j,\uparrow} \\
\Psi_{j,\downarrow}\\
\chi_{j,\uparrow}\\
\chi_{j,\downarrow}
\end{pmatrix},    
\end{align} 
where we introduce the following symbols,
\begin{equation}
\begin{split}
    H&_\pm\equiv
    \begin{pmatrix}
        -\frac{i\Gamma_0}{2} & -J_{\inter,\pm}\delta_--J_{\intra,\mp} \\
        -J_{\inter,\mp}\delta_+-J_{\intra,\pm} & -\frac{i\Gamma_0}{2} \\
    \end{pmatrix},\\
\label{Eq:Hu_Hd}
&J_{\inter, \pm}=t_{\inter} \pm \frac{g_{\inter}}{4},
\quad 
J_{\intra, \pm }=t_{\intra} \pm \frac{g_{\intra}}{4}, 
\end{split}
\end{equation}
and the vectors, 
\begin{equation}
\begin{split} 
    \Psi_{j,\sigma}^{T} =
\begin{pmatrix}
    A_{j,\sigma} & B_{j,\sigma}  
\end{pmatrix},
\quad 
    \chi_{j,\sigma}^{T}
=\begin{pmatrix}
    C_{j,\sigma} & D_{j,\sigma} 
\end{pmatrix}, 
\end{split}
\end{equation}
with the transpose $T$ and the unknown  amplitudes $A_{j,\sigma}$, $B_{j,\sigma}$, $C_{j,\sigma}$ and $D_{j,\sigma}$.
In the above, we have taken the ansatz, $\delta_{+}=A_{j+1,\sigma}/A_{j,\sigma}=C_{j+1,\sigma}/C_{j,\sigma}$ and $\delta_{-}=B_{j-1,\sigma}/B_{j,\sigma}=D_{j-1,\sigma}/D_{j,\sigma}$ when we map the problem to a matrix form. 
We can block-diagonalize the above 8-by-8 matrix into the following form, 
\begin{equation}
\begin{pmatrix}
H_+ & \Delta_0 I_{2 \times 2} & 0 & 0 \\
\Delta_0 I_{2 \times 2} & -H_+ & 0 & 0 \\
0 & 0 & H_- & -\Delta_0 I_{2 \times 2} \\
0 & 0 & -\Delta_0 I_{2 \times 2} & -H_-
\end{pmatrix} ,
\label{Eq:GBZ_matrix}
\end{equation}
where the full energy spectra can be represented as $\pm\sqrt{E_{+,\lambda}^2 + \Delta_0^2}$ and $\pm\sqrt{E_{-,\lambda}^2 + \Delta_0^2}$, with the index $\lambda \in \{ +, - \}$ and the eigenvalues $E_{\pm,\lambda}$ of the blocks $H_{\pm}$.
In what follows, we specify the boundary condition and find the corresponding spectra.

\subsection{PBC spectra}

Under the PBC, one can set $\delta_{\pm}=e^{\pm ika_{0}}$ and  obtain 
\begin{align}
E_{\pm,\lambda}^{\rm pbc} (k) = & -\frac{i\Gamma_0}{2} +\lambda
\Bigg[
\left( J_{\intra,\mp} + J_{\inter,\pm}e^{-ika_0} \right) \nonumber \\
&
\times \left( J_{\intra,\pm} + J_{\inter,\mp}e^{ika_0} \right)
\Bigg]^{\frac{1}{2}} .
\end{align}
It can be shown that the above expression of the full spectrum,
$\pm\sqrt{ \big[E_{\pm,\lambda}^{\rm pbc} (k) \big]^2 + \Delta_0^2}$, recovers  Eq.~\eqref{Eq:general-conditions} in the main text.

\subsection{OBC spectra}

\begin{figure}[h]
    \includegraphics[width=0.99\linewidth]{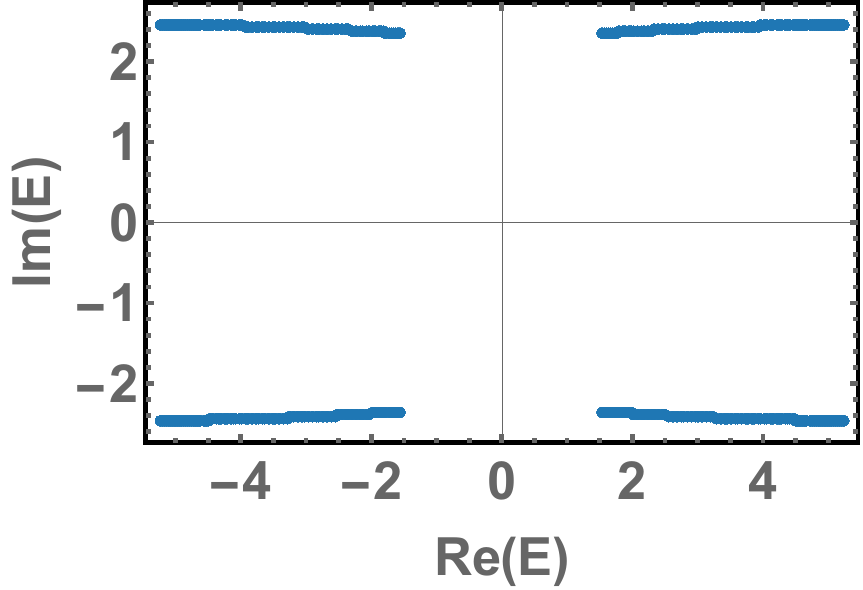}
    \caption{OBC energy spectra, $\pm \sqrt{E_{+,\lambda}^2 + \Delta_0^2}$, calculated from  Eq.~\eqref{Eq:Hu_energy_spectra} for the parameter set II~A.
    See Table~\ref{Table:Parameters} for the adopted values of the full parameter set.    
    } 
    \label{Fig:AII_OBC_spactra_seperate}
\end{figure}

To calculate the OBC spectra, we perform imaginary gauge transformation for the $H_{+}$ and $H_{-}$ blocks separately. We first focus on the matrix $H_{+}$ and transform the basis by adopting
\begin{align}
B_{j,\uparrow} &= \sqrt{\frac{J_{\intra,+}}{J_{\intra,-}}}B^{\prime}_{j,\uparrow} , \\
\delta_{\pm} &= \sqrt{\frac{J_{\intra,\pm}}{J_{\intra,\mp}}\frac{J_{\inter,\pm}}{J_{\inter,\mp}}}\delta^{\prime}_{\pm}. 
\end{align}
With this procedure, the block $H_+$ becomes
\begin{equation}
H_+' = \begin{pmatrix}
h_{11} & h_{12} \\
h_{21} & h_{22}
\end{pmatrix},
\label{Eq:recurrence_relation_Eu}
\end{equation}
where the matrix elements are defined as
\begin{align}
h_{11} &= \frac{-i\Gamma_0}{2},\\
h_{12} &= -\sqrt{J_{\inter,+}J_{\inter,-}}\delta^{\prime}_{-}-\sqrt{J_{\intra,-}J_{\intra,+}}, \\
h_{21} &=  -\sqrt{J_{\inter,-}J_{\inter,+}}\delta^{\prime}_{+} -\sqrt{J_{\intra,+}J_{\intra,-}}, \\
h_{22} &= \frac{-i\Gamma_0}{2},
\end{align}
with basis $(A_{j,\uparrow},B^{\prime}_{j,\uparrow})$,
and has the eigenvalues, 
\begin{equation}
\begin{split}
    E_{+,\lambda}=-i&\frac{\Gamma_{0}}{2}
    + \lambda \Big[ J_{\inter,+}J_{\inter,-}+J_{\intra,+}J_{\intra,-}\\
&+\sqrt{J_{\intra,+}J_{\intra,-}J_{\inter,+}J_{\inter,-}}(\delta^{\prime}_{+}+\delta^{\prime}_{-}) \Big]^{1/2} .
\label{Eq:Hu_energy_spectra}
\end{split}
\end{equation} 
After the mapping, we can directly calculate the OBC spectrum by setting $\delta^{\prime}_{ \pm}$  to $e^{ \pm i k a_0}$. In Fig.~\ref{Fig:AII_OBC_spactra_seperate}, we show the energy spectra corresponding to parameter set II~A. Here we plot the four bands, $\pm\sqrt{ E_{+,\lambda}  ^2+\Delta_0^2}$, showing spectra consistent with Fig.~\ref{Fig:phase_and_band_diagram_case1}.

So far, we have focused on the $H_{+}$ block. 
For the $H_{-}$ block, we perform the following imaginary gauge transformation,
\begin{align}
B_{j,\downarrow} &= \sqrt{\frac{J_{\intra,-}}{J_{\intra,+}}}B^{\prime}_{j,\downarrow} , \\
\delta_{\pm} &= \sqrt{\frac{J_{\intra,\mp}}{J_{\intra,\pm}}\frac{J_{\inter,\mp}}{J_{\inter,\pm}}}\delta^{\prime\prime}_{\pm}. 
\end{align}
One thus follows a similar procedure to obtain the energy bands,
$\pm\sqrt{E_{-,\lambda}^2+\Delta_0^2}$, which turn out to be identical to the spectra of $H_{+}$.

\section{PBC spectrum of the block Hamiltonian in Eq.~\eqref{Eq:H_blocks-for-winding} }

\label{Appendix:Winding}

\begin{figure*}
    \centering
    \includegraphics[width=0.99\linewidth]{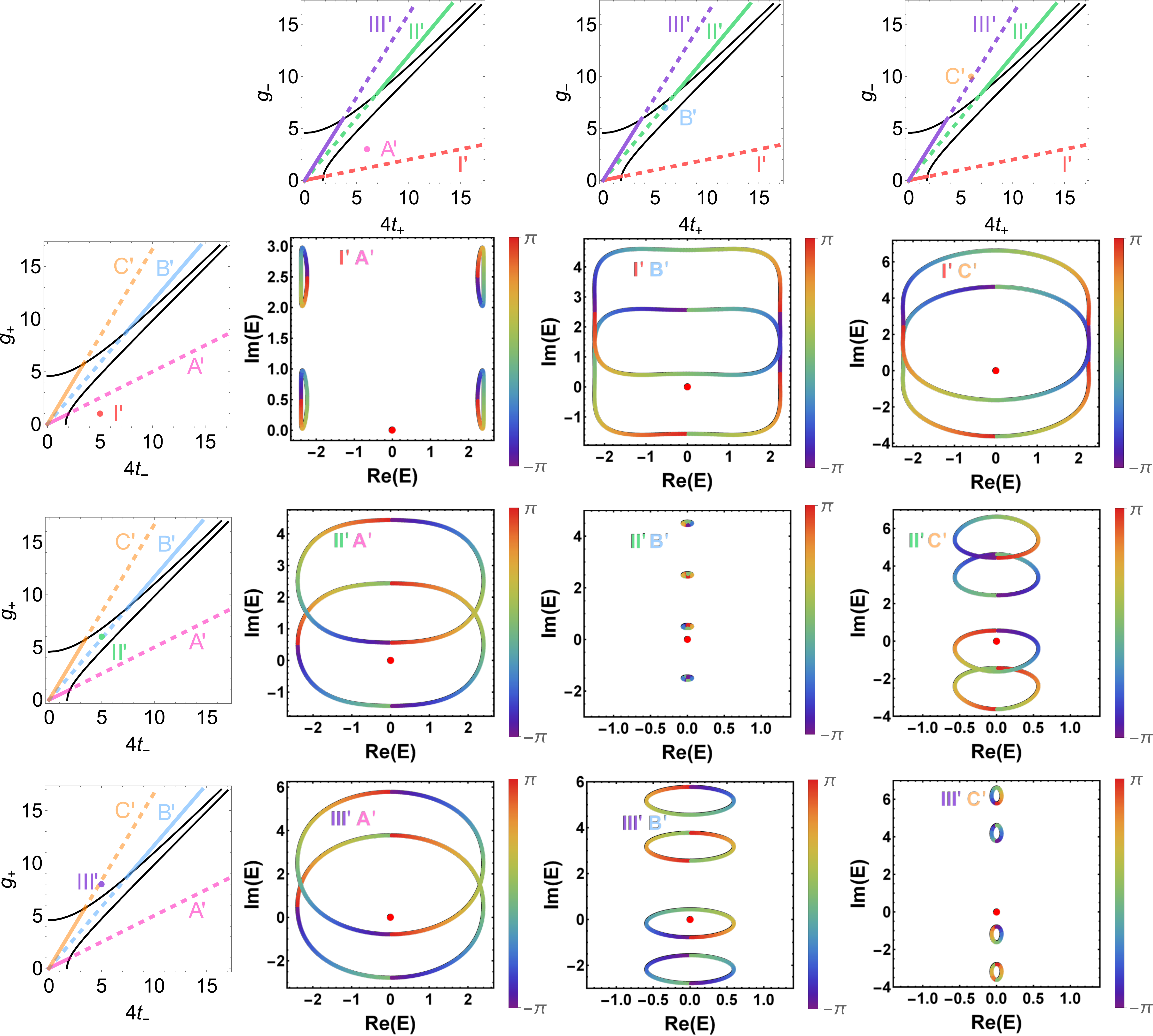}
    \caption{PBC  spectra for   $H_{+,+}$ defined in Eq.~\eqref{Eq:H_blocks-for-winding} with the adopted parameter values identical to those in Fig.~\ref{Fig:phase_and_band_diagram_general_perturbation}. 
    In each panel, the color map marks the spectral trajectory corresponding to the $ka_{0}$ value going from $-\pi$ to $\pi$, and the red dot marks the origin in the complex energy plane.
    See Table~\ref{Table:Parameters} for the adopted values of the full parameter sets. 
    } 
    \label{Fig:Block_band}
\end{figure*}

In this section we discuss the PBC spectra of the block $H_{+,+}$ defined in Eq.~\eqref{Eq:H_blocks-for-winding}.
In Fig.~\ref{Fig:Block_band} we show its PBC spectra, adopting parameter values corresponding to the spectra of Fig.~\ref{Fig:phase_and_band_diagram_general_perturbation}.  
The computed value from  Eq.~\eqref{Eq:winding} consistently matches the number of the spectral trajectory winds around the origin across the panels in Fig.~\ref{Fig:Block_band}. The parameter regions with a nonzero winding number are also consistent with those hosting Majorana zero modes.

\section{Adopted parameter sets throughout this work }
\label{Appendix:parameter}

In Table~\ref{Table:Parameters}, we list the adopted values of the parameter sets for the numerics throughout this work.

\begin{table}[th]
\centering
\caption{Parameter sets adopted for the numerics throughout this work.
}

\begin{tabular}{l l c c c c}
\hline
\hline
\multicolumn{2}{ c }{ \textbf{Figure}}
  & \multicolumn{4}{ c }{ \textbf{Parameter values}}\\ 
  &   &\textbf{$(4t_{+},g_{-})$} & \textbf{$(4t_{-},g_{+})$} & \textbf{$(\Gamma_{+},\Gamma_{-},\Delta_{0})$} & 
\textbf{$\delta h_{x}$} \\
\hline
Fig.~\ref{fig:energy_spectry_case(iib)} & Panel~(c) & $(4,3)$ &  $(2,1.5)$ & $(5,0,1)$ & 0 \\ 
& Panel~(d) & $(4,4)$ &  $(2,2)$ & $(5,0,1)$ & 0 \\
& Panel~(e) & $(4,6)$ & $(2,3)$ & $(5,0,1)$ & 0 \\ 
& Panel~(f) & $(4,7.6)$ & $(2,3.8)$ & $(5,0,1)$ & 0 \\ 
& Panel~(g) & $(4,9)$ & $(2,4.5)$ &  $(5,0,1)$ & 0 \\ 
& Panel~(h) & $(4,15)$ & $(2,7.5)$ & $(5,0,1)$ & 0 \\ \hline
Fig.~\ref{fig:energy_spectrum_(iso_b)} & 
Panel~(a) & $(4,3)$ & $(4,3)$ & $(4,0,3)$ & 0  \\ 
(see also & Panel~(b) & $(4,9)$ & $(4,9)$ & $(5,0,1)$ & $0$ \\ 
footnote~\textnormal{\footnotemark[1]}.)& Panel~(c) & $(4,4.5)$ & $(4,4.5)$ & $(4,0,3)$ & $0$ \\ 
& Panel~(d) & $(4,5)$ & $(4,5)$ & $(5,0,5)$ & $0$ \\ \hline
 Fig.~\ref{Fig:phase_and_band_diagram_case1_perturbation}
 & Set~A & $(12,3)$ & -- & $(5,0,1)$ & $0.01$ \\ 
(see also   
 & Set~B & $(12,13)$ & -- & $(5,0,1)$ & $0.01$ \\ 
footnote~\textnormal{\footnotemark[2]}.)
& Set~C & $(12,15)$ & -- & $(5,0,1)$ & $0.01$ \\ 
 & Set~I & -- & $(4,4)$ & $(5,0,1)$ & $0.01$ \\
 & Set~II & -- & $(4,6)$ & $(5,0,1)$ & $0.01$ \\ 
 & Set~III & -- & $(4,9)$ & $(5,0,1)$ & $0.01$ \\ \hline
 Fig.~\ref{Fig:phase_and_band_diagram_case2_perturbation}
 
 & Set~D & $(12,8)$ & -- & $(4,0,2)$ & $0.01$ \\ 
 (see also   
 & Set~E & $(12,14)$ & -- & $(4,0,2)$ & $0.01$   \\
 footnote~\textnormal{\footnotemark[3]}.)
 & Set~F & $(12,16)$ & -- & $(4,0,2)$ & $0.01$   \\
 & Set~IV & -- & $(4,3)$ & $(4,0,2)$ & $0.01$   \\
 & Set~V & -- & $(4,6)$ & $(4,0,2)$ & $0.01$  \\
 & Set~VI & -- & $(4,10)$ & $(4,0,2)$ & $0.01$  \\
 \hline
Fig.~\ref{Fig:B,I,per,state,n=200} & & $(12,3)$  & $(4,6)$  & $(5,0,1)$ & $0.01$ \\
\hline
 Fig.~\ref{Fig:phase_and_band_diagram_general_perturbation}
 & 
 Set~A$^\prime$ & $(6,3)$ & -- & $(3,2,1)$ & $0.01$ \\ 
 (See also & Set~B$^\prime$ & $(6,7)$ & -- & $(3,2,1)$ & $0.01$ \\ 
footnote~\textnormal{\footnotemark[4]}.)  & Set~C$^\prime$ & $(6,10)$ & -- & $(3,2,1)$ & $0.01$ \\ 
 & Set~I$^\prime$ & -- & $(5,1)$ & $(3,2,1)$ & $0.01$ \\
 & Set~II$^\prime$ & -- & $(5,6)$ & $(3,2,1)$ & $0.01$ \\ 
 & Set~III$^\prime$ & -- & $(5,8)$ & $(3,2,1)$ & $0.01$ \\ \hline
 Fig.~\ref{fig:energy_spectrum_general_(iso_a)} &  & $(6,6)$ & $(5,5)$ & $(2,3,1)$ & $0.01$ \\  \hline
 Fig.~\ref{fig:general_gamma_gap_close_case(ii)_spectra} & 
 Panel~(a) & $(5,2)$ & $(2.5,1)$ & $(3,2,1)$ & $0$ \\ 
& Panel~(b) & $(5,4)$ & $(2.5,2)$ & $(3,2,1)$ & $0$ \\ 
& Panel~(c) & $(5,6)$ & $(2.5,3)$ & $(3,2,1)$ & $0$ \\
& Panel~(d) & $(5,9)$ & $(2.5,4.5)$ & $(3,2,1)$ & $0$ \\ 
& Panel~(e) & $(5,12)$ & $(2.5,6)$ & $(3,2,1)$ & $0$ \\ 
\hline 
Fig.~\ref{fig:energy_spectra_GammaA=-GammaB} & Set~G & $(8,5)$ & -- & $(0,5,1)$ & $0.01$ \\ 
 & Set~H & $(8,10)$ & -- & $(0,5,1)$ & $0.01$ \\ 
 & Set~VII & -- & $(6,3)$ & $(0,5,1)$ & $0.01$\\ 
 & Set~VIII & -- & $(6,5)$ & $(0,5,1)$ & $0.01$ \\ 
\hline 
Fig.~\ref{Fig:AII_OBC_spactra_seperate} &   & $(12,3)$ & $(4,6)$ & $(5,0,1)$ & $0$ \\ 
\hline \hline
\end{tabular}
\footnotetext[1]{The parameter sets used here are identical to those in Fig.~\ref{fig:energy_spectrum_(iso_b)_disorder1}--\ref{fig:energy_spectrum_(iso_b)_disorder2}, except for additional random dissipation terms.}
\footnotetext[2]{The parameter sets used here are identical to those in Fig.~\ref{Fig:phase_and_band_diagram_case1}, except for $\delta h_x = 0$ in the latter.}
\footnotetext[3]{The parameter sets used here are identical to those in Fig.~\ref{Fig:phase_and_band_diagram_case2}, except for $\delta h_x = 0$ in the latter.}
\footnotetext[4]{The parameter sets used here are identical to those in Fig.~\ref{Fig:Block_band}, except for $\delta h_x = 0$ in the latter.}
\label{Table:Parameters}
\end{table}

\end{document}